# The State of AI Ethics
## June 2020

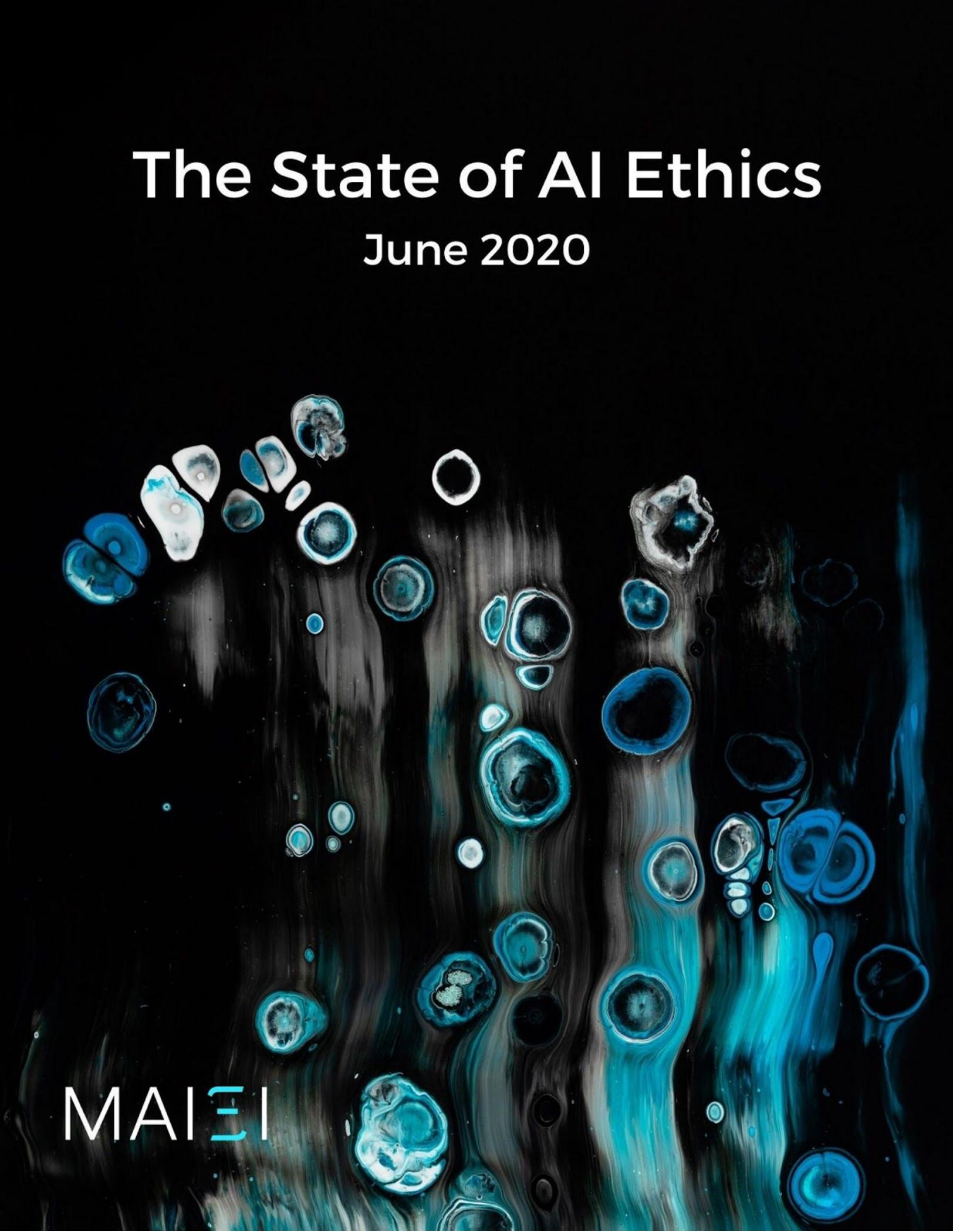

This report was prepared by the Montreal AI Ethics Institute — an international non-profit research institute helping people understand the societal impacts of AI and equipping them to take action. **Learn more at montrealethics.ai**

The content in this report is based on our weekly AI Ethics newsletter and other ongoing research at the institute.
**Subscribe to the newsletter at aiethics.substack.com**

This work is licensed under a **Creative Commons Attribution 4.0 International License**.

Primary contact for the report: **Abhishek Gupta (abhishek@montrealethics.ai)**

The full team behind the report:

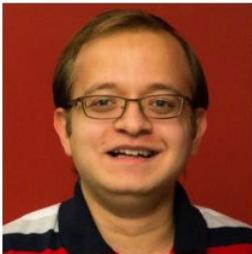

Abhishek Gupta
FOUNDER

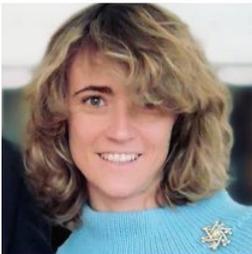

Marianna Ganapini
RESEARCHER (PHD)

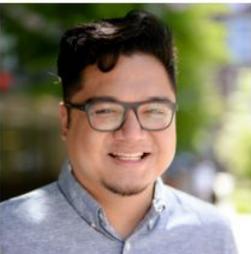

Renjie Butalid
CO-FOUNDER

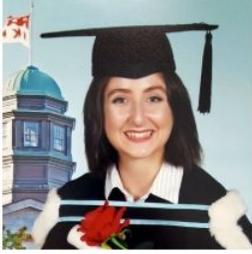

Camylle Lanteigne
RESEARCHER, WORKSHOP FACILITATOR

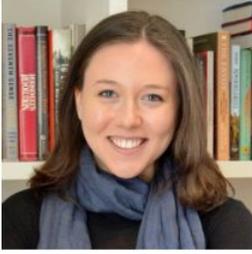

Allison Cohen
RESEARCHER

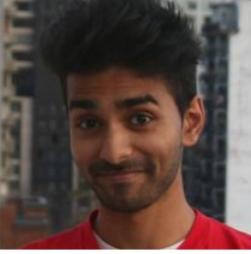

Mo Akif
DIRECTOR OF COMMUNICATIONS

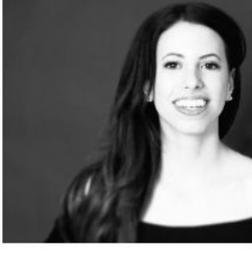

Tania De Gasperis
RESEARCHER, WORKSHOP FACILITATOR

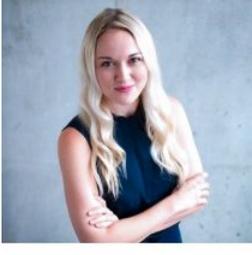

Victoria Heath
RESEARCHER

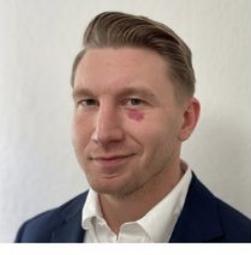

Erick Galinkin
RESEARCHER

# Table of Contents

# *Note that all the original sources we have used for this report are listed at [the end.](#) The work in the following pages combines summarization of the material supplemented with insights from the research staff.

# Introduction

These past few months have been especially challenging, and the deployment of technology in ways hitherto untested at an unrivaled pace has left the internet and technology watchers aghast. Artificial intelligence has become the byword for technological progress and is being used in everything from helping us combat the COVID-19 pandemic to nudging our attention in different directions as we all spend increasingly larger amounts of time online.

It has never been more important that we keep a sharp eye out on the development of this field and how it is shaping our society and interactions with each other. With this inaugural edition of the **State of AI Ethics** we hope to bring forward the most important developments that caught our attention at the Montreal AI Ethics Institute this past quarter. Our goal is to help you navigate this ever-evolving field swiftly and allow you and your organization to make informed decisions.

This pulse-check for the state of discourse, research, and development is geared towards researchers and practitioners alike who are making decisions on behalf of their organizations in considering the societal impacts of AI-enabled solutions.

We cover a wide set of areas in this report spanning ***Agency and Responsibility, Security and Risk, Disinformation, Jobs and Labor, the Future of AI Ethics***, and more. Our staff has worked tirelessly over the past quarter surfacing signal from the noise so that you are equipped with the right tools and knowledge to confidently tread this complex yet consequential domain.

To stay up-to-date with the work at MAIEI, including our public competence building, we encourage you to stay tuned on https://montrealethics.ai which has information on all of our research.

We hope you find this useful and look forward to hearing from you!

Wishing you well,
Abhishek Gupta
Founder, Montreal AI Ethics Institute

# 1. Agency and Responsibility

## Go Deep: Research Summaries

### Robot Rights? Let's Talk About Human Welfare Instead

The debate when ethicists ask for rights to be granted to robots is based on notions of biological chauvinism and that if robots display the same level of agency and autonomy, not doing so would not only be unethical but also cause a setback for the rights that were denied to disadvantaged groups. By branding robots as slaves and implying that they don't deserve rights has fatal flaws in that they both use a term, slave, that has connotations that have significantly harmed people in the past and also that dehumanization of robots is not possible because it assumes that they are not human to begin with.

While it may be possible to build a sentient robot in the distant future, in such a case there would be no reason to not grant it rights but until then, real, present problems are being ignored for imaginary future ones. The relationship between machines and humans is tightly intertwined but it's not symmetrical and hence we must not confound the "being" part of human beings with the characteristics of present technological artifacts.

Technologists assume that since there is a dualism to a human being, in the sense of the mind and the body, then it maps neatly such that the software is the mind and the robot body maps to the physical body of a human, which leads them to believe that a sentient robot, in our image, can be constructed, it's just a very complex configuration that we haven't completely figured out yet. The more representative view of thinking about robots at present is to see them as objects that inhabit our physical and social spaces.

Objects in our environment take on meaning based on the purpose they serve to us, such as a park bench meaning one thing to a skateboarder and another to a casual park visitor. Similarly, our social interactions are always situated within a larger ecosystem and that needs to be taken into consideration when thinking about the interactions between humans and objects. In other words, things are what they are, because of the way they configure our social practices and how technology extends the biological body.Our conception of human beings, then, is that we are and have always been fully embedded and enmeshed with our designed surroundings, and that we are critically dependent on tl embeddedness for sustaining ourselves.

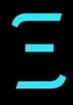

Because of this deep embedding, instead of seeing the objects around us merely as machines or on the other end as 'intelligent others', we must realize that they are very much a part of ourselves because of the important role they play in defining both our physical and social existence.

Some argue that robots take on a greater meaning when they are in a social context like care robots and people might be attached to them, yet that is quite similar to the attachment one develops to other artifacts like a nice espresso machine or a treasured object handed down for generations. They have meaning to the person but that doesn't mean that the robot, as present technology, needs to be granted rights.

While a comparison to slaves and other disenfranchised groups is made when robots are denied rights because they are seen as 'less' than others, one mustn't forget that it happens to be the case that it is so because they are perceived as instruments and means to achieve an end. By comparing these groups to robots, one dehumanizes actual human beings. It may be called anthropocentric to deny rights to robots but that's what needs to be done: to center on the welfare of humans rather than inanimate machines.

An interesting analogue that drives home the point when thinking about this is the Milgram Prison experiment where subjects who thought they had inflicted harms on the actors, who were a part of the experiment, were traumatized even after being told that the screams they heard were from the actors. From an outside perspective, we may say that no harm was done because they were just actors but to the person who was the subject of the experiment, the experience was real and not an illusion and it had real consequences. In our discussion, the robot is an actor and if we treat it poorly, then that reflects more so on our interactions with other artifacts than on whether robots are "deserving" of rights or not. Taking care of various artifacts can be thought of as something that is done to render respect to the human creators and the effort that they expended to create it.

Discussion of robot rights for an imaginary future that may or may not arrive takes away focus and perhaps resources from the harms being done to real humans today as part of the AI systems being built with bias and fairness issues in them. Invasion of privacy, bias against the disadvantaged, among other issues are just some of the few already existing harms that are being leveled on humans as intelligent systems percolate into the everyday fabric of social and economic life.

From a for-profit perspective, such systems are poised and deployed with the air of boosting the bottom line without necessarily considering the harms that emerge as a consequence. In pro-social contexts, they are seen as a quick fix solution to inherently messy and complex problems.

The most profound technologies are those that disappear into the background and in subtle ways shape and form our existence. We already see that with intelligent systems pervading many aspects of our lives. So we're not as much in threat from a system like Sophia which is a rudimentary chatbot hidden behind a facade of flashy machinery but more so from Roomba which impacts us more and could be used as a tool to surveil our homes. Taking ethical concerns seriously means considering the impact of weaving in automated technology into daily life and how the marginalized are disproportionately harmed.

In the current dominant paradigm of supervised machine learning, the systems aren't truly autonomous, there is a huge amount of human input that goes into enabling the functioning of the system, and thus we actually have human-machine systems rather than just pure machinic systems. The more impressive the system seems, the more likely that there was a ton of human labor that went into making it possible. Sometimes, we even see systems that started off with a different purpose such as reCAPTCHA that are used to prevent spam being refitted to train ML systems. The building of AI systems today doesn't just require highly skilled human labor but it must be supplemented with mundane jobs of labeling data that are poorly compensated and involve increasingly harder tasks as, for example, image recognition systems become more powerful, leading to the labeling of more and more complex images which require greater effort. This also frames the humans doing the low skilled work squarely in the category of being dehumanized because of them being used as a means to an end without adequate respect, compensation and dignity.

An illustrative example where robots and welfare of humans comes into conflict was when a wheelchair user wasn't able to access the sidewalk because it was blocked by a robot and she mentioned that without building for considering the needs of humans, especially those with special needs, we'll have to make debilitating compromises in our shared physical and social spaces. Ultimately, realizing the goals of the domain of AI ethics needs to reposition our focus on humans and their welfare, especially when conflicts arise between the "needs" of automated systems compared to those of humans.

# Go Wide: Article Summaries

## We Asked an A.I. to Write a Column for Us. The Results Were Wild

What happens when AI starts to take over the more creative domains of human endeavour? Are we ready for a future where our last bastion, the creative pursuit, against the rise of machines is violently snatched away from us? In a fitting start to feeling bereft in the times of global turmoil, this article starts off with a story created by a machine learning model called GPT-2 that utilizes training data from more than 8 million documents online and predicts iteratively the next word in a sentence given a prompt. The story is about "Life in the Time of Coronavirus" that paints a desolate and isolating picture of a parent who is following his daily routine and feels different because of all the changes happening around them. While the short story takes weird turns and is not completely coherent, it does give an eerie feeling that blurs the line between what could be perceived as something written by a human compared to that by a machine.

A news-styled article on the use of facial recognition systems for law enforcement sounds very believable if presented outside of the context of the article. The final story, a fictional narrative, presents a fractured, jumpy storyline of a girl with a box that has hallucinatory tones to its storytelling. The range of examples from this system is impressive but it also highlights how much further these systems have to go before they can credibly take over jobs. That said, there is potential to spread disinformation via snippets like the second example we mention and hence, something to keep in mind as you read things online.

## Computers Do Not Make Art, People Do

Technology, in its widest possible sense, has been used as a tool to supplement the creative process of an artist, aiding them in exploring the adjacent possible in the creative phasespace. For decades we've had computer scientists and artists working together to create software that can generate pieces of art that are based on procedural rules, random perturbations of the audience's input and more. Off late, we've had an explosion in the use of AI to do the same, with the whole ecosystem being accelerated as people collide with each other serendipitously on platforms like Twitter creating new art at a very rapid pace. But, a lot of people have been debating whether these autonomous systems can be attributed artistic agency and if they can be called artists in their own right. The author here argues that it isn't the case because even with the push into using technology that is more automated than other tools we've used in the past, there is more to be said abc the artistic process than the simple mechanics of creating the artwork. Drawing on art history and other domains, there is an argument to be made as to what art really is - there are strong arguments in support of it playing a role in servicing social relationships between two entities. We, as humans, already do that with things like exchanging gifts, romance, conversation and other forms of social engagement where the goal is to alter the social relationships. Thus, the creative process is more so a co-ownership oriented model where the two entities are jointly working together to create something that alters the social fabric between them.

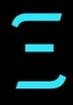

As much as we'd like to think some of the AI-enabled tools today have agency, that isn't necessarily the case when we pop open the hood and see that it is ultimately just software that for the most part still relies heavily on humans setting goals and guiding it to perform tasks. While human-level AI might be possible in the distant future, for now the AI-enabled tools can't be called artists and are merely tools that open up new frontiers for exploration. This was the case with the advent of the camera that de-emphasized the realistic paint form and spurred the movement towards modern art in a sense where the artists are more focused on abstract ideas that enable them to express themselves in novel ways. Art doesn't even have to be a tangible object but it can be an experience that is created. Ultimately, many technological innovations in the past have been branded as having the potential to destroy the existing art culture but they've only given birth to new ideas and imaginings that allow people to express themselves and open up that expression to a wider set of people.

# 2. Bias and Algorithmic Injustice

## Go Deep: Research Summaries

### Learning to Diversify from Human Judgments – Research Directions and Open Challenges

Ranking and retrieval systems for presenting content to consumers are geared towards enhancing user satisfaction, as defined by the platform companies which usually entails some form of profit-maximization motive, but they end up reflecting and reinforcing societal biases, disproportionately harming the already marginalized.

In fairness techniques applied today, the outcomes are focused on the distributions in the result set and the categorical structures and the process of associating values with the categories is usually de-centered. Instead, the authors advocate for a framework that does away with rigid, discrete, and ascribed categories and looks at subjective ones derived from a large pool of diverse individuals. Focusing on visual media, this work aims to bust open the problem of underrepresentation of various groups in this set that can render harm on to the groups by deepening social inequities and oppressive world views. Given that a lot of the content that people interact with online is governed by automated algorithmic systems, they end up influencing significantly the cultural identities of people.

While there are some efforts to apply the notion of diversity to ranking and retrieval systems, they usually look at it from an algorithmic perspective and strip it of the deep cultural and contextual social meanings, instead choosing to reference arbitrary heterogeneity. Demographic parity and equalized odds are some examples of this approach that apply the notion of social choice to score the diversity of data. Yet, increasing the diversity, say along gender lines, falls into the challenge of getting the question of representation right, especially trying to reduce gender and race into discrete categories that are one-dimensional, third-party and algorithmically ascribed.

The authors instead propose sourcing this information from the individuals themselves such that they have the flexibility to determine if they feel sufficiently represented in the result set. This is contrasted with the degree of sensitive attributes that are present in the result sets which is what prior approaches have focused on. From an algorithmic perspective, the authors advocate for the use of a technique called determinantal point process (DPP) that assigns a higher probability score to sets that have higher spreads based on a predefined distan metric.

How DPP works is that for items that the individual feels represents them well, the algorithm clusters those points closer together, for points that they feel don't represent them well, it moves those away from the ones that represent them well in the embedding space. Optimizing for the triplet loss helps to achieve the goals of doing this separation.

But, the proposed framework still leaves open the question of sourcing in a reliable manner these ratings from the individuals about what represents and doesn't represent them well and then encoding them in a manner that is amenable to being learned by an algorithmic system.

While large-scale crowdsourcing platforms which are the norm in seeking such ratings in the machine learning world, given that their current structuring precludes raters' identities and perceptions from consideration, this framing becomes particularly challenging in terms of being able to specify the rater pool. Nonetheless, the presented framework provides an interesting research direction such that we can obtain more representation and inclusion in the algorithmic systems that we build.

## Suckers List: How Allstate's Secret Auto Insurance Algorithm Squeezes Big Spenders

In Maryland, Allstate, an auto insurer, filed with the regulators that the premium rates needed to be updated because they were charging prices that were severely outdated. They suggested that not all insurance premiums be updated at once but instead follow recommendations based on an advanced algorithmic system that would be able to provide deeper insights into the pricing that would be more appropriate for each customer based on the risk that they would file a claim. This was supposed to be based on a constellation of data points collected by the company from a variety of sources.

Because of the demand from the regulators for documentation supporting their claim, they submitted thousands of pages of documentation that showed how each customer would be affected, a rare window into the pricing model which would otherwise have been veiled under privacy and trade secret arguments. A defense that is used by many companies that utilize discriminatory pricing strategies using data sourced beyond what they should be using to make pricing decisions.

The State of AI Ethics, June 2020        13

According to the investigating journalists, the model boiled down to somethi[ng] quite simple: the more money you had and the higher your willingness to not budge from the company, the more the company would try to squeeze from you in terms of premiums.

Driven by customer retention and profit motives, the company pushed increases on those that they knew could afford them and would switch to save dollars. But, for those policies that had been overpriced, they offered less than 0.5% in terms of a discount limiting their downsides while increases were not limited, often going up as high as 20%.

While they were unsuccessful in getting this adopted in Maryland where it was deemed discriminatory, the model has been approved for use in several states thus showing that opaque models can be deployed not just in high-tech industries but anywhere to provide individually tailored pricing to extract away as much of the consumer surplus as possible based on the purportedly undisclosed willingness of the customer to pay (as would be expressed by their individual demand curves which aren't directly discernible to the producer).

Furthermore, the insurers aren't mandated to make disclosures of how they are pricing their policies and thus, in places where they should have offered discounts, they've only offered pennies on the dollar, disproportionately impacting the poorest for whom a few hundred dollars a year can mean having sufficient meals on the table.

Sadly, in the places where their customer retention model was accepted, the regulators declined to answer why they chose to accept it, except in Arkansas where they said such pricing schemes aren't discriminatory unless the customers are grouped by attributes like race, color, creed or national origin. This takes a very limited view of what price discrimination is, harkening back to an era where access to big data about the consumer was few and far between. In an era dominated by data brokers that compile thick and rich data profiles on consumers, price discriminaton extends far beyond the basic protected attributes and can be tailored down to specificities of each individual.

Other companies in retail goods and online learning services have been following this practice of personalized pricing for many years, often defending it as the cost of doing in business when they based the pricing on things like zip codes, which are known proxies for race and paying capacity. Personalized pricing is different from dynamic pricing, as seen when booking plane tickets, which is usually based on the timing of purchase whereas here the prices are based on attributes that are specific to the customer which they often don't have any control over.

The State of AI Ethics, June 2020                                                                                                       14

A 2015 Obama Administration report mentioned that, "Differential pricing insurance markets can raise serious fairness concerns, particularly when major risk factors are outside an individual customer's control." Why the case of auto insurance is so much more predatory than, say buying stationery supplies, is that it is mandatory in almost all states and not having the vehicle insured can lead to fines, loss of licenses and even incarceration. Transport is an essential commodity for people to get themselves to work, children to school and a whole host of other daily activities.

In Maryland, the regulators had denied the proposal by Allstate to utilize their model but in official public records, the claim is marked as "withdrawn" rather than "denied" which the regulators claim makes no internal difference but Allstate used this difference to get their proposal past the regulators in several other states. They had only withdrawn their proposal after being denied by the regulators in Maryland.

The National Association of Insurance Commissioners mentioned that most regulators don't have the right data to be able to meaningfully evaluate rate revision proposals put forth by insurers and this leads to approvals without review in a lot of cases. Even the data journalists had to spend a lot of time and effort to discern what the underlying models were and how they worked, essentially summing up that the insurers don't necessarily lie but don't give you all the information unless you know to ask the right set of questions.

Allstate has defended its price optimization strategy, called Complementary Group Rating (CGR) as being more objective, and based on mathematical rigor, compared to the ad-hoc, judgemental pricing practices that have been followed before, ultimately citing better outcomes for their customers. But, this is a common form of what is called "mathwashing" in the AI ethics domain where discriminatory solutions are pushed as fair under the veneer of mathematical objectivity.

Regulators in Florida said that setting prices based on the "modeled reaction to rate changes" was "unfairly discriminatory." Instead of being cost-based, as is advocated by regulators for auto-insurance premiums because they support an essential function, Allstate was utilizing a model that was heavily based on the likelihood of the customer sticking with them even in the face of price rises which makes it discriminatory. These customers are price-inelastic and hence don't change their demand much even in the face of dramatic price changes.

Consumer behaviour when purchasing insurance policies for the most part remains static once they've made a choice, often never changing insurers over the course of their lifetime which leads them to not find the optimal price for themselves. This is mostly a function of the fact that the decisions are loaded with having to judge complex terms and conditions across a variety of providers and the customers are unwilling to have to go through the exercise again and again at short intervals.

Given the opacity of the pricing models today, it is almost impossible to find out what the appropriate pricing should be for a particular customer and hence the most effective defense is to constantly check for prices from the competitors. But, this unduly places the burden on the shoulders of the consumer.

## Machine Learning Fairness – Lessons Learned

Google had announced its AI principles on building systems that are ethical, safe and inclusive, yet as is the case with so many high level principles, it's hard to put them into practice unless there is more granularity and actionable steps that are derived from those principles. Here are the principles:

- Be socially beneficial
- Avoid creating or reinforcing unfair bias
- Be built and tested for safety
- Be accountable to people
- Incorporate privacy design principles
- Uphold high standards of scientific excellence
- Be made available for uses that accord with these principles

This talk focused on the second principle and did just that in terms of providing concrete guidance on how to translate this into everyday practice for design and development teams.

Humans have a history of making product design decisions that are not in line with the needs of everyone. Examples of the crash dummy and band-aids mentioned above give some insight into the challenges that users face even when the designers and developers of the products and services don't necessarily have ill intentions. Products and services shouldn't be designed such that they perform poorly for people due to aspects of themselves that they can't change.

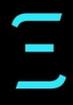

For example, when looking at the Open Image dataset, searching for imag marked with wedding indicate stereotypical Western weddings but those from other cultures and parts of the world are not tagged as such. From a data perspective, the need for having more diverse sources of data is evident and the Google team made an effort to do this by building an extension to the Open Images data set by providing users from across the world to snap pictures from their surroundings that captured diversity in many areas of everyday life. This helped to mitigate the problem that a lot of open image data sets have in being geographically skewed.

Biases can enter at any stage of the ML development pipeline and solutions need to address them at different stages to get the desired results. Additionally, the teams working on these solutions need to come from a diversity of backgrounds including UX design, ML, public policy, social sciences and more.

So, in the area of fairness by data which is one of the first steps in the ML product lifecycle and it plays a significant role in the rest of the steps of the lifecycle as well since data is used to both train and evaluate a system. Google Clips was a camera that was designed to automatically find interesting moments and capture them but what was observed was that it did well only for a certain type of family, under particular lighting conditions and poses. This represented a clear bias and the team moved to collect more data that better represented the situations for a variety of families that would be the target audience for the products. Quickdraw was a fun game that was built to ask users to supply their quickly sketched hand drawings of various commonplace items like shoes.

The aspiration from this was that given that it was open to the world and had a game element to it, it would be utilized by many people from a diversity of backgrounds and hence the data so collected would have sufficient richness to capture the world. On analysis, what they saw was that most users had a very particular concept of a shoe in mind, the sneaker which they sketched and there were very few women's shoes that were submitted. What this example highlighted was that data collection, especially when trying to get diverse samples, requires a very conscious effort that can account for what the actual distribution the system might encounter in the world and make a best effort attempt to capture their nuances. Users don't use systems exactly in the way we intend them to, so reflect on who you're able to reach and not reach with your system and how you can check for blindspots, ensure that there is some monitoring for how data changes over time and use these insights to build automated tests for fairness in data.

The second approach that can help with fairness in ML systems is looking measurement and modeling. The benefits of measurement are that it can be tracked over time and you can test for both individuals and groups at scale for fairness. Different fairness concerns require different metrics even within the same product. The primary categories of fairness concerns are disproportionate harms and representational harms. The Jigsaw API provides a tool where you can input a piece of text and it tells you the level of toxicity of that piece of text. An example in the earlier version of the system rated sentences of the form "I am straight" as not toxic while those like "I am gay" as toxic. So what was needed to be able to see what was causing this and how it could be addressed.

By removing the identity token, they monitored for how the prediction changed and then the outcomes from that measurement gave indications on where the data might be biased and how to fix it. An approach can be to use block lists and removals of such tokens so that sentences that are neutral are perceived as such without imposing stereotypes from large corpora of texts. These steps prevent the model from accessing information that can lead to skewed outcomes. But, in certain places we might want to brand the first sentence as toxic if it is used in a derogatory manner against an individual, we require context and nuance to be captured to make that decision. Google undertook Project Respect to capture positive identity associations from around the world as a way of improving data collection and coupled that with active sampling (an algorithmic approach that samples more from the training data set in areas where it is under performing) to improve outputs from the system.

Another approach is to create synthetic data that mimics the problematic cases and renders them in a neutral context. Adversarial training and updated loss functions where one updates a model's loss function to minimize difference in performance between groups of individuals can also be used to get better results. In their updates to the toxicity model, they've seen improvements, but this was based on synthetic data on short sentences and it is still an area of improvement. Some of the lessons learned from the experiments carried out by the team:

- Test early and test often

- Develop multiple metrics (quantitative and qualitative measures along with user testing is a part of this) for measuring the scale of each problem

- Possible to take proactive steps in modeling that are aware of production constraints

From a design perspective, think about fairness in a more holistic sense and bu communication lines between the user and the product. As an example, Turkish is a gender neutral language, but when translating to English, sentences take on gender along stereotypes by attributing female to nurse and male to doctor. Say we have a sentence, "Casey is my friend", given no other information we can't infer what the gender of Casey is and hence it is better to present that choice to the user from a design perspective because they have the context and background and can hence make the best decision. Without that, no matter how much the model is trained to output fair predictions, they will be erroneous without the explicit context that the user has. Lessons learned from the experiments include:

- Context is key

- Get information from the user that the model doesn't have and share information with the user that the model has and they don't

- How do you design so the user can communicate effectively and have transparency so that can you get the right feedback?

- Get feedback from a diversity of users

- See the different ways in how they provide feedback, not every user can offer feedback in the same way

- Identify ways to enable multiple experiences

- We need more than a theoretical and technical toolkit, there needs to be rich and context-dependent experience

Putting these lessons into practice, what's important is to have consistent and transparent communication and layering on approaches like datasheets for data sets and model cards for model reporting will aid in highlighting appropriate uses for the system and where it has been tested and warn of potential misuses and where the system hasn't been tested.

## Algorithmic Injustices Towards a Relational Ethics

The paper starts by setting the stage for the well understood problem of building truly ethical, safe and inclusive AI systems that are increasingly leveraging ubiquitous sensors to make predictions on who we are and how we might behave.

But, when these systems are deployed in socially contested domains, for example "normal" behaviour where loosely we can think of normal as that defined by the majority and treating everything else as anomalous, then they don't make value-free judgements and are not amoral in their operations.

By viewing the systems as purely technical, the solutions to address these problems are purely technical which is where most of the fairness research has focused and it ignores the context of the people and communities where these systems are used. The paper serves to question the foundations of these systems and to take a deeper look at unstated assumptions in the design and development of the systems. It urges the readers, and the research community at large, to consider this from the perspective of relational ethics. It makes 4 key suggestions:

- Center the focus of development on those within the community that will face a disproportionate burden or negative consequences from the use of the system

- Instead of optimizing for prediction, it is more important to think about how we gain a fundamental understanding of why we're getting certain results which might be arising because of historical stereotypes that were captured as a part of the development and design of the system

- The systems end up creating a social and political order and then reinforcing it, meaning we should involve a larger systems based approach to designing the systems

- Given that the terms of bias, fairness, etc evolve over time and what's acceptable at some time becomes unacceptable later, the process asks for constant monitoring, evaluation and iteration of the design to most accurately represent the community in context.

At MAIEI, we've advocated for an interdisciplinary approach leveraging the citizen community spanning a wide cross section to best capture the essence of different issues as closely as possible from those who experience them first hand. Placing the development of an ML system in context of the larger social and political order is important and we advocate for taking a systems design approach (see A Primer in Systems Thinking by Donna Meadows) which creates two benefits: one is that several ignored externalities can be considered and second to involve a wider set of inputs from people who might be affected by the system and who understand how the system will sit in the larger social and political order in which it will be deployed. Also, we particularly enjoyed the point on requiring a constant iterative process to the development and deployment of AI systems borrowing from cybersecurity research on how security of the system is not done and over with requiring constant monitoring and attention to ensure the safety of the system.

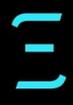

## Social Biases in NLP Models as Barriers for Persons with Disabilities

Underrepresentation of disabilities in datasets and how they are processed in NLP tasks is an important area of discussion that is often not studied empirically in the literature that primarily focuses on other demographic groups. There are many consequences of this, especially as it relates to how text related to disabilities is classified and has impacts on how people read, write, and seek information about this.

Research from the World Bank indicates that about 1 billion people have disabilities of some kind and often these are associated with strong negative social connotations. Utilizing 56 linguistic expressions as they are used in relation to disabilities and classifying them into recommended and non-recommended uses (following the guidelines from Anti-Defamation League, ACM SIGACCESS, and ADA National Network), the authors seek to study how automated systems classify phrases that indicate disability and whether usages split by recommended vs. non-recommended uses make a difference in how these snippets of text are perceived.

To quantify the biases in the text classification models, the study uses the method of perturbation. It starts by collecting instances of sentences that have naturally occurring pronouns he and she. Then, they replace them with the phrases indicating disabilities as identified in the previous paragraph and compare the change in the classification scores in the original and perturbed sentences. The difference indicates how much of an impact the use of a disability phrase has on the classification process.

Using the Jigsaw tool that gives the toxicity score for sentences, they test these original and perturbed sentences and observe that the change in toxicity is lower for recommended phrases vs. the non-recommended ones. But, when disaggregated by categories, they find that some of them elicit a stronger response than others. Given that the primary use of such a model might in the case of online content moderation (especially given that we now have more automated monitoring happening as human staff has been thinning out because of pandemic related closures), there is a high rate of false positives where it can suppress content that is non-toxic and is merely discussing disability or replying to other hate speech that talks about disability.

To look at sentiment scores for disability related phrases, the study looks at the popular BERT model and adopts a template-based fill-in-the-blank analysis. Given a query sentence with a missing word, BERT produces a ranked list of words that can fill the blank. Using a simple template perturbed with recommended disability phrases, the study then looks at how the predictions from the BERT model change when disability phrases are used in the sentence. What is observed is that a large percentage of the words that are predicted by the model have negative sentiment scores associated with them. Since BERT is used quite widely in many different NLP tasks, such negative sentiment scores can have potentially hidden and unwanted effects on many downstream tasks.

Such models are trained on large corpora, which are analyzed to build "meaning" representations for words based on co-occurrence metrics, drawing from the idea that "you shall know a word by the company it keeps". The study used the Jigsaw Unintended Bias in Toxicity Classification challenge dataset which had a mention of a lot of disability phrases. After balancing for different categories and analyzing toxic and non-toxic categories, the authors manually inspected the top 100 terms in each category and found that there were 5 key types: condition, infrastructure, social, linguistic, and treatment. In analyzing the strength of association, the authors found that condition phrases had the strongest association, and was then followed by social phrases that had the next highest strongest association. This included topics like homelessness, drug abuse, and gun violence all of which have negative valences. Because these terms are used when discussing disability, it leads to a negative shaping of the way disability phrases are shaped and represented in the NLP tasks.

The authors make recommendations for those working on NLP tasks to think about the socio-technical considerations when deploying such systems and to consider the intended, unintended, voluntary, and involuntary impacts on people both directly and indirectly while accounting for long-term impacts and feedback loops.

Such indiscriminate censoring of content that has disability phrases in them leads to an underrepresentation of people with disabilities in these corpora since they are the ones who tend to use these phrases most often. Additionally, it also negatively impacts the people who might search for such content and be led to believe that the prevalence of some of these issues are smaller than they actually are because of this censorship. It also has impacts on reducing the autonomy and dignity of these people which in turn has a larger implication of how social attitudes are shaped.

# Go Wide: Article Summaries

## The Second Wave of Algorithmic Accountability

The article dives into explaining how the rising interest in ensuring fair, transparent, ethical AI systems that are held accountable via various mechanisms advocated by research in legal and technical domains constitutes the "first wave" of algorithmic accountability that challenges existing systems. Actions as a part of this wave need to be carried out incessantly with constant vigilance of the deployment of AI systems to avoid negative social outcomes. But, we also need to challenge why we have these systems in the first place, and if they can be replaced with something better. As an example, instead of making the facial recognition systems more inclusive, given the fact that they cause social stratification perhaps they shouldn't be used at all. A great point made by the article is that under the veneer of mainstream economic and AI rationalizations, we obscure broken social systems which ultimately harm society at a more systemic level.

## The Unnatural Ethics of AI Could Be Its Undoing

The Trolley Problem is a widely touted ethical and moral dilemma wherein a person is asked to make a split-second choice to save one or more than one life based on a series of scenarios where the people that need to be saved have different characteristics including their jobs, age, gender, race, etc. In recent times, with the imminent arrival of self-driving cars, people have used this problem to highlight the supposed ethical dilemma that the vehicle system might have to grapple with as it drives around.

This article makes a point about the facetious nature of this thought experiment as an introduction to ethics for people that will be building and operating such autonomous systems. The primary argument being that it's a contrived situation that is unlikely to arise in the real-world setting and it distracts from other more pressing concerns in AI systems. Moral judgments are relativistic and depend on cultural values of the geography where the system is deployed. The Nature paper cited in the article showcases the differences in how people respond to this dilemma.

There is an eeriness to this whole experimental setup, the article gives some examples on how increasingly automated environments, devoid of human social interactions and language, are replete with the clanging and humming of machines that give an entirely inhuman experience. For most systems, they are going to be a reflection of the biases and stereotypes that we have in the world, captured in the system because of the training and development paradigms of AI systems today. We'd need to make changes and bring in diversity to t development process, creating awareness of ethical concerns, but the Trolley Problem isn't the most effective way to get started on it.

## This Dating App Exposes the Monstrous Bias of Algorithms

Most of us have a nagging feeling that we're being forced into certain choices when we interact with each other on various social media platforms. But, is there a way that we can grasp that more viscerally where such biases and echo chambers are laid out bare for all to see? The article details an innovative game design solution to this problem called Monster Match that highlights how people are trapped into certain niches on dating websites based on AI-powered systems like collaborative filtering. Striking examples of that in practice are how your earlier choices on the platform box you into a certain category based on what the majority think and then recommendations are personalized based on that smaller subset. What was observed was that certain racial inequalities from the real world are amplified on platforms like these where the apps are more interested in keeping users on the platform longer and making money rather than trying to achieve the goal as advertised to their users. More than personal failings of the users, the design of the platform is what causes failures in finding that special someone on the platform. The creators of the solution posit that through more effective design interventions, there is potential for improvement in how digital love is realized, for example, by offering a reset button or having the option to opt-out of the recommendation system and instead relying on random matches. Increasingly, what we're going to see is that reliance on design and other mechanisms will yield better AI systems than purely technical approaches in improving socially positive outcomes.

## Catherine D'Ignazio: 'Data Is Never a Raw, Truthful Input – and It Is Never Neutral'

The article presents the idea of data feminism which is described as the intersection between feminism and data practices. The use of big data in today's dominant paradigm of supervised machine learning lends itself to large asymmetries that reflect the power imbalances in the real world. The authors of the new book Data Feminism talk about how data should not just speak for itself, for behind the data, there are a large number of structures and assumptions that bring it to the stage where they are collated into a dataset.

They give examples of how sexual harassment numbers, while mandated to be reported to a central agency from college campuses might not be very accurate because they rely on the atmosphere and degree of comfort that those campuses promote which in turn influences how close the reported numbers will be to the actual cases. The gains and losses from the use of big data are not distributed evenly and the losses disproportionately impact the marginalized.

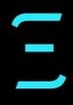

There are a number of strategies that can be used to mitigate the harms from such flawed data pipelines. Not an exhaustive list but it includes the suggestion of having more exposure for technical students to the social sciences and moving beyond having just a single ethics class as a check mark for having educated the students on ethics. Secondly, having more diversity in the people developing and deploying the AI systems would help spot biases by asking the hard questions about both the data and the design of the system. The current COVID-19 numbers might also suffer from similar problems because of how medical systems are structured and how people who don't have insurance might not utilize medical facilities and get themselves tested thus creating an underrepresentation in the data.

## Racial Disparities in Automated Speech Recognition

This recent work highlights how commercial speech recognition systems carry inherent bias because of a lack of representation from diverse demographics in the underlying training datasets. What the researchers found was that even for identical sentences spoken by different racial demographics, the systems had widely differing levels of performance. As an example, for black users, the error rates were much higher than those for white users which probably had something to do with the fact that there is specific vernacular language used by black people which wasn't adequately represented in the training dataset for the commercial systems.

This pattern has a tendency to be amplifying in nature, especially for systems that aren't frozen and continue to learn with incoming data. A vicious cycle is born where because of poor performance from the system, black people will be disincentivized from using the system because it takes a greater amount of work to get the system to work for them thus lowering utility. As a consequence of lower use, the systems get fewer training samples from black people thus further aggravating the problem. This leads to amplified exclusionary behavior mirroring existing fractures along racial lines in society. As a starting point, collecting more representative training datasets will aid in mitigating at least some of the problems in these systems.

## Working to Address Algorithmic Bias? Don't Overlook the Role of Demographic Data

Algorithmic bias at this point is a well-recognized problem with many people working on ways to address issues, both from a technical and policy perspective. There is potential to use demographic data to serve better those who face algorithmic discrimination but the use of such data is a challenge because of ethical and legal concerns. Primarily, a lot of jurisdictions don't allow for the capture and use of protected class attributes or sensitive data for the fear of their misuse. Even within jurisdictions, there is a patchwork of recommendations which makes compliance difficult. Even with all this well established, proxy attributes can be used to predict the protected data and in a sense, according to some legislations, they become protected data themselves and it becomes hard to extricate the non-sensitive data from the sensitive data. Because of such tensions and the privacy intrusions on data subjects when trying to collect demographic data, it is hard to align and advocate for this collection of data over the other requirements within the organization, especially when other bodies and leadership will look to place privacy and legal compliance over bias concerns.

Even if there was approval and internal alignment in collecting this demographic data, if there is voluntary provision of this data from data subjects, we run the risk of introducing a systemic bias that obfuscates and mischaracterizes the whole problem. Accountability will play a key role in evoking trust from people to share their demographic information and proper use of it will be crucial in ongoing success. Potential solutions are to store this data with a non-profit third-party organization that would meter out the data to those who need to use it with the consent of the data subject.

To build a better understanding, Partnership on AI is adopting a multistakeholder approach leveraging diverse backgrounds, akin to what the Montreal AI Ethics Institute does, that can help inform future solutions that will help to address the problems of algorithmic bias by the judicious use of demographic data.

## AI Advances to Better Detect Hate Speech

Detection and removal of hate speech is particularly problematic, something that has been exacerbated as human content moderators have been scarce in the pandemic related measures as we covered here. So are there advances in NLP that we could leverage to better automate this process? Recent work from Facebook AI Research shows some promise. Developing a deeper semantic understanding across more subtle and complex meanings and working across a variety of modalities like text, images and videos will help to more effectively combat t problem of hate speech online. Building a pre-trained universal representation of content for integrity problems and improving and utilizing post-level, self-supervised learning to improve whole entity understanding has been key in improving hate speech detection. While there are clear guidelines on hate speech, when it comes to practice there are numerous challenges that arise from multi-modal use, differences in cultures and context, differences in idioms, language, regions, and countries. This poses challenges even for human reviewers who struggle with identifying hate speech accurately.

A particularly interesting example shared in the article points out how text which might seem ambiguous when paired with an image to create a meme can take a whole new meaning which is often hard to detect using traditional automated tooling. There are active efforts from malicious entities who craft specific examples with the intention of evading detection which further complicates the problem. Then there is the counterspeech problem where a reply to hate speech that contains the same phrasing but is framed to counter the arguments presented can be falsely flagged to be brought down which can have free speech implications.

The relative scarcity of examples of hate speech in its various forms in relation to the larger non-hate speech content also poses a challenge for learning, especially when it comes to capturing linguistic and cultural nuances. The new method proposed utilizes focal loss which aims to minimize the impact of easy-to-classify examples on the learning process which is coupled with gradient blending which computes an optimal blend of modalities based on their overfitting patterns. The technique called XLM-R builds on BERT by using a new pretraining recipe called RoBERTa that allows training on orders of magnitude more data for longer periods of time. Additionally, NLP performance is improved by learning across languages using a single encoder that allows learning to be transferred across languages. Since this is a self-supervised method, they can train on large unlabeled datasets and have also found some universal language structures that allow vectors with similar meanings across languages to be closer together.

## Algorithms Associating Appearance and Criminality Have a Dark Past

Facial recognition technology (FRT) continues to get mentions because of the variety of ways that it can be misused across different geographies and contexts. With the most recent case where FRT is used to determine criminality, it brings up an interesting discussion around why techniques that have no basis in science, those which have been debunked time and time again keep resurfacing and what we can do to better educate researchers on their moral responsibilities in pursuing such work. The author of this article gives some historical context for where phrenology started, pointing to the work of Francis Galton who used t "photographic composite method" to try and determine characteristics of one's personality from a picture. Prior, measurements of skull size and other facial features wasn't deemed as a moral issue and the removal of such techniques from discussion was done on the objection that claims around the localization of different brain functions was seen as antithetical to the unity of the soul according to Christianity.

The authors of the paper that is being discussed in the article saw only empirical concerns with the work that they put forth and didn't see any of the moral shortcomings that were pointed out. Additionally, they justified the work as being only for scientific curiosity. They also failed to realize the various statistical biases introduced in the collection of data as to the disparate rates of arrests, and policing, the perception of different people by law enforcement, juries, and judges and historical stereotypes and biases that confound the data that is collected.Thus, the labeling itself is hardly value-neutral. More so, the authors of the study framed criminality as an innate characteristic rather than the social and other circumstances that lead to crime.

Especially when a project like this resurrects class structures and inequities, one must be extra cautious of doing such work on the grounds of "academic curiosity". The author of this article thus articulates that researchers need to take their moral obligations seriously and consider the harm that their work can have on people. While simply branding this as phrenology isn't enough, the author mentions that identifying and highlighting the concerns will lead to more productive conversations.

## Beware of These Futuristic Background Checks

An increase in demand for workers for various delivery services and other gig work has accelerated the adoption of vetting technology like those that are used to do background checks during the hiring process. But, a variety of glitches in the system such as sourcing out-of-date information to make inferences, a lack of redressal mechanisms to make corrections, among others has exposed the flaws in an overreliance on automated systems especially in places where important decisions need to be made that can have a significant impact on a person's life such as employment.

Checkr, the company that is profiled in this article claims to use AI to scan resumes, compare criminal records, analyze social media accounts, and examine facial expressions during the interview process. During a pandemic, when organizations are short-staffed and need to make rapid decisions, Checkr offers a way to streamline the process, but this comes at a cost. Two supposed benefits that they offer are that they are able to assess a match between the criminal record and t person being actually concerned, something that can especially be fraught with errors in cases where the person has a common name. Secondly, they are also able to correlate and resolve discrepancies in the different terms that may be used for crimes across different jurisdictions.

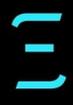

A person spoke about his experience with another company that did these AI-powered background checks utilizing his public social media information and bucketed some of his activity into categories that were too coarse and unrepresentative of his behaviour, especially when such automated judgements are made without a recourse to correct, this can negatively affect the prospects of being hired. Another point brought up in the article is that social media companies might themselves be unwilling to tolerate scraping of their users' data to do this sort of vetting which against their terms of use for access to the APIs. Borrowing from the credit reporting world, the Fair Credit Reporting Act in the US offers some insights when it mentions that people need to be provided with a recourse to correct information that is used about them in making a decision and that due consent needs to be obtained prior to utilizing such tools to do a background check. Though it doesn't ask for any guarantees of a favorable outcome post a re-evaluation, at least it does offer the individual a bit more agency and control over the process.

# 3. Disinformation

## Go Deep: Research Summaries

### The Toxic Potential of YouTube's Feedback Loop

On YouTube everyday, more than a billion hours of video are watched everyday where approximately 70% of those are watched by automated systems that then provide recommendations on what videos to watch next for human users in the column on the side. There are more than 2 billion users on the YouTube platform so this has a significant impact on what the world watches. Guillaume had started to notice a pattern in the recommended videos which tended towards radicalizing, extreme and polarizing content which were underlying the upward trend of watch times on the platform.

On raising these concerns with the team, at first there were very few incentives for anyone to address issues of ethics and bias as it related to promoting this type of content because they feared that it would drive down watch time, the key business metric that was being optimized for by the team. So maximizing engagement stood in contrast to the quality of time that was spent on the platform.

The vicious feedback loop that it triggered was that as such divisive content performed better, the AI systems promoted this to optimize for engagement and subsequently content creators who saw this kind of content doing better created more of such content in the hopes of doing well on the platform. The proliferation of conspiracy theories, extreme and divisive content thus fed its own demand because of a misguided business metric that ignored social externalities. Flat earthers, anti-vaxxers and other such content creators perform well because the people behind this content are a very active community that spend a lot of effort in creating these videos, thus meeting high quality standards and further feeding the toxic loop. Content from people like Alex Jones and Trump tended to perform well because of the above reasons as well.

Guillaume's project AlgoTransparency essentially clicks through video recommendations on YouTube to figure out if there are feedback loops. He started this with the hopes of highlighting latent problems in the platforms that continue to persist despite policy changes, for example with YouTube attempting to automate the removal of reported and offensive videos. He suggests that the current separation of the policy and engagement algorithm leads to problems like gaming of the platform algorithm by motivated state actors that seek to disrupt democratic processes of a foreign nation. The platforms on the other hand ha very few incentives to make changes because the type of content emerging from such activity leads to higher engagement which ultimately boosts their bottom line.

Guillaume recommends having a combined system that can jointly optimize for both thus helping to minimize problems like the above. A lot of the problems are those of algorithmic amplification rather than content curation. Many metrics like number of views, shares, and likes don't capture what needs to be captured. For example, the types of comments, reports filed, and granularity of why those reports are filed. That would allow for a smarter way to combat the spread of such content. However, the use of such explicit signals compared to the more implicit ones like number of views comes at the cost of breaking the seamlessness of the user experience. Again we run into the issue of a lack of motivation on part of the companies to do things that might drive down engagement and hurt revenue streams.

The talk gives a few more examples of how people figured out ways to circumvent checks around the reporting and automated take-down mechanisms by disabling comments on the videos which could previously be used to identify suspicious content. An overarching recommendation made by Guillaume in better managing a more advanced AI system is to understand the underlying metrics that the system is optimizing for and then envision scenarios of what would happen if the system had access to unlimited data.

Thinking of self-driving cars, an ideal scenario would be to have full conversion of the traffic ecosystem to one that is autonomous leading to fewer deaths but during the transition phase, having the right incentives is key to making a system that will work in favor of social welfare. If one were to imagine a self-driving car that shows ads while the passenger is in the car, it would want to have a longer drive time and would presumably favor longer routes and traffic jams thus creating a sub-optimal scenario overall for the traffic ecosystem. On the other hand, a system that has the goal of getting from A to B as quickly and safely as possible wouldn't fall into such a trap. Ultimately, we need to design AI systems such that they help humans flourish overall rather than optimize for monetary incentives which might run counter to the welfare of people at large.

# Go Wide: Article Summaries

## How Spreaders of Misinformation Acquire Influence Online

The article provides a taxonomy of communities that spread misinformation online and how they differ in their intentions and motivations. Subsequently, different strategies can be deployed in countering the disinformation originating from these communities. There isn't a one-size-fits-all solution that would have been the case had the distribution and types of the communities been homogenous. The degree of influence that each of the communities wield is a function of 5 types of capital: economic, social, cultural, time and algorithmic, definitions of which are provided in the article. Understanding all these factors is crucial in combating misinformation where different capital forms can be used in different proportions to achieve the desired results, something that will prove to be useful in addressing disinformation around the current COVID-19 situation.

## Want to Find a Misinformed Public? Facebook's Already Done It

The social media platform offers a category of pseudoscience believers which advertisers can purchase and target. According to The Markup, this category has 78 million people in it and attempts to purchase ads targeting this category were approved quite swiftly. There isn't any information available as to who has purchased ads targeting this category. The journalist team was able to find at least one advertiser through the "Why am I seeing this ad?" option and they reached out to that company to investigate and they found that the company hadn't selected the pseudoscience category but it had been auto-selected by Facebook for them. Facebook allows users the option to change the interests that are assigned to each user but it is not something that many people know about and actively monitor.

Some other journalists had also unearthed controversy-related categories that amplified messages and targeted people who might be susceptible to such kind of misinformation. With the ongoing pandemic, misinformation is propagating at a rapid rate and there are many user groups that continue to push conspiracy theories. Other concerns around being able to purchase ads to spread misinformation related to potential cures and remedies for the coronavirus continue to be approved. With the human content moderators being asked to stay home (as we covered here) and an increasing reliance on untested automated solutions, it seems that this problem will continue to plague the platform.

The State of AI Ethics, June 2020    32

## Say Goodbye to the Information Age: It's All About Reputation Now

There isn't a dearth of information available online, one can find confirmatory evidence to almost any viewpoint since the creation and dissemination of information has been democratized by the proliferation of the internet and ease of use of mass-media platforms. So in the deluge of information, what is the key currency that helps us sift through all the noise and identify the signal? This article lays out a well articulated argument for how reputation and being able to assess it is going to be a key skill that people will need to have in order to effectively navigate the information ecosystem effectively. We increasingly rely on other people's judgement of content (akin to how MAIEI analyzes the ecosystem of AI ethics and presents you with a selection), coupled with algorithmically-mediated distribution channels, we are paradoxically disempowered by more information and paralyzed into inaction and confusion without a reputable source to curate and guide us.

There are many conspiracy theories, famous among them that we never visited the Moon, Flat Earth and more recently that 5G is causing the spread of the coronavirus. As rational readers, we tend to dismiss this as misinformation yet we don't really spend time to analyze the evidence that these people present to support their claims. To a certain extent, our belief that we did land on the Moon depends on our trust in NASA and other news agencies that covered this event yet we don't venture to examine the evidence first-hand.

More so, with highly specialized knowledge becoming the norm, we don't have the right tools and skills to even be able to analyze the evidence and come to meaningful conclusions. So, we must rely on those who provide us with this information. Instead of analyzing the veracity of a piece of information, the focus of a mature digital citizen needs to be on being able to analyze the reputation pathway of that information, evaluate the agendas of the people that are disseminating the information and critically analyze the intentions of the authorities of the sources.

How we rank different pieces of information arriving to us via our social networks need to be appraised for this reputation and source tracing, in a sense a second-order epistemology is what we need to prepare people for. In the words of Hayek, "civilization rests on the fact that we all benefit from the knowledge that we do not possess." Our cyber-world can become civilized by evaluating this knowledge that we don't possess critically when mis/disinformation can spread just as easily as accurate information.

## How "Truth Decay" is Harming America's Coronavirus Recovery

A very clear way to describe the problem plaguing the US response to the coronavirus, the phenomenon of truth decay is not something new but has happened many times in the past when trust in key institutions deteriorated and led to a diffused response to the crisis at hand, extending the recovery period beyond what would be necessary if there was a unified response. In the US, the calls for reopening the economy, following guidance on using personal protective equipment, and other recommendations is falling along partisan lines. The key factor causing this is how the facts and data are being presented differently to different audiences. While this epidemic might have been the perfect opportunity for bringing people together, because it affects different segments of society differently, it hasn't been what everyone expected it to be.

At the core is the rampant disagreement between different factions on facts and data. This is exacerbated by the blurring of facts and opinions. In places like newsrooms and TV shows, there is an intermingling of the two which makes it harder for everyday consumers to discern fact from opinion. The volume of opinion has gone up compared to facts and people's declining trust in public health authorities and other institutions is also aggravating the problem. Put briefly, people are having trouble finding the truth and don't know where to go looking for it.

This is also the worst time to be losing trust in experts; with a plethora of information available online, people are feeling unnecessarily empowered that they have the right information, comparable to that of experts. Coupled with a penchant for confirming their own beliefs, there is little incentive for people to fact-check and refer to multiple sources of information. When different agencies come out with different recommendations and there are policy changes in the face of new information, something that is expected given that this is an evolving situation, people's trust in these organizations and experts erodes further as they see them as flip-flopping and not knowing what is right. Ultimately, effective communication along with a rebuilding of trust will be necessary if we're to emerge from this crisis soon and restore some sense of normalcy.

# 4. Disinformation: Solutions

## Go Deep: Research Summaries

### The Deepfake Detection Challenge: Insights and Recommendations for AI and Media Integrity

Synthetic media is any media (text, image, video, audio) that is generated by an AI system or that is synthesized. On the other hand, non-synthetic media is one that is crafted by humans using a panoply of techniques, including tools like Photoshop.

Detecting synthetic media alone doesn't solve the media integrity challenges, especially as the techniques get more sophisticated and trigger an arms race between detection and evasion methods. These methods need to be paired with other existing techniques that fact checkers and journalists already use in determining whether something is authentic or synthesized. There are also pieces of content that are made through low tech manipulations like the Nancy Pelosi video from 2019 which showed her drunk but in reality it was just a slowed down video.

Other such manipulations include simpler things like putting fake and misleading captions below the true video and people without watching the whole thing are misled into believing what is summarized in the caption. In other cases, the videos might be value neutral or informative even when they are generated so merely detecting something as being generated doesn't suffice. A meaningful way to utilize automated tools is a triaging utility that flags content to be reviewed by humans in a situation where it is not possible to manually review everything on the platform.

While tech platforms can build and utilize tools that help them with these tasks, the adjacent possible needs of the larger ecosystem need to be kept in mind such that they can be served at the same time, especially for those actors that are resource-constrained and don't have the technical capabilities to build it themselves. The tools need to be easy to use and shouldn't have high friction such that they become hard to integrate into existing workflows. Through open sourcing and licensing, the tools can be made available to the wider ecosystem but it can create the opportunity for adversaries to strengthen their methods as well. This can be countered by responsible disclosure as we'll cover below.

For any datasets created as a part of this challenge and otherwise to aid detection, one must ensure that it captures sufficient diversity in terms of environment and other factors and reflects the type of content that might be encountered in the world. The scoring rules need to be such that they minimize gaming and overfitting and capture the richness of variation that a system might encounter. For example most datasets today in this domain aim to mitigate the spread of pornographic material. They also need to account for the vastly different frequencies of occurrence of authentic and generated content.

Solutions in this domain involve an inherent tradeoff between pro-social use and potential malicious use for furthering the quality of inauthentic content. The release of tools should be done in a manner that enhances pro-social use while creating deterrents for malicious use. The systems should be stress-tested by doing red team-blue team exercises to enhance robustness because this is inherently an adversarial exercise. Such challenges should be held often to encourage updating of techniques because it is a fast evolving domain where progress happens in the span of a few months.

Results from such detection need to be accessible to the public and stakeholders and explanations for the research findings should be made available alongside the challenge to encourage better understanding by those that are trying to make sense of the digital content. Responsible disclosure practices will be crucial in enabling the fight against disinformation to have the right tools while deterring adversaries from utilizing the same tools to gain an advantage. A delayed release mechanism where the code is instantly made available to parties in a non-open source manner while the research and papers are made public with the eventual release of the code as well after a 6-12 months delay which would help with the detectors having a headstart over the adversaries.

Such detection challenges can benefit from extensive multi-stakeholder consultations which require significant time and effort so budget for that while crafting and building such challenges. Some of the allocation of prize money should be towards better design from a UX and UI perspective. It should also include explainability criteria so that non-technical users are able to make sense of the interventions and highlights of fake content such as bounding boxes around regions of manipulations. The process of multi-stakeholder input should happen at an early stage allowing for meaningful considerations to be incorporated and dataset design that can be done appropriately to counter bias and fairness problems.

Finally, strong, trusting relationships are essential to the success of the process and require working together over extended periods to have the hard conversations with each other. It is important to have clear readings ahead of meetings that everyone has to complete so that discussions come from an informed place. Spending time scoping and coming to clearer agreement about projects goals and deliverables at the beginning of the process is also vital to success.

**Technology-Enabled Disinformation: Summary, Lessons, and Recommendations**

There is a distinction between misinformation and disinformation – misinformation is the sharing of false information unintentionally where no harm is intended whereas disinformation is false information that is spread intentionally with the aims of causing harm to the consumers. This is also referred to as information pollution and fake news. It has massive implications that have led to real harms for people in many countries with one of the biggest examples being the polarization of views in the 2016 US Presidential elections.

Meaningful solutions to this will only emerge when we have researchers from both technical and social sciences backgrounds working together to gain a deeper understanding of the root causes. This isn't a new problem and has existed for a very long time, it's just that with the advent of technology and more people being connected to each other we have a much more rapid dissemination of the false information and modern tools enable the creation of convincing fake images, text and videos, thus amplifying the negative effects.

Some of the features that help to delve deeper into the study of how mis/disinformation spreads are:

- Democratization of content creation: with practically anyone now having the ability to create and publish content, information flow has increased dramatically and there are few checks for the veracity of content and even fewer mechanisms to limit the flow rate of information.

- Rapid news cycle and economic incentives: with content being monetized, there is a strong incentive to distort information to evoke a response from the reader such that they click through and feed the money-generating apparatus.

- Wide and immediate reach and interactivity: by virtue of almost the entire globe being connected, content quickly reaches the furthest corners of the planet. More so, content creators are also able to, through quantitative experiments, determine what kind of content performs well and then tailor that to feed the needs of people.

- Organic and intentionally created filter bubbles: the selection of who follow along with the underlying plumbing of the platforms permits for the creation of echo chambers that further strengthen polarization and do little to encourage people to step out and have a meaningful exchange of ideas.

- Algorithmic curation and lack of transparency: the inner workings of platforms are shrouded under the veil of IP protections and there is little that is well-known about the manipulative effects of the platforms on the habits of content consumers.

- Scale and anonymity of online accounts: given the weak checks for identity, people are able to mount "sybil" attacks that leverage this lack of strong identity management and are able to scale their impact through the creation of content and dispersion of content by automated means like bot accounts on the platform.

What hasn't changed even with the introduction of technology are the cognitive biases which act as attack surfaces for malicious actors to inject mis/disinformation. This vulnerability is of particular importance in the examination and design of successful interventions to combat the spread of false information. For example, the confirmation bias shows that people are more likely to believe something that conforms with their world-view even if they are presented with overwhelming evidence to the contrary. In the same vein, the backfire effect demonstrates how people who are presented with such contrary evidence further harden their views and get even more polarized thus negating the intention of presenting them with balancing information.

In terms of techniques, the adversarial positioning is layered into three tiers with spam bots that push out low-quality content, quasi-bots that have mild human supervision to enhance the quality of content and pure human accounts that aim to build up a large following before embarking on spreading the mis/disinformation.

From a structural perspective, the alternate media sources often copy-paste content with source attribution and are tightly clustered together with a marked separation with other mainstream media outlets. On the consumer front, there is research that points to the impact that structural deficiencies in the platforms, say Whatsapp where source gets stripped out in sharing information, create not only challenges for researchers trying to study the ecosystem but also exacerbate the local impact effect whereby a consumer trusts things coming from friends much more so than other potentially more credible sources from an upstream perspective.

Existing efforts to study the ecosystem require a lot of manual effort but there hope in the sense that there are some tools that help automate the analysis. As an example, we have the Hoaxy tool, a tool that collects online mis/disinformation and other articles that are fact-checking versions. Their creators find that the fact-checked articles are shared much less than the original article and that curbing bots on a platform has a significant impact.

There are some challenges with these tools in the sense that they work well on public platforms like Twitter but on closed platforms with limited ability to deploy bots, automation doesn't work really well. Additionally, even the metrics that are surfaced need to be interpreted by researchers and it isn't always clear how to do that.

The term 'deepfake' originated in 2017 and since then a variety of tools have been released such as Face2Face that allow for the creation of reanimations of people to forge identity, something that was alluded to in this paper here on the evolution of fraud. While being able to create such forgeries isn't new, what is new is that this can be done now with a fraction of the effort, democratizing information pollution and casting aspersions on legitimate content as one can always argue something was forged.

Online tracking of individuals, which is primarily used for serving personalized advertisements and monetizing the user behaviors on websites can also be used to target mis/disinformation in a fine-grained manner. There are a variety of ways this is done through third-party tracking like embedding of widgets to browser cookies and fingerprinting. This can be used to manipulate vulnerable users and leverage sensitive attributes gleaned from online behaviors that give malicious actors more ammunition to target individuals specifically. Even when platforms provide some degree of transparency on why users are seeing certain content, the information provided is often vague and doesn't do much to improve the understanding for the user.

Earlier attempts at using bots used simplistic techniques such as tweeting at certain users and amplifying low-credibility information to give the impression that something has more support than it really does but recent attempts have become more sophisticated: social spambots. These slowly build up credibility within a community and then use that trust to sow disinformation either automatically or in conjunction with a human operator, akin to a cyborg.

Detection and measurement of this problem is a very real concern and researchers have tried using techniques like social network graph structure, account data and posting metrics, NLP on content and crowdsourcing analysis. From a platform perspective, they can choose to analyze the amount of time spent browsing posts vs. the time spent posting things.

There is an arms race between detection and evasion of bot accounts: sometimes even humans aren't able to detect sophisticated social bots. Additionally, there are instances where there are positive and beneficial bots such as those that aggregate news or help coordinate disaster response which further complicates the detection challenge. There is also a potential misalignment in incentives since the platforms have an interest in having higher numbers of accounts and activity since it helps boost their valuations while they are the ones that have the maximum amount of information to be able to combat the problem.

This problem of curbing the spread of mis/disinformation can be broken down into two parts: enabling detection on the platform level and empowering readers to select the right sources. We need a good definition of what fake news is, one of the most widely accepted definitions is that it is something that is factually false and intentionally misleading. Framing a machine learning approach here as an end-to-end task is problematic because it requires large amounts of labelled data and with neural network based approaches, there is little explanation offered which makes downstream tasks harder.

So we can approach this by breaking it down into subtasks, one of which is verifying the veracity of information. Most current approaches use human fact-checkers but this isn't a scalable approach and automated means using NLP aren't quite proficient at this task yet. There are attempts to break down the problem even further such as using stance detection to see if information presented agrees, disagrees or is unrelated to what is mentioned in the source. Other approaches include misleading style detection whereby we try to determine if the style of the article can offer clues to the intent of the author but that is riddled with problems of not having necessarily a strong correlation with a misleading intent because the style may be pandering to hyperpartisanship or even if it is neutral that doesn't mean that it is not misleading.

Metadata analysis looking at the social graph structure, attributes of the sharer and propagation path of the information can lend some clues as well. While all these attempts have their own challenges and in the arms race framing, there is a constant battle between attack and defense, even if the problem is solved, we still have human cognitive biases which muddle the impacts of these techniques. UX and UI interventions might serve to provide some more information as to combating those.

As a counter to the problems encountered in marking content as being "disputed" which leads to the implied truth effect leading to larger negative externalities, an approach is to show "related" articles when something is disputed and then u that as an intervention to link to fact-checking websites like Snopes. Other in-platform interventions include the change from Whatsapp to show "forwarded" next to messages so that people had a bit more insight into the provenance of the message because there was a lot of misinformation that was being spread in private messaging. There are also third-party tools like SurfSafe that are able to check images as people are browsing against other websites where they might have appeared and if they haven't appeared in many places, including verified sources, then the user can infer that the image might be doctored.

Education initiatives by the platform companies for users to spot misinformation are a method to get people to become more savvy. There have also been attempts to assign nutrition labels to sources to list their slant, tone of the article, timeliness of the article and the experience of the author which would allow a user to make a better decision on whether or not to trust an article. Platforms have also attempted to limit the spread of mis/disinformation by flagging posts that encourage gaming of the sharing mechanisms on the platform, for example, downweighting posts that are "clickbait".

The biggest challenges in the interventions created by the platforms themselves are that they don't provide sufficient information as to make the results scientifically reproducible. Given the variety of actors and motivations, the interventions need to be tailored to be able to combat them such as erecting barriers to the rate of transmission of mis/disinformation and demonetization for actors with financial incentives but for state actors, detection and attribution might be more important. Along with challenges in defining the problem, one must look at socio-technical solutions because the problem has more than just the technical component, including the problem with human cognitive biases.

Being an inherently adversarial setting, it is important to see that not all techniques being used by the attackers are sophisticated, some simple techniques when scaled are just as problematic and require attention. But, given that this is constantly evolving, detecting disinformation today doesn't mean that we can do so successfully tomorrow. Additionally, disinformation is becoming more personalized, more realistic and more widespread.

There is a misalignment in incentives as explored earlier in terms of what the platforms want and what's best for users but also that empowering users to the point of them being just skeptical of everything isn't good either, we need to be able to trigger legitimate and informed trust in the authentic content and dissuade them away from the fake content.

Among the recommendations proposed by the authors are: being specific abc what a particular technological or design intervention means to achieve, breaking down the technological problems into smaller, concrete subproblems that have tractable solutions and then recombining them into the larger pipeline. We must also continue to analyze the state of the ecosystem and tailor defenses such that they can combat the actors at play. Additionally, rethinking of the monetary incentives on the platform can help to dissuade some of the financially-motivated actors.

Educational interventions that focus on building up knowledge so that there is healthy skepticism and learning how to detect markers for bots, the capabilities of technology to create fakes today and discussions in "public squares" on this subject are crucial yet we mustn't place too much of a burden on the end-user that distracts them from their primary task which is interaction with others on the social network. If that happens, people will just abandon the effort. Additionally, designing for everyone is crucial, if the interventions, such as installing a browser extension, are complicated, then one can only reach the technically-literate people and everyone else gets left out.

On the platform end, apart from the suggestions made above, they should look at the use of design affordances that aid the user in judging the veracity, provenance and other measures to discern legitimate information vs. mis/disinformation. Teaming up with external organizations that specialize in UX/UI research will aid in understanding the impacts of the various features within the platform. Results from such research efforts need to be made public and accessible to non-technical audiences. Proposed solutions also need to be interdisciplinary to have a fuller understanding of the root causes of the problem. Also, just as we need tailoring for the different kinds of adversaries, it is important to tailor the interventions to the various user groups who might have different needs and abilities.

The paper also makes recommendations for policymakers, most importantly that the work in regulations and legislations be grounded in technical realities that are facing the ecosystem so that they don't undershoot or overshoot the needs for successfully combating mis/disinformation. For users, there are a variety of recommendations provided in the references but notably being aware of our own cognitive biases and having a healthy degree of skepticism and checking information against multiple sources before accepting it as legitimate are the most important ones.

# Go Wide: Article Summaries

## Here's How Social Media Can Combat the Coronavirus 'Infodemic'

Disinformation is harmful even during times when we aren't going through large scale changes but this year the US has elections, the once in a decade census, and the COVID-19 pandemic. Malicious agents are having a field day disbursing false information, overwhelming people with a mixture of true and untrue pieces of content. The article gives the example of a potential lockdown and people reflecting on their experience with the Boston Marathon bombings including stockpiling essentials out of panic. This was then uncovered to have originated from conspiracy theorists, but in an environment where contact with the outside world has become limited and local touch points such as speaking with your neighbor have dwindled, we're struggling with our ability to combat this infodemic.

Social media is playing a critical role in getting information to people but if it's untrue, we end up risking lives especially if it's falsehoods on how to protect yourself from contracting a disease. But wherever there is a challenge lies a corresponding opportunity: social media companies have a unique window into discovering issues that a local population is concerned about and it can, if used effectively, be a source for providing crisis response to those most in need with resources that are specific and meaningful.

## Study: 'Accuracy Nudge' Could Curtail COVID-19 Misinformation Online

When it comes to disinformation spreading, there isn't a more opportune time than now with the pandemic raging where people are juggling several things to manage and cope with lifestyle and work changes. This has increased the susceptibility of people to sharing news and other information about how to protect themselves and their loved ones from COVID-19. As the WHO has pointed out, we are combating both a pandemic and an infodemic at the same time.

What's more important is that this might be the time to test out design and other interventions that might help curb the spread of disinformation. This study highlighted how people shared disinformation more often than they believed its veracity. In other words, when people share content, they care more about what they stand to gain (social reward cues) for sharing the content than whether the content they're sharing is accurate or not. To combat this, the researchers

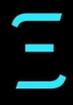
embarked on an experiment to see if asking users to check whether something was true before sharing — a light accuracy nudge, would change their behaviour.

While there was a small positive effect in terms of them sharing disinformation less when prompted to check for accuracy, the researchers pointed out that the downstream effects could be much larger because of the amplification effects of how content propagates on social media networks. It points to a potentially interesting solution that might be scalable and could help fight against the spread of disinformation.

## Going Viral: How to Boost the Spread of Coronavirus Science on Social Media

The WHO has mentioned the infodemic as being one of the causes that is exacerbating the pandemic as people follow differing advice on what to do. Communication by authorities has been persistent but at times ineffective and this article dives into how one could enhance the visibility of credible information by governments, health authorities and scientists so that the negative impacts of the infodemic can be curbed. But, spewing scientific facts from a soapbox alone isn't enough — one is competing with all the other pieces of information and entertainment for attention and that needs to be taken into account. One of the key findings is that starting a dialogue helps more than just sending a one-way communiqué. Good science communication relies on the pillars of storytelling, cutting through the jargon and making the knowledge accessible.

While online platforms are structured such that polarization is encouraged through the algorithmic underpinnings of the system, we should not only engage when there is something that we disagree with, instead taking the time to amplify good science is equally important. Using platform-appropriate messaging, tailoring content to the audience and not squabbling over petty details, especially when they don't make a significant impact on the overall content helps to push out good science signals in the ocean of information pollution.

Clickbait-style headlines do a great job of hooking in people but when leading people into making a certain assumption and then debunking it, you stand the risk of spreading misinformation if someone doesn't read the whole thing, so in trying to make headlines engaging, it is important to consider what might be some unintended consequences if someone didn't read past the subtitle. Science isn't just about the findings, the process only gets completed when we have effective communication to the larger audience of the results, and now more than ever, we need accurate information to overpower the pool of misinformation out there.

## How Artificial Intelligence Can Save Journalism

There is a potential for AI to automate repetitive tasks and free up scarce resources towards more value-added tasks. With a declining business model and tough revenue situations, newsrooms and journalism at large are facing an existential crisis. Cutting costs while still keeping up high standards of reporting will require innovation on the part of newsrooms to adapt emerging technologies like AI.

For example, routine tasks like reporting on sports scores from games and giving updates on company earnings calls is already something that is being done by AI systems in several newsrooms around the world. This frees up time for journalists to spend their efforts on things like long-form journalism, data-driven and investigative journalism, analysis and feature pieces which require human depth and creativity. Machine translation also offers a handy tool making the work of journalists accessible to a wider audience without them having to invest in a lot of resources to do the translations themselves. This also brings up the possibility of smaller and resource-constrained media rooms to use their limited resources for doing in-depth pieces while reaching a wider audience by relying on automation.

Transcription of audio interviews so that reporters can work on fact-checking and other associated pieces also helps bring stories to fruition faster, which can be a boon in the rapidly changing environment. In the case of evolving situations like the pandemic, there is also the possibility of using AI to parse through large reams of data to find anomalies and alert the journalist of potential areas to cover. Complementing human skills is the right way to adopt AI rather than thinking of it as the tool that replaces human labor.

## Study: Facebook's Fake News Labels Have a Fatal Flaw

The article gives an explanation for why truth labels on stories are not as effective as we might think them to be because of something called the implied truth effect. Essentially, it states that when some things are marked as explicitly false and other false stories aren't, people believe them to be true even if they are outright false because of the lack of a label. Fact checking all stories manually is an insurmountable task for any platform and the authors of the study mention a few approaches that could potentially mitigate the spread of false content but none are a silver bullet. There is an ongoing and active community that researches how we might more effectively dispel disinformation but it's nascent and with the proliferation of AI systems, more work needs to be done in this arms race of building tools vs increasing capabilities of systems to generate believable fake content.

# 5. Humans and AI

## Go Deep: Research Summaries

### AI-Mediated Exchange Theory

This paper by Xiao Ma and Taylor W. Brown puts forth a framework that extends the well studied Social Exchange Theory (SET) to study human-AI interactions via mediation mechanisms. The authors make a case for how current research needs more interdisciplinary collaboration between technical and social science scholars stemming from a lack of shared taxonomy that places research in similar areas on separate grounds. They propose two axes of human/AI and micro/macro perspectives to visualize how researchers might better collaborate with each other. Additionally, they make a case for how AI agents can mediate transactions between humans and create potential social value as an emergent property of those mediated transactions.

As the pace of research progress quickens and more people from different fields engage in work on the societal impacts of AI, it is essential that we build on top of each other's work rather than duplicating efforts. Additionally, because of conventional differences in how research is published and publicized in the social sciences and technical domains, there's often a shallowness in the awareness of the latest work being done at the intersection of these two domains. What that means is that we need a shared taxonomy that allows us to better position research such that potential gaps can be discovered and areas of collaboration can be identified. The proposed two axes structure in the paper goes some distance in helping to bridge this current gap.

AI systems are becoming ever more pervasive in many aspects of our everyday lives and we definitely see a ton of transactions between humans that are mediated by automated agents. In some scenarios, they lead to net positive for society when they enable discovery of research content faster as might be the case for medical research being done to combat covid-19 but there might be negative externalities as well where they can lead to echo chambers walling off content from a subset of your network on social media platforms thus polarizing discussions and viewpoints. A better understanding of how these interactions can be engineered to skew positive will be crucial as AI agents get inserted to evermore aspects of our lives, especially ones that will have a significant impact on our lives.

We also foresee an emergence of tighter interdisciplinary collaboration that c shed light on these inherently socio-technical issues which don't have unidimensional solutions. With the rising awareness and interest from both social and technical sciences, the emerging work will be both timely and relevant to addressing challenges of the societal impacts of AI head on. As a part of the work being done at MAIEI we push for each of our undertakings to have an interdisciplinary team as a starting point towards achieving this mandate.

## Health Care, Capabilities, and AI Assistive Technologies

Most concerns when aiming to use technology within healthcare are along the lines of replacing human labor and the ones that are used in aiding humans to deliver care don't receive as much attention. With the ongoing pandemic, we've seen this come into the spotlight as well and this paper sets the stage for some of the ethical issues to watch out for when thinking about using AI-enabled technologies in the healthcare domain and how to have a discussion that is grounded in concrete moral principles.

An argument put forth to counter the use of AI solutions is that they can't "care" deeply enough about the patients and that is a valid concern, after all machines don't have empathy and other abilities required to have an emotional exchange with humans. But, a lot of the care work in hospitals is routine and professionalism asks for maintaining a certain amount of emotional distance in the care relationship. Additionally, in places where the ratio of patients/carers is high, they are unable to provide personalized attention and care anyways. In that respect, human-provided care is already "shallow" and the author cites research where care that is too deep actually hurts the carer when the patients become better and move out of their care or die. Thus, if this is the argument, then we need to examine more deeply our current care practices.

The author also posits that if this is indeed the state of care today, then it is morally less degrading to be distanced by a machine than by a human. In fact, the use of AI to automate routine tasks in the rendering of medical care will actually allow human carers to focus more on the emotional and human aspects of care.

Good healthcare, supposedly that provided by humans doesn't have firm grounding in the typical literature on the ethics of healthcare and technology. It's more so a list of things not to do but not positive guidance on what this kind of good healthcare looks like. Thus, the author takes a view that it must, at the very least, respect, promote and preserve the dignity of the patient.

Yet, this doesn't provide concrete enough guidance and we can expand on this to say that dignity is a) treating the patient as a human b) treating them as a part of a culture and community and c) treating them as a unique human. To add ev more concreteness, the author borrows from the work done in economics on the capabilities approach. This capabilities approach states that having the following 10 capabilities in their entirety is necessary for a human to experience dignity in their living: life, bodily health, bodily integrity, being able to use your senses, imaginations and thoughts, emotions, practical reasoning, affiliation, other species, play, and control over one's environment. This list applied to healthcare gives us a good guideline for what might constitute the kind of healthcare that we deem should be provided by humans, with or without the use of technology.

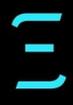

Now, the above list might seem too onerous for healthcare professionals but we need to keep in mind that healthcare to achieve a good life as highlighted by the capabilities approach things that are dependent on things beyond just the healthcare professionals and thus, the needs as mentioned above need to be distributed accordingly. The threshold for meeting them should be high but not so high that they are unachievable.

Principles are only sufficient for giving us some guidance for how to act in difficult situations or ethical dilemmas, they don't determine the outcome because they are only one element in the decision making process. We have to rely on the context of the situation and the moral surroundings of it. The criteria proposed are to be used in moral deliberation and should address whether the criterion applies to the situation, is it satisfied and is it sufficiently met (which is in reference to the threshold).

With the use of AI-enabled technology, privacy is usually cited as a major concern but the rendering of care is decidedly a non-private affair, imagine a scenario where the connection facilitated by technology allows for meeting the social and emotional needs of a terminal patient, if there is a situation where the use of technology allows for a better and longer life, then in these cases there can be an argument for sacrificing privacy to meet the needs of the patient. Ultimately, a balance needs to be struck between the privacy requirements and other healthcare requirements and privacy should not be blindly touted as the most important requirement.

Framing the concept of the good life with a view of restoring, maintaining and enhancing the capabilities of the human, one mustn't view eudaimonia as happiness but rather the achievement of the capabilities listed because happiness in this context would fall outside of the domain of ethics. Additionally, the author proposes the Care Experience Machine thought experiment that can meet all the care needs of a patient and asks the question if it would be morally wrong to plug in a patient into such a machine. While intuitively it might seem wrong, we struggle when trying to come up with concrete objections. As long as the patient feels cared for and has, from an objective standpoint, their care needs met, becomes hard to contest how such virtual care might differ from real care that is provided by humans.

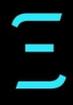

If one can achieve real capabilities, such as the need to have freedom of movement and interaction with peers outside of their care confinement and virtual reality technology enables that, then the virtual good life enhances the real good life – a distinction that becomes increasingly blurred as technology progresses.

Another moral argument put forward in determining whether to use technology-assisted healthcare is if it is too paternalistic to determine what is best for the patient. In some cases where the patient is unable to make decisions that restore, maintain and enhance their capabilities, such paternalism might be required but it must always be balanced with other ethical concerns and keeping in mind the capabilities that it enables for the patient.

When we talk about felt care and how to evaluate whether care rendered is good or not, we should not only look at the outcomes of the process through which the patient exits the healthcare context but also the realization of some of the capabilities during the healthcare process. To that end, when thinking about felt care, we must also take into account the concept of reciprocity of feeling which is not explicitly defined in the capabilities approach but nonetheless forms an important aspect of experiencing healthcare in a positive manner from the patient's perspective.

In conclusion, it is important to have an in-depth evaluation of technology assisted healthcare that is based on moral principles and philosophy, yet resting more on concrete arguments rather than just the high-level abstracts as they provide little practical guidance in evaluating different solutions and how they might be chosen to be used in different contexts. An a priori dismissal of technology in the healthcare domain, even when based on very real concerns like breach of privacy in the use of AI solutions which require a lot of personal data, begets further examination before arriving at a conclusion.

## **Go Wide: Article Summaries**

### **Ancient Animistic Beliefs Live On in Our Intimacy With Tech**

The article brings up some interesting points around how we bond with things that are not necessarily sentient and how our emotions are not discriminating when it comes to reducing loneliness and imprinting on inanimate objects. People experience surges in oxytocin as a consequence of such a bonding experience which further reinforces the relationship. This has effects for how increasing sentient-appearing AI systems might be used to manipulate humans into a "relationship" and potentially steer them towards making purchases, for example via chatbot interfaces by evoking a sense of trust. The article also makes a point about how such behaviour is akin to animism and in a sense forms a response to loneliness in the digital realm, allowing us to continue to hone our empathy skills for where they really matter, with other human beings.

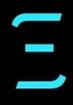

## Empathy Machine: Humans Communicate Better After Robots Show Their Vulnerable Side

With more and more of our conversations being mediated by AI-enabled systems online, it is important to see if robots can be harnessed to affect positive behaviour change in our interactions with each other. While there have been studies that demonstrate the positive impact that robots can have on influencing individual behaviour, this study highlighted how the presence of robots can influence human to human interactions as well. What the researchers found was that having a robot that displayed positive and affective behavior triggered more empathy from humans towards other humans as well as other positive behaviors like listening more and splitting speaking time amongst members more fairly. This is a great demonstration of how robots can be used to improve our interactions with each other. Another researcher pointed out that a future direction of interest would be to see how repeated exposure to such robot interactions can influence behaviour and if the effects so produced would be long-lasting even in the absence of the robot to participate in the interactions.

## At The Limits of Thought

Since time immemorial there has been a constant tussle between making predictions and being able to understand the underlying fundamentals of how those predictions worked. In the era of big data, those tensions are exacerbated as machines become more inscrutable while making predictions using ever-more higher-dimensional data which lies beyond intuitive understanding of humans. We try to reason through some of that high-dimensional data by utilizing techniques that either reduce the dimensions or visualize into 2- or 3-dimensions which by definition will tend to lose some fidelity. Bacon had proposed that humans should utilize tools to gain a better understanding of the world around them - until recently where the physical processes of the world matched quite well with our internal representations, this wasn't a big concern. But a growing reliance on tools means that we rely more on what is made possible by the tools as they measure and model the world.

Statistical intelligence and models often get things right but often they are host to reconstruction as to how they arrived at certain predictions. Models provide for abstractions of the world and often don't need to follow exactly the real-world equivalents. For example, while the telescope allows us to peer far into the distance, its construction doesn't completely mimic a biological eye. More so, radio telescopes that don't follow optics at all give us a unique view into distant objects which are just not possible if we rely solely on optical observations.

Illusions present us with a window into the limits of our perceptual systems and bring into focus the tension between the reality and what we think is the reality. Through a variety of examples like the Necker Cube, one can demonstrate that our perception and reality can often have gaps. A statistical analogue is the Simpson's Paradox where insights gleaned from one dataset are completely reversed when analyzed at a different scale or by combining multiple datasets. Accuracy paradoxes do something similar where underrepresentation in a dataset of a minority leads to poor performance of the predictions for those minorities, what is often dubbed as algorithmic bias.

"In just the same way that prediction is fundamentally bounded by sensitivity of measurement and the shortcomings of computation, understanding is both enhanced and diminished by the rules of inference."

In language models, we've seen that end-to-end deep learning systems that are opaque to our understanding perform quite a bit better than traditional machine translation approaches that rest on decades of linguistic research. This bears some resemblance to Searle's Chinese Room experiment where if we just look at the inputs and the outputs, there isn't a guarantee that the internal workings of the system work in exactly the way we expect them to.

"The most successful forms of future knowledge will be those that harmonise the human dream of understanding with the increasingly obscure echoes of the machine."

## A.I. Engineers Should Spend Time Training Not Just Algorithms, But Also The Humans Who Use Them

Abhishek Gupta (founder of the Montreal AI Ethics Institute) was featured in Fortune where he detailed his views on AI safety concerns in RL systems, the "token human" problem, and automation surprise among other points to pay attention to when developing and deploying AI systems. Especially in situations where these systems are going to be used in critical scenarios, humans operating in tandem with these systems and utilizing them as decision inputs need to gain a deeper understanding of the inherent probabilistic nature of the predictions from

The State of AI Ethics, June 2020                                                                                                       51

these systems and make decisions that take it into consideration rather than blindly trusting recommendations from an AI system because they have been accurate in 99% of the scenarios.

## Using Multimodal Sensing to Improve Awareness in Human-AI Interaction

With increasing capabilities of AI systems, and established research that demonstrates how human-machine combinations operate better than each in isolation, this paper presents a timely discussion on how we can craft better coordination between human and machine agents with the aim of arriving at the best possible understanding between them. This will enhance trust levels between the agents and it starts with having effective communication. The paper discusses how framing this from a human-computer interaction (HCI) approach will lead to achieving this goal. This is framed with intention-, context-, and cognition-awareness being the critical elements which would be responsible for the success of effective communication between human and machine agents.

## Different Intelligibility for Different Folks

Intelligibility is a notion that is worked on by a lot of people in the technical community who seek to shed a light on the inner workings of systems that are becoming more and more complex. Especially in the domains of medicine, warfare, credit allocation, judicial systems and other areas where they have the potential to impact human lives in significant ways, we seek to create explanations that might illuminate how the system works and address potential issues of bias and fairness.

However, there is a large problem in the current approach in the sense that there isn't enough being done to meet the needs of a diverse set of stakeholders who require different kinds of intelligibility that is understandable to them and helps them meet their needs and goals. One might argue that a deeply technical explanation ought to suffice and others kinds of explanations might be derived from that but it makes them inaccessible to those who can't parse well the technical details, often those who are the most impacted by such systems. The paper offers a framework to situate the different kinds of explanations such that they are able to meet the stakeholders where they are at and provide explanations that not only help them meet their needs but ultimately engender a higher level of trust from them by highlighting better both the capabilities and limitations of the systems.

The State of AI Ethics, June 2020                                                                 52

## Aligning AI to Human Values Means Picking the Right Metrics

AI value alignment is typically mentioned in the context of long-term AGI systems but this also applies to the narrow AI systems that we have today. Optimizing for the wrong metric leads to things like unrealistic and penalizing work schedules, hacking attention on video platforms, charging more money from poorer people to boost the bottomline and other unintended consequences.

Yet, there are attempts by product design and development teams to capture human well-being as metrics to optimize for. "How does someone feel about how their life is going?" is a pretty powerful question that gives a surprising amount of insight into well-being distanced from what might be influencing them at the moment because it makes them pause and reflect on what matters to them. But, capturing this subjective sentiment as a metric in an inherently quantitative world of algorithms is unsurprisingly littered with mines.

A study conducted by Facebook and supported by external efforts found that passive use of social media triggered feelings of ennui and envy while active use including interactions with others on the network led to more positive feelings. Utilizing this as a guiding light, Facebook strove to make an update that would be more geared towards enabling meaningful engagement rather than simply measuring the number of likes, shares and comments. They used user panels as an input source to determine what constituted meaningful interactions on the platform and tried to distill this into the well-being metrics. Yet, this suffered from several flaws, namely that the evaluation of this change was not publicly available and was based on the prior work comparing passive vs. active use of social media.

This idea of well-being optimization extends to algorithmic systems beyond social media platforms, for example, with how gig work might be better distributed on a platform such that income fluctuations are minimized for workers who rely on it as a primary source of earnings. Another place could be amending product recommendations to also capture environmental impacts such that consumers can incorporate that into their purchasing decisions apart from just the best price deals that they can find.

Participatory design is going to be a key factor in the development of these metrics; especially given the philosophy of "nothing about us without us" as a north star to ensure that there isn't an inherent bias in how well-being is optimized for. Often, we'll find that proxies will need to stand in for actual well-being in which case it is important to ensure that the metrics are not static and are revised in consultation with users at periodic intervals. Tapping into the process of double loop learning, an organization can not only optimize for value towards its shareholders but also towards all its other stakeholders. While purely quantitati metrics have obvious limitations when trying to capture something that is inherently subjective and qualitative, we need to attempt something in order to start and iterate as we go along.

## Why Lifelong Learning is the International Passport to Success

In a world where increasing automation of cognitive labor due to AI-enabled systems will dramatically change the future of labor, it is now more important than ever that we start to move away from a traditional mindset when it comes to education. While universities in the previous century rightly provided a great value in preparing students for jobs, as jobs being bundle of tasks and those tasks rapidly changing with some being automated, we need to focus more on training students for things that will take much longer to automate, for example working with other humans, creative and critical thinking and driving innovation based on insights and aggregating knowledge across a diversity of fields. Lifelong learning serves as a useful model that can impart some of these skills by breaking up education into modules that can be taken on an "at will" basis allowing people to continuously update their skills as the landscape changes. Students will go in and out of universities over many years which will bring a diversity of experiences to the student body, encouraging a more close alignment with actual skills as needed in the market. While this will pose significant challenges to the university system, innovations like online learning and certifications based on replenishment of skills like in medicine could overcome some of those challenges for the education ecosystem.

## You Can't Fix Unethical Design by Yourself

Individual actions are powerful, they create bottom-up change and empower advocates with the ability to catalyze larger change. But, when we look at products and services with millions of users where designs that are inherently unethical become part of everyday practice and are met with a slight shrug of the shoulders resigning to our fates, we need a more systematized approach that is standardized and widely practiced. Ethics in AI is having its moment in the spotlight with people giving talks and conferences focusing on it as a core theme yet it falls short of putting the espoused principles into practice.

More often than not, you have individuals, rank and file employees who go out of their way, often on personal time, to advocate for the use of ethical, safety and inclusivity in the design of systems, sometimes even at the risk of their employment. While such efforts are laudable, they lack widespread impact and awareness that is necessary to move the needle, we need leaders at the top who can affect sweeping changes to adopt these guidelines not just in letter but in spirit and then transmit them as actionable policies to their workforce. It needs to

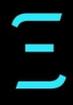
arrive at a point where people advocating for this change don't need to do so frc a place of moral and ethical obligations which customers can dispute but from a place of policy decisions which force disengagement for non-adherence to these policies. We need to move from talk to action not just at a micro but at a macro scale.

# 6. Jobs and Labor

## Go Deep: Research Summaries

### The Wrong Kind of AI? Artificial Intelligence and the Future of Labor Demand

Do increasing efficiency and social benefits stand in exclusion to each other when it comes to automation technology? With the development of the "right" kind of AI, this doesn't have to be the case. AI is a general purpose technology that has wide applications and being offered as a platform, it allows others to build advanced capabilities on top of existing systems creating an increasingly powerful abstraction layer with every layer.

According to the standard approach in economics, a rise in productivity is often accompanied with an increase in the demand for labor and hence a rise in wages along with standards of living. But, when there is a decoupling between the deployment of technology and the associated productivity accrual, it can lead to situations where we see more output but not a corresponding increase in the standards of living as the benefits accrue to capital owners rather than wage-earning labor which is distanced from the production lifecycle. This unevenness in the distribution of gains causes losses of jobs in one sector while increasing productivity in others, often masking effects at an aggregate level through the use of purely economic focused indicators like GDP growth rates.

The authors expound on how the current wave of automation is highly focused on labor replacement driven by a number of factors. When this comes in the form of automation that is just as good as labor but not significantly better, we get the negative effects as mentioned before, that is a replacement of labor without substantial increases in the standards of living. Most of these effects are felt by those in the lower rungs of the socio-economic ladder where they don't have alternate avenues for employment and ascent. A common message is that we just have to wait as we did in the case of the industrial revolution and new jobs will emerge that we couldn't have envisioned which will continue to fuel economic prosperity for all. This is an egregious comparison that overlooks that the current wave of automation is not creating simultaneous advances in technology that allow the emergence of a new class of tasks within jobs for which humans are well-suited. Instead, it's increasingly moving into domains that were strongholds of human skills that are not easily repeatable or reproducible. What we saw in the past was an avenue made available to workers to move out of low skill tasks agriculture to higher skill tasks in manufacturing and services.

Some examples of how AI development can be done the "right" way to create social benefits:

- In education, we haven't seen a significant shift in the way things are done for close to 200 years. It has been shown that different students have different learning styles and can benefit from personalized attention. While it is infeasible to do so in a traditional classroom model, AI offers the potential to track metrics on how the student interacts with different material, where they make mistakes, etc., offering insights to educators on how to deliver a better educational experience. This is accompanied by an increase in the demand for teachers who can deliver different teaching styles to match the learning styles of students and create better outcomes.

- A similar argument can be made in the field of healthcare where AI systems can allow medical staff to spend more time with patients offering them personalized attention for longer while removing the need for onerous and drudgery in the form of menial tasks like data entry.

- Industrial robots are being used to automate the manufacturing line often cordoning off humans for safety reasons. Humans are also decoupled from the process because of a difference in the level of precision that machines can achieve compared to humans. But we can get the best of both worlds by combining human flexibility and critical thinking to address problems in an uncertain environment with the high degree of preciseness of machines by creating novel interfaces, for example, by using Augmented Reality.

An important distinction that the authors point out in the above examples is that they are not merely the job of enablers, humans that are used to train machines in a transitory fashion, but those that genuinely complement machine skills.

There are market failures when it comes to innovation and in the past, governments have helped mitigate those failures via public-private partnerships that led to the creation of fundamental technologies like the internet. But, this has decreased over the past two decades because of smaller amount of resources being invested by the government in basic research and the technology revolution becoming centered in Silicon Valley which has a core focus on automation that replaces labor, and with that bias and their funding of university and academic studies, they are causing the best minds of the next generation to have the same mindset.

Markets are also known to struggle when there are competing paradigms and once one pulls ahead, it is hard to switch to another paradigm even if it might be more productive thus leading to an entrenchment of the dominant paradigm. The social opportunity cost of replacing labor is lower than the cost of labor, pushing the ecosystem towards labor replacing automation. Without accounting for these externalities, the ecosystem has little incentive to move towards the right kind of AI. This is exacerbated by tax incentives imposing costs on labor while providing a break on the use of capital. Additionally, areas where the right kind of AI can be developed don't necessarily fall into the cool domain of research and thus aren't prioritized by the research and development community. Let's suppose large advances were made in AI for health care. This would require accompanying retraining of support staff aside from doctors, and the high level bodies regulating the field would impose resistance, thus slowing down the adoption of this kind of innovation.

Ultimately, we need to lean on a holistic understanding of the way automation is going to impact the labor market and it will require human ingenuity to shape the social and economic ecosystems such that they create net positive benefits that are as widely distributed as possible. Relying on the market to figure this out on its own is a recipe for failure.

## Go Wide: Article Summaries

### AI Is Coming for Your Most Mind-Numbing Office Tasks

The labor impacts of AI require nuance in discussion rather than fear-mongering that veers between over-hyping and downplaying concerns when the truth lies somewhere in the middle. In the current paradigm of supervised machine learning, AI systems need a lot of data before becoming effective at their automation tasks. The bottom rung of this ladder consists of robotic process automation that merely tracks how humans perform a task (say, by tracking the clicks of humans as they go about their work) and ape them step by step for simple tasks like copying and pasting data across different places. The article gives an example of an organization that was able to minimize churn in their employees by more than half because of a reduction in data drudgery tasks like copying and pasting data across different systems to meet legal and compliance obligations. Economists point out that white-collar jobs like these and those that are middle-tier in terms of skills that require little training are at the highest risk of automation. While we're still ways away from AI taking up all jobs, there is a slow march starting from automating the most menial tasks, potentially freeing us up to do more value-added work.

## Tech's Shadow Workforce Sidelined, Leaving Social Media to the Machines

With a rising number of people relying on social media for the news, the potential for hateful content and misinformation spreading has never been higher. Content moderation on platforms like Facebook and YouTube is still largely a human endeavor where there are legions of contract workers that spend their days reviewing whether different pieces of content meet the community guidelines of the platform. Due to the spread of the pandemic and offices closing down, a lot of these workers have been asked to leave (they can't do this work from home as the platform companies explained because of privacy and legal reasons), leaving the platforms in the hands of automated systems.

The efficacy of these systems has always been questionable and as some examples in the article point out, they've run amok taking down innocuous and harmful content alike, seeming to not have very fine-tuned abilities. The problem with this is that legitimate sources of information, especially on subjects like COVID-19, are being discouraged because of their content being taken down and having to go through laborious review processes to have their content be approved again. While this is the perfect opportunity to experiment with the potential of using automated systems for content moderation given the traumatic experience that humans have to undergo as a part of this job, the chasms that need to be bridged still remain large between what humans have to offer and what the machines are capable of doing at the moment.

## Here's What Happens When an Algorithm Determines Your Work Schedule

Workplace time management and accounting are common practices but for those of us who work in places where schedules are determined by automated systems, they can have many negative consequences, a lot of which could be avoided if employers paid more attention to the needs of their employees. Clopening is the notion where an employee working at a retail location is asked to not only close the location at the end of the day but also arrive early the next day to open the location. This among other practices like breaks that are scheduled down to the minute and on-call scheduling (something that was only present in the realm of emergency services) wreak havoc on the physical and mental health of employees. In fact, employees surveyed have even expressed willingness to take pay cuts to have greater control over their schedules.

In some places with ad-hoc scheduling, employees are forced to be spontaneous with their home responsibilities like taking care of their children, errands, etc. While some employees try to swap shifts with each other, often even that becomes hard to do because others are also in similar situations. Some systems track customer demand and reduce pay for hours worked tied to that leading to added uncertainty even with their paychecks. During rush seasons, employees might be scheduled for back to back shifts ignoring their needs to be with families, something that a human manager could empathize with and accommodate for.

Companies supplying this kind of software hide behind the disclaimer that they don't take responsibility for how their customers use these systems which are often black-box and inscrutable to human analysis. This is a worrying trend that hurts those who are marginalized and those who require support when juggling several jobs just to make ends meet. Relying on automation doesn't absolve the employers of their responsibility towards their employees.

## Automation Will Take Jobs but AI Will Create Them

While the dominant form of discussion around the impacts of automation have been that it will cause job losses, this work from Kevin Scott offers a different lens into how jobs might be created by AI in the Rust Belt in the US where automation and outsourcing have been gradually stripping away jobs. Examples abound of how entrepreneurs and small business owners with an innovative mindset have been able to leverage advances in AI, coupling them with human labor to repurpose their businesses from areas that are no longer feasible to being profitable.

Precision farming utilizes things like drones with computer vision capabilities to detect hotspots with pests, disease, etc. in the big farms that would otherwise require extensive manual labor which would limit the size of the farms. Self-driving tractors and other automated tools also augment human effort to scale operations. The farm owners though highlight the opaqueness and complexity of such systems which make them hard to debug and fix themselves which sometimes takes away from the gains.

On the other hand, in places like nursing homes that get reimbursed based on the resource utilization rates by their residents, tools using AI can help minimize human effort in compiling data and let them spend more of their effort on human contact which is not something that AI succeeds on yet. While automation has progressed rapidly, the gains haven't been distributed equally.

In other places where old factories were shut down, some of them are now being utilized by ingenious entrepreneurs to bring back manufacturing jobs that cleverly combine human labor with automation to deliver high-quality, custom products to large enterprises. Thus, there will be job losses from automation but the onus lies with us in steering the progress of AI towards economic and ethical values that we believe in.

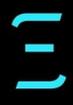

# 7. Law and Governance

## Go Deep: Research Summaries

### What's Next for AI Ethics, Policy, and Governance? A Global Overview

In this ongoing piece of work, the authors present the landscape of ethical documents that has been flooded with guidelines and recommendations coming from a variety of sectors including government, private organizations, and NGOs. Starting with a dive into the stated and unstated motivations behind the documents, the reader is provided with a systematic breakdown of the different documents prefaced with the caveat that where the motivations are not made explicit, one can only make a best guess based on the source of origin and people involved in its creation. The majority of the documents from the governmental agencies were from the Global North and western countries which led to a homogeneity of issues that were tackled and the recommendations often touted areas of interest that were specific to their industry and economical make up. This left research and development areas of interest like tourism and agriculture largely ignored which continue to play a significant role in the Global South. The documents from the former category were also starkly focused on gaining a competitive edge, which was often stated explicitly, with a potential underlying goal of attracting scarce, high-quality AI talent which could trigger brain drain from other countries that are not currently the dominant players in the AI ecosystem. Often, they were also positioning themselves to gain an edge and define a niche for themselves, especially in the case of countries that are non-dominant and thus overemphasizing the benefits while downplaying certain negative consequences that might arise from widespread AI use, like the displacement and replacement of labor.

For documents from private organizations, they mostly focused on self and collective regulation in an effort to pre-empt stringent regulations from taking effect. They also strove to tout the economic benefits to society at large as a way of de-emphasizing the unintended consequences. A similar dynamic as in the case of government documents played out here where the interests of startups and small and medium sized businesses were ignored and certain mechanisms proposed would be too onerous for such smaller organizations to implement this further entrenching the competitive advantage of larger firms.

The NGOs on the other hand seemed to have the largest diversity both in terms the participatory process of creation and the scope, granularity, and breadth of issues covered which gave technical, ethical, and policy implementation details making them actionable. Some documents like the Montreal Declaration for Responsible AI were built through an extensive public consultation process and consisted of an iterative and ongoing approach that the Montreal AI Ethics Institute contributed to as well. The IEEE document leverages a more formal standards making approach and consists of experts and concerned citizens from different parts of the world contributing to its creation and ongoing updating.

The social motivation is clearly oriented towards creating larger societal benefits, internal motivation is geared towards bringing about change in the organizational structure, external strategic motivation is often towards creating a sort of signaling to showcase leadership in the domain and also interventional to shape policy making to match the interests of those organizations.

Judging whether a document has been successful is complicated by a couple of factors: discerning what the motivations and the goals for the document were, and the fact that most implementations and use of the documents is done in a pick-and-choose manner complicating attribution and weight allocation to specific documents. Some create internal impacts in terms of adoption of new tools, change in governance, etc., while external impacts often relate to changes in policy and regulations made by different agencies. An example would be how the STEM education system needs to be overhauled to better prepare for the future of work. Some other impacts include altering customer perception of the organization as one that is a responsible organization which can ultimately help them differentiate themselves.

At present, we believe that this proliferation of ethics documents represents a healthy ecosystem which promotes a diversity of viewpoints and helps to raise a variety of issues and suggestions for potential solutions. While there is a complication caused by so many documents which can overwhelm people looking to find the right set of guidelines that helps them meet their needs, efforts such as the study being done in this paper amongst other efforts can act as guideposts to lead people to a smaller subset from which they can pick and choose the guidelines that are most relevant to them.

### AI Governance: A Holistic Approach to Implement Ethics in AI

The white paper starts by highlighting the existing tensions in the definitions of AI as there are many parties working on advancing definitions that meet their needs. One of the most commonly accepted ones frames AI systems as those that are able to adapt their behavior in response to interactions with the world independent of human control. Also, another popular framing is that AI is something that mimics human intelligence and is constantly shifting as a goal post as what was once perceived as AI, when sufficiently integrated and accepted in society becomes everyday technology.

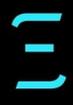

One thing that really stands out in the definitions section is how ethics are defined, which is a departure from a lot of other such documents. The authors talk about ethics as a set of principles of morality where morality is an assemblage of rules and values that guide human behavior and principles for evaluating that behavior. They take a neutral stand on the definition, a far cry from framing it as a positive inclination of human conduct to allow for diversity in embedding ethics into AI systems that are in concordance with local context and culture.

AI systems present many advantages which most readers are now already familiar given the ubiquity of AI benefits as touted in everyday media. One of the risks of AI-enabled automation is the potential loss of jobs, the authors make a comparison with some historical cases highlighting how some tasks and jobs were eliminated creating new jobs while some were permanently lost. Many reports give varying estimates for the labor impacts and there isn't yet a clear consensus on the actual impacts that this might have on the economy.

From a liability perspective, there is still debate as to how to account for the damage that might be caused to human life, health and property by such systems. In a strict product liability regime like Europe, there might be some guidance on how to account for this, but most regimes don't have specific liability allocations for independent events and decisions meaning users face coverage gaps that can expose them to significant harms.

By virtue of the complexity of deep learning systems, their internal representations are not human-understandable and hence lack transparency, which is also called the black box effect. This is harmful because it erodes trust from the user perspective, among other negative impacts.

Social relations are altered as more and more human interactions are mediated and governed by machines. We see examples of that in how our newsfeeds are curated, toys that children play with, and robots taking care of the elderly. This decreased human contact, along with the increasing capability of machine systems, examples of which we see in how disinformation spreads, will tax humans in constantly having to evaluate their interactions for authenticity or worse, relegation of control to machines to the point of apathy.

Since the current dominant paradigm in machine learning is that of supervised machine learning, access to data is crucial to the success of the systems and in cases where there aren't sufficient protections in place for personal data, it can le to severe privacy abuses. Self determination theory states that autonomy of humans is important for proper functioning and fulfillment, so an overreliance on AI systems to do our work can lead to loss of personal autonomy, which can lead to a sense of digital helplessness. Digital dementia is the cognitive equivalent where relying on devices for things like storing phone numbers, looking up information, etc. will over time lead to a decline in cognitive abilities.

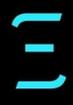

The echo chamber effect is fairly well studied, owing to the successful use of AI technologies to promulgate disinformation to the masses during the US Presidential elections of 2016. Due to the easy scalability of the systems, the negative effects are multiplicative in nature and have the potential to become run-away problems.

Given that AI systems are built on top of existing software and hardware, errors in the underlying systems can still cause failures at the level of AI systems. More so, given the statistical nature of AI systems, behaviour is inherently stochastic and that can cause some variability in response which is difficult to account for. Flash crashes in the financial markets are an example of this. For critical systems that require safety and robustness, there is a lot that needs to be done for ensuring reliability.

Building ethics compliance by design can take a bottom-up or top-down approach, the risk with a bottom-up approach is that by observing examples of human behaviour and extracting ethics principles from that, instead of getting things that are good for people, you get what's common. Hence, the report advocates for a top-down approach where desired ethical behavior is directly programmed into the system.

Casuistic approaches to embedding ethics into systems would work well in situations where there are simple scenarios, such as in healthcare when the patient has a clear directive of do-not-resuscitate. But, in cases where there isn't one and where it is not possible to seek a directive from the patient, such an approach can fail and it requires that programmers either in a top-down manner embed rules or the system learns from examples. Though, in a high-stakes situation like healthcare, it might not be ideal to rely on learning from examples because of skewed and limited numbers of samples.

A dogmatic approach would also be ill-advised where a system might slavishly follow a particular school of ethical beliefs which might lead it to make decisions that might be unethical in certain scenarios. Ethicists utilize several schools of thought when addressing a particular situation to arrive at a balanced decision. It will be crucial to consult with a diversity of stakeholders such that the nuances of different situations can be captured well. The WEF is working with partners come up with an "ethical switch" that will imbue flexibility on the system such that it can operate with different schools of thought based on the demands of the situation.The report also proposes the potential of utilizing a guardian AI system that can monitor other AI systems to check for compliance with different sets of AI principles.

Given that AI systems operate in a larger socio-technical ecosystem, we need to tap into fields like law and policy making to come up with effective ways of integrating ethics into AI systems, part of which can involve creating binding legal agreements that tie in with economic incentives.While policy making and law are often seen as slow to adapt to fast changing technology, there are a variety of benefits to be had, for example higher customer trust for services that have adherence to stringent regulations regarding privacy and data protection. This can serve to be a competitive advantage and counter some of the negative innovation barriers imposed by regulations. Another point of concern with these instruments is that they are limited by geography which leads to a patchwork of regulation that might apply on a product or service that spans several jurisdictions. Some other instruments to consider include: self-regulation, certification, bilateral investments treaties, contractual law, soft law, agile governance, etc.

The report highlights the initiatives by IEEE and WEF in creating standards documents. The public sector through its enormous spending power can enhance the widespread adoption of these standards such as by utilizing them in procurement for AI systems that are used to interact with and serve their citizens. The report also advocates for the creation of an ethics board or Chief Values Officer as a way of enhancing the adoption of ethical principles in the development of products and services.

For vulnerable segments of the population, for example children, there need to be higher standards of data protection and transparency that can help parents make informed decisions about which AI toys to bring into their homes. Regulators might play an added role of enforcing certain ethics principles as part of their responsibility. There also needs to be broader education for AI ethics for people that are in technical roles.

Given that there are many negative applications of AI, it shouldn't preclude us from using AI systems for positive use cases, a risk assessment and prudent evaluation prior to use is a meaningful compromise. That said, there are certain scenarios where AI shouldn't be used at all and that can be surfaced through the risk or impact assessment process.

There is a diversity of ethical principles that have been put forth by vario organizations, most of which are in some degree of accordance with local laws, regulations, and value sets. Yet, they share certain universal principles across all of them. One concern highlighted by the report is on the subject of how even widely accepted and stated principles of human rights can be controversial when translated into specific mandates for an AI system. When looking at AI-enabled toys as an example, while they have a lot of privacy and surveillance related issues, in countries where there isn't adequate access to education, these toys could be seen as a medium to impart precision education and increase literacy rates. Thus, the job of the regulator becomes harder in terms of figuring out how to balance the positive and negative impacts of any AI product. A lot of it depends on the context and surrounding socio-economic system as well.

Given the diversity in ethical values and needs across communities, an approach might be for these groups to develop and apply non-binding certifications that indicate whether a product meets the ethical and value system of that community. Since there isn't a one size fits all model that works, we should aim to have a graded governance structure that has instruments in line with the risk and severity profile of the applications.

Regulation in the field of AI thus presents a tough challenge, especially given the interrelatedness of each of the factors. The decisions need to be made in light of various competing and sometimes contradictory fundamental values. Given the rapid pace of technological advances, the regulatory framework needs to be agile and have a strong integration into the product development lifecycle. The regulatory approach needs to be such that it balances speed so that potential harms are mitigated with overzealousness that might lead to ineffective regulations that stifle innovation and don't really understand well the technology in question.

## Beyond a Human Rights Based Approach To AI Governance: Promise, Pitfalls and Plea

AI is currently enjoying a summer of envy after having gone through a couple of winters of disenchantment, with massive interest and investments from researchers, industry and everyone else there are many uses of AI to create societal benefits but they aren't without their socio-ethical implications. AI systems are prone to biases, unfairness and adversarial attacks on their robustness among other real-world deployment concerns. Even when ethical AI systems are deployed for fostering social good, there are risks that they cater to only a particular group to the detriment of others.

Moral relativism would argue for a diversity of definitions as to what constitut good AI which would depend on the time, context, culture and more. This would be reflected in market decisions by consumers who choose products and services that align with their moral principles but it poses a challenge for those trying to create public governance frameworks for these systems. This dilemma would push regulators towards moral objectivism which would try and advocate for a single set of values that are universal making the process of coming up with a shared governance framework easier. A consensus based approach utilized in crafting the EC Trustworthy AI guidelines settled on human rights as something that everyone can get on board with.

Given the ubiquity in the applicability of human rights, especially with their legal enshrinement in various charters and constitutions, they serve as a foundation to create legal, ethical and robust AI as highlighted in the EC Trustworthy AI guidelines. Stressing on the importance of protecting human rights, the guidelines advocate for a Trustworthy AI assessment in case that an AI system has the potential to negatively impact the human rights of an individual, much like the better established data protection impact assessment requirement under the GDPR. Additional requirements are imposed in terms of ex-ante oversight, traceability, auditability, stakeholder consultations, and mechanisms of redress in case of mistakes, harms or other infringements.

The universal applicability of human rights and their legal enshrinement also renders the benefits of established institutions like courts whose function is to monitor and enforce these rights without prejudice across the populace. But they don't stand uncontested when it comes to building good AI systems; they are often seen as too Western, individualistic, narrow in scope and abstract to be concrete enough for developers and designers of these systems. Some arguments against this are that they go against the plurality of value sets and are a continued form of former imperialism imposing a specific set of values in a hegemonic manner. But, this can be rebutted by the signing of the original Universal Declaration of Human Rights that was done by nations across the world in an international diplomatic manner. However, even despite numerous infringements, there is a normative justification that they ought to be universal and enforced.

While human rights might be branded as too individual focused, potentially creating a tension between protecting the rights of individuals to the detriment of societal good, this is a weak argument because stronger protection of individual rights has knock-on social benefits as free, healthy and well-educated (among other individual benefits) creates a net positive for society as these individuals are better aware and more willing to be concerned about societal good.

While there are some exceptions to the absolute nature of human rights, most are well balanced in terms of providing for the societal good and the good of others while enforcing protections of those rights. Given the long history of enforcement and exercises in balancing these rights in legal instruments, there is a rich jurisprudence on which people can rely when trying to assess AI systems.

While human rights create a social contract between the individual and the state, putting obligations on the state towards the individual but some argue that they don't apply horizontally between individuals and between an individual and a private corporation. But, increasingly that's not the case as we see many examples where the state intervenes and enforces these rights and obligations between an individual and a private corporation as this falls in its mandate to protect rights within its jurisdiction.

The abstract nature of human rights, as is the case with any set of principles rather than rules, allows them to be applied to a diversity of situations and to hitherto unseen situations as well. But, they rely on an ad-hoc interpretation when enforcing them and are thus subjective in nature and might lead to uneven enforcement across different cases. Under the EU, this margin of appreciation is often criticized in the sense that it leads to weakening and twisting of different principles but this deferment to those who are closer to the case actually allows for a nuanced approach which would be lost otherwise.

On the other hand we have rules which are much more concrete formulations and thus have a rigid definition and limited applicability which allows for uniformity but it suffers from inflexibility in the face of novel scenarios.

Yet, both rules and principles are complementary approaches and often the exercise of principles over time leads to their concretization into rules under existing and novel legal instruments.

While human rights can thus provide a normative, overarching direction for the governance of AI systems, they don't provide the actual constituents for an applicable AI governance framework. For those that come from a non-legal background, often technical developer and designers of AI systems, it is essential that they understand their legal and moral obligations to codify and protect these rights in the applications that they build. The same argument cuts the other way, requiring a technical understanding of how AI systems work for legal practitioners such that they can meaningfully identify when breaches might have occurred. This is also important for those looking to contest claims of breaches of their rights in interacting with AI systems.

This kind of enforcement requires a wide public debate to ensure that they f within accepted democratic and cultural norms and values within their context. While human rights will continue to remain relevant even in an AI systems environment, there might be novel ways in which breaches might occur and those might need to be protected which require a more thorough understanding of how AI systems work. Growing the powers of regulators won't be sufficient if there isn't an understanding of the intricacies of the systems and where breaches can happen, thus there is more of a need to enshrine some of those responsibilities in law to enforce this by the developers and designers of the system.

## **Go Wide: Article Summaries**

### **This Is The Year Of AI Regulations**

Given the large public awareness and momentum that built up around the ethics, safety and inclusion issues in AI, we will certainly see a lot more concrete actions around this in 2020. The article gives a few examples of Congressional Hearings on these topics and advocates for the industry to come up with some standards and definitions to aid the development of meaningful regulations. Currently, there isn't a consensus on these definitions and it leads to varying approaches addressing the issues at different levels of granularity and angles. What this does is create a patchwork of incoherent regulations across domains and geographies that will ultimately leave gaps in effectively mitigating potential harms from AI systems that can span beyond international borders. While there are efforts underway to create maps of all the different attempts of defining principle sets, we need a more coordinated approach to bring forth regulations that will ultimately protect consumer safety.

# 8. Privacy

## Go Deep: Research Summaries

**Apps Gone Rogue: Maintaining Personal Privacy in an Epidemic**

In containing an epidemic the most important steps include quarantine and contact tracing for more effective testing. While before, this process of contact tracing was hard and fraught with errors and omissions, relying on memories of individuals, we now carry around smartphones which allow for ubiquitous tracking ability that is highly accurate. But such ubiquity comes with invasion of privacy and possible limits on freedoms of citizens. Such risks need to be balanced with public interest in mind while using enhanced privacy preserving techniques and any other measures that center citizen welfare in both a collective and individual sense.

For infections that can be asymptomatic in the early days, like the COVID-19, it is essential to have contact tracing, which identifies all people that came in close contact with an infected person and might spread the infection further. This becomes especially important when you have a pandemic at hand, burdening the healthcare system and testing every person is infeasible.An additional benefit of contact tracing is that it mitigates resurgence of the peaks of infection.

R0 determines how quickly a disease will spread and is dependent on three factors (period of infection, contact rate and mode of transmission) out of which the first and third are fixed so we're only left with control over the contact rate.With an uptake of an application that facilitates contact tracing, the amount of reduction in contact rate is an increasing return because of the number of people that might come in contact with an infected person and thus, we get a greater reduction of R0 in terms of percentage compared to the percentage uptake of the application in the population. Ultimately, reducing R0 to below 1 leads to a slowdown in the spread of the infection thus helping the healthcare system cope up with the sudden stresses that are brought on by pandemic peaks.

One of the techniques that governments or agencies responsible for public health use is broadcasting in which the information of diagnosed carriers is made public via various channels but it carries severe issues like exposing private information of individuals and businesses where they might have been which can trigger stigma, ostracization and unwarranted punitive harm. It also suffers from the problem of people needing to access this source of information of their own volition and then

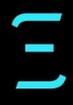

self-identify (and remember correctly) if they've been in recent contact with diagnosed carrier.

Selective broadcasting is a more restricted form of the above where information about diagnosed carriers is shared to a select group of individuals based on location proximity in which case the user's location privacy would have to be compromised and in another vector of dissemination, messages are sent to all users but filtered on device for their specific location and is not reported back to the broadcaster. But, the other second-order negative effects remain the same as broadcasting. Both though require the download of an application which might decrease the uptake of it by people.

Unicasting is when messages are sent tailored specifically to each user and they require the download of an app which needs to be able to track timestamps and location and has severe consequences in terms of government surveillance and abuse.

Participatory sharing is a method where diagnosed carriers voluntarily share their information and thus have more data control but it still relies on individual action both on the sender and receiver and its efficacy is questionable at best. There is also a risk of abuse by malicious actors to spread misinformation and seed chaos in society via false alarms.

Private Kit: Safe Paths is an open-source solution developed by MIT that allows for contact tracing in a privacy preserving way. It utilizes the encrypted location trail of a diagnosed carrier who chooses to share that with public health agencies and then other users who are also using the solution can pull this data and via their own logged location trail get a result of they've been in close contact with a diagnosed carrier. In the later phases of development of this solution, the developers will enable a mix of participatory sharing and unicasting to further prevent possible data access by third parties including governments for surveillance purposes.

Risks of contact tracing include possible public identification of the diagnosed carrier and severe social stigma that arises as a part of that. Online witch hunts to try and identify the individual can often worsen the harassment and include spreading of rumors about their personal lives. The privacy risks for both individuals and businesses have potential for severe harm, especially during times of financial hardship, this might be very troublesome.

Privacy risks also extend to non-users because of proximal information that can be derived from location trails, such as employees that work at particular businesses that were visited by a diagnosed carrier. It can also bring upon the same stigma and ostracization to the family members of these people.

Without meaningful alternatives, especially in health and risk assessment during a pandemic, obtaining truly informed consent is a real challenge that doesn't yet have any clear solutions.

Along with information, be it through any of the methods identified above, it is very important to provide appropriate context and background to the alerts to prevent misinformation and panic from spreading especially for those with low health, digital and media literacy. On the other hand, some might not take such alerts seriously and increase the risk for public health by not following required measures such as quarantine and social distancing.

Given the nature of such solutions, there is a significant risk of data theft from crackers as is the case for any application that collects sensitive information like health status and location data. The solutions can also be used for fraud and abuse, for example, by blackmailing business owners and demanding ransom, failing to pay which they would falsely post information that they're diagnosed carriers and have visited their place of business.

Contact tracing technology requires the use of a smartphone with GPS and some vulnerable populations might not always have such devices available like the elderly, homeless and people living in low-income countries who are at high risk of infection and negative health outcomes. Ensuring that technology that works for all will be an important piece to mitigating the spread effectively.

There is an inherent tradeoff between utility from the data provided and the privacy of the data subjects. Compromises may be required for particularly severe outbreaks to manage the spread.

The diagnosed carriers are the most vulnerable stakeholders in the ecosystem of contact tracing technology and they require the most protection. Adopting open-source solutions that are examinable by the wider technology ecosystem can engender public trust. Additionally, having proper consent mechanisms in place and exclusion of the requirement of extensive third party access to the location data can also help allay concerns. Lastly, time limits on the storage and use of the location trails will also help address privacy concerns and increase uptake and use of the application in supporting public health measures.

For geolocation data that might affect businesses, especially in times of economic hardship, information release should be done such that they are informed prior to the release of the information but there is little else in current methods that can both protect privacy and at the same time provide sufficient data utility.

For those without access to smartphones with GPS, providing them with sor information on contact tracing can still help their communities. But, one must present information in a manner that accounts for variation in health literacy levels so an appropriate response is elicited from the people. Alertness about potential misinformation and educational awareness are key during times of crises to encourage people to have measured responses following the best practices as advised by health agencies rather than those based on fear mongering by ill informed and/or malicious actors.

Encryption and other cybersecurity best practices for data security and privacy are crucial for the success of the solution. Time limits on holding data for COVID-19 is recommended at 14-37 days, the period of infection, but for an evolving pandemic one might need it for longer for more analysis. Tradeoffs need to be made between privacy concerns and public health utility. Different agencies and regions are taking different approaches with varying levels of efficacy and only time will tell how this change will be best managed. It does present an opportunity though for creating innovative solutions that both allow for public sharing of data and also reduce privacy intrusions.

## Maximizing Privacy and Effectiveness in COVID-19 Apps

While the insights presented in this piece of work are ongoing and will continue to be updated, we felt it important to highlight the techniques and considerations compiled by the OpenMined team as it is one of the few places that adequately capture, in a single place, most of the technical requirements needed to build a solution that respects fundamental rights while balancing them with public health outcomes as people rush to make AI-enabled apps to combat COVID-19. Most articles and research work coming out elsewhere are very scant and abstract in the technical details that would be needed to meet the ideals of respecting privacy and enabling health authorities to curb the spread of the pandemic.

The four key techniques that will help preserve and respect rights as more and more people develop AI-enabled applications to combat COVID-19 are: on-device data storage and computation, differential privacy, encrypted computation and privacy-preserving identity verification.

The primary use cases, from a user perspective, for which apps are being built are to get: proximity alerts, exposure alerts, information on planning trips, symptom analysis and demonstrate proof of health. From a government and health authorities perspective, they are looking for: fast contact tracing, high-precision self-isolation requests, high-precision self-isolation estimation, high-precision symptomatic citizen estimation and demonstration of proof of health.

While public health outcomes are at the top of the mind for everyone, the abo use cases are trying to achieve the best possible tradeoff between economic impacts and epidemic spread. Using the techniques highlighted in this work, it is possible to do so without having to erode the rights of citizens.

This living body of work is meant to serve as a high-level guide along with resources to enable both app developers and verifiers implement and check for privacy preservation which has been the primary pushback from citizens and civil activists. Evoking a high degree of trust from people will improve adoption of the apps developed and hopefully allow society and the economy to return to normal sooner while mitigating the harmful effects of the epidemic.

There is a fair amount of alignment in the goals of both individuals and the government with the difference being that the government is looking at aggregate outcomes for society. Some of the goals shared by governments across the world include: preventing the spread of the disease, eliminating the disease, protecting the healthcare system, protecting the vulnerable, adequately and appropriately distributing resources, preventing secondary breakouts, minimizing economic impacts and panic.

The need for digital contact tracing is important because manual interventions are usually highly error prone and rely on human memory to trace how the person might have come in contact with. The requirement for high-precision self-isolation requests will avoid the need for geographic quarantines where everyone in an area is forced to self-isolate which leads to massive disruptions in the economy and can stall the delivery of essential services like food, electricity and water. The additional benefits of high-precision self-isolation is that it can help create an appropriate balance between economic harms and epidemic spread.

High-precision symptomatic citizen estimation is a useful application in that it allows for more fine-grained estimation of the number of people that might be affected beyond what the test results indicate which can further strengthen the precision of other measures that are undertaken. A restoration of normalcy in society is going to be crucial as the epidemic starts to ebb, in this case, having proof of health that helps to determine the lowest risk individuals will allow for them to participate in public spaces again further bolstering the supply of essential services and relieving the burden from a small subset of workers who are participating.

To service the needs of both what the users want and what the government wants, we need to be able to collect the following data: historical and current absolute location, historical and current relative position and verified group identity, where group refers to any demographic that the government might be interested in, for example, age or health status.

To create an application that will meet these needs, we need to collect data from a variety of sources, compute aggregate statistics on that data and then set up some messaging architecture that communicates the results to the target population. The toughest challenges lie in the first and second parts of the process above, especially to do the second part in a privacy-preserving manner.

For historical and current absolute location, one of the first options considered by app developers is to record GPS data in the background. Unfortunately, this doesn't work on iOS devices and even then has several limitations including coarseness in dense, urban areas and usefulness only after the app has been running on the user device for some time because historical data cannot be sourced otherwise. An alternative would be to use Wi-Fi router information which can give more accurate information as to whether someone has been self-isolating or not based on whether they are connected to their home router. There can be historical data available here which makes it more useful though there are concerns with lack of widespread Wi-Fi connectivity in rural areas and tracking when people are outside homes. Other ways of obtaining location data could be from existing apps and services that a user uses – for example, history of movements on Google Maps which can be parsed to extract location history. There is also historical location data that could be pieced together from payments history, cars that record location information and personal cell tower usage data.

Historical and current relative data is even more important to map the spread of the epidemic and in this case, some countries like Singapore have deployed Bluetooth broadcasting as a means of determining if people have been in close proximity. The device broadcasts a random number (which could change frequently) which is recorded by devices passing by close to each other and in case someone is tested positive, this can be used to alert people who were in close proximity to them. Another potential approach highlighted in the article is to utilize gyroscope and ambient audio hashes to determine if two people might have been close together, though Bluetooth will provide more consistent results. The reason to use multiple approaches is the benefit of getting more accurate information overall since it would be harder to fake multiple signals.

Group membership is another important aspect where the information can be used to finely target messaging and calculating aggregate statistics. But, for some types of group membership, we might not be able to rely completely on self-reported data. For example, health status related to the epidemic would require verification from an external third-party such as a medical institution or testing facility to minimize false information.

The State of AI Ethics, June 2020                                                                                  76

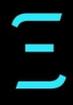

There are several privacy preserving techniques that could be applied to application given that you have: confirmed COVID-19 patient data in a cloud, all other user data on each individual's device, and data on both the patients and the users including historical and current absolute and relative locations and group identifier information.

Private set intersections can be used to calculate whether two people were in proximity to each other based on their relative and absolute location information. Private set intersection operates similarly to normal set intersection to find elements that are common between two sets but does so without disclosing any private information from either of the sets. This is important because performing analysis even on pseudonymized data without using privacy preservation can leak a lot of information.

Differential privacy is another critical technique to be utilized, DP consists of providing mathematical guarantees (even against future data disclosures) that analysis on the data will not reveal whether or not your data was part of the dataset. It asserts that from the analysis, one is not able to learn anything about your data that they wouldn't have been able to learn from other data about you. Google's battle-tested C++ library is a great resource to start along with the Python wrapper created by the OpenMined team.

To address the need for verified group identification, one can utilize the concept of a private identity server. It essentially functions as a trusted intermediary between a user that wants to provide a claim and another party that wants to verify the claim. It functions by querying a service from which it can verify whether the claim is true and then serve that information up to the party wishing to verify the claim without giving away personal data. While it might be hard to trust a single intermediary, this role can be decentralized to provide for obtaining a higher degree of trust by relying on a consensus mechanism.

# Go Wide: Article Summaries

## A Crisis of Ethics in Technology Innovation

Building on theory from management studies by Christensen et al. the authors of this article dive into how leaders of tech organizations, especially upstarts that are rapid in the disruption of incumbents should approach the accompanying responsibilities that come with a push into displacing existing paradigms of how an industry works. When there is a decoupling of different parts of the value chain in how a service is delivered, often the associated protections that apply to the entire pipeline fall by the wayside because of distancing from the end user and a diffusion of responsibility across multiple stakeholders in the value chain.

While end users driven innovation will seek to reinforce such models, regulations and protections are never at the top of such demands and they create a burden on the consumers once they realize that things can go wrong and negatively affect them. The authors advocate for the leaders of the companies to proactively employ a systems thinking approach to identify different parts that they are disrupting, how that might affect users, what would happen if they become the dominant player in the industry and then apply lessons learned from such an exercise to pre-emptively design safeguards into the system to mitigate unintended consequences.

## How Much Privacy Are You Entitled to During a Pandemic?

Many countries are looking at utilizing existing surveillance and counter-terrorism tools to help track the spread of the coronavirus and are urging tech companies and carriers to assist with this. The US is looking at how they can tap into location data from smartphones, following in the heels of Israel and South Korea that have deployed similar measures. While extraordinary measures might be justified given the time of crisis we're going through, we mustn't lose sight of what behaviors we are normalizing as a part of this response against the pandemic. Russia and China are also using facial recognition technologies to track movements of people, while Iran is endorsing an app that might be used as a diagnosis tool.

Expansion of the boundaries of surveillance capabilities and government powers is something that is hard to reign back in once a crisis is over. In some cases, like the signing of the Freedom Act in the USA reduced government agency data collection abilities that were expanded under the Patriot Act. But, that's not always the case and even so, the powers today exceed those that existed prior to the enactment of the Patriot Act. What's most important is to ensure that decisions policy makers take today keep in mind the time limits on such expansion of powers and do trigger a future privacy crisis because of it.

## Translating a Surveillance Tool into a Virus Tracker for Democracies

While no replacement for social distancing, a virus tracking tool putting into practice the technique of contact tracing is largely unpalatable to Western democracies because of expectations of privacy and freedom of movement. A British effort underway to create an app that meets democratic ideals of privacy and freedom while also being useful in collecting geolocation data to aid in the virus containment efforts. It is based on the notion of participatory sharing, relying on people's sense of civic duty to contribute their data in case they test positive.

While in the USA, discussions between the administration and technology companies has focused on large scale aggregate data collection, in a place like the UK with a centralized healthcare system, there might be higher trust levels in sharing data with the government. While the app doesn't require uptake by everyone to be effective, but a majority of the people would need to use it to bring down the rate of spread. The efficacy of the solution itself will rely on being able to collect granular location data from multiple sources including Bluetooth, Wi-Fi, cell tower data, and app check-ins.

## Don't Like Dystopian Surveillance? Flatten the Coronavirus Curve

A lot of high level CDC officials are advising that if people in the USA don't follow best practices of social distancing, sheltering in place, and washing hands regularly, the outbreak will not have peaked and the infection will continue to spread, especially hitting those who are the most vulnerable including the elderly and those with pre-existing conditions. On top of the public health impacts, there are also concerns of growing tech-enabled surveillance which is being seriously explored as an additional measure to curb the spread.

While privacy and freedom rights are enshrined in the constitution, during times of crisis, government and justice powers are expanded to allow for extraordinary measures to be adopted to restore the safety of the public. This is one of those times and the US administration is actively exploring options in partnership with various governments on how to effectively combat the spread of the virus including the use of facial recognition technology. This comes shortly after the techlash and a potential bipartisan movement to curb the degree of data collection by large firms, which seem to have come to a halt as everyone scrambles to battle the coronavirus.

## As Coronavirus Surveillance Escalates, Personal Privacy Plummets

Regional governments are being imbued with escalated powers to override local legislations in an effort to curb the spread of the virus. The article provides details on efforts by various countries across the world, yet we only have preliminary data on the efficacy of each of those measures and we require more time before being able to judge which of them is the most effective. That said, in a pandemic that is fast spreading, we don't have the luxury of time and must make decisions as quickly as possible using the information at hand, perhaps using guidance from prior crises.

But, what we've seen so far is minimal coordination from agencies across the world and that's leading to ad-hoc, patchy data use policies that will leave the marginalized more vulnerable. Strategies that use public disclosure of those that have been tested positive in the interest of public health are causing harm to the individuals and other individuals that are close to them such as their families. As experienced by a family in New York, online vigilantes attempted to harass the individuals while their family pleaded and communicated measures that they had taken to isolate themselves to safeguard others. Unfortunately, the virus might be bringing out the worst in all of us.

## How to Cover Your Tracks Every Time You Go Online

An increasing number of tools and techniques are being used to track our behaviour online and while some may have potential benefits, for example, the use of contact tracing to potentially improve public health outcomes, if this is not done in a privacy-preserving manner, there can be severe implications for your data rights. But, barring special circumstances like the current pandemic, there are a variety of simple steps that you can take to protect your privacy online. These range from simple steps like using an incognito browser window which doesn't store any local information about your browsing on your device to using things like VPNs which protect snooping of your browsing patterns even from your ISP.

When it comes to using the incognito function of your browser, if you're logged into a service online, there isn't any protection though it does prevent storing cookies on your device. With VPNs, there is an implicit trust placed in the provider of that service to not store logs of your browsing activity. An even more secure option is to use a privacy-first browser like Tor which routes your traffic requests through multiple locations making tracking hard. There is also an OS built around this called TailsOS that offers tracking protection from the device perspective as well not leaving any trace on the host machine allowing you to boot up from a USB.

The EFF also provides a list of tools that you can use to get a better grip on your privacy as you browse online.

## Who's Allowed to Track My Kids Online?

Under the Children's Online Privacy Protection Act, the FTC levied its largest fine yet of $170m on YouTube last year for failing to meet requirements of limiting personal data collection for children under the age of 13. Yet, as many advocates of youth privacy point out, the fines, though they appear to be large, don't do enough to deter such personal data collection. They advocate for a stronger version of the Act while requiring more stringent enforcement from the FTC which has been criticized for slow responses and a lack of sufficient resources. While the current Act requires parental consent for children below 13 to be able to utilize a service that might collect personal data, there is no verification performed on the self-declared age provided at the time of sign up which weakens the efficacy of this requirement. Secondly, the sharp threshold of 13 years old immediately thrusts children into an adult world once they cross that age and some people are advocating for a more graduated approach to the application of privacy laws.

## Chinese Citizens Are Racing Against Censors to Preserve Coronavirus Memories on GitHub

Given that such a large part of the news cycle is dominated by the coronavirus, we tend to forget that there might be censors at work that are systematically suppressing information in an attempt to diminish the seriousness of the situation. Some people are calling GitHub the last piece of free land in China and have utilized access to it to document news stories and people's first hand experiences in fighting the virus before they are scrubbed from local platforms like WeChat and Weibo. They hope that such documentation efforts will not only shed light on the reality and on the ground situation as it unfolds but also give everyone a voice and hopefully provide data to others who could use it to track the movement of the virus across the country. Such times of crisis bring out creativity and this attempt highlights our ability as a species to thrive even in a severely hostile environment.

## Can I Opt Out of Facial Scans at the Airport?

There is a clear economic and convenience case to be made (albeit for the majority, not for those that are judged to be minorities by the system and hence get subpar performance from the system) where you get faster processing and boarding times when trying to catch a flight. Yet, for those that are more data-privacy minded, there is an option to opt-out though leveraging that option doesn't necessarily mean that the alternative will be easy, as the article points out, travelers have experienced delays and confusion from the airport staff. Often, the alternatives are not presented as an option to travelers giving a false impression that people have to submit to facial recognition systems. Some civil rights and ethics researchers tested the system and got varying mileage out of their experiences but urge people to exercise the option to push back against technological surveillance.

**With Painted Faces, Artists Fight Facial Recognition Tech**

London is amongst a few cities that has seen public deployment of live facial recognition technology by law enforcement with the aim of increasing public safety. But, more often than not, it is done so without public announcement and an explanation as to how this technology works, and what impacts it will have on people's privacy. As discussed in an article by MAIEI on smart cities, such a lack of transparency erodes public trust and affects how people go about their daily lives. Several artists in London as a part of regaining control over their privacy and to raise awareness are using the technique of painting adversarial patterns on their faces to confound facial recognition systems. They employ highly contrasting colors to mask the highlights and shadows on their faces and practice pattern use as created and disseminated by the CVDazzle project that advocates for many different styles to give the more fashion-conscious among us the right way to express ourselves while preserving our privacy. Such projects showcase a rising awareness for the negative consequences of AI-enabled systems and also how people can use creative solutions to combat problems where laws and regulations fail them.

# 9. Security and Risk

## Go Deep: Research Summaries

### Adversarial Machine Learning – Industry Perspectives

There is mounting evidence that organizations are taking seriously the threats arising from malicious actors geared towards attacking ML systems. This is supported by the fact that organizations like ISO and NIST are building up frameworks for guidance on securing ML systems, that working groups from the EU have put forth concrete technical checklists for the evaluating the trustworthiness of ML systems and that ML systems are becoming key to the functioning of organizations and hence they are inclined to protect their crown jewels.

The organizations surveyed as a part of this study spanned a variety of domains and were limited to those that have mature ML development. The focus was on two personas: ML engineers who are building these systems and security incident responders whose task is to secure the software infrastructure including the ML systems. Depending on the size of the organization, these people could be in different teams, same team or even the same person. The study was also limited to intentional malicious attacks and didn't investigate the impacts of naturally occurring adversarial examples, distributional shifts, common corruption and reward hacking.

Most organizations that were surveyed as a part of the study were found to primarily be focused on traditional software security and didn't have the right tools or know-how in securing against ML attacks. They also indicated that they were actively seeking guidance in the space. Most organizations were clustered around concerns regarding data poisoning attacks which was probably the case because of the cultural significance of the Tay chatbot incident. Additionally, privacy breaches were another significant concern followed by concerns around model stealing attacks that can lead to the loss of intellectual property. Other attacks such as attacking the ML supply chain and adversarial examples in the physical domain didn't catch the attention of the people that were surveyed as a part of the study.

One of the gaps between reality and expectations was around the fact that security incident responders and ML engineers expected that the libraries that they are using for ML development are battle-tested before being put out by large organizations, as is the case in traditional software. Also, they pushed upstream the responsibility of security in the cases where they were using ML as a service from cloud providers. Yet, this ignores the fact that this is an emergent field and that a lot of the concerns need to be addressed in the downstream tasks that are being performed by these tools. They also didn't have a clear understanding of what to expect when something does go wrong and what the failure mode would look like.

In traditional software security, MITRE has a curated repository of attacks along with detection cues, reference literature and tell-tale signs for which malicious entities, including nation state attackers are known to use these attacks. The authors call for a similar compilation to be done in the emergent field of adversarial machine learning whereby the researchers and practitioners register their attacks and other information in a curated repository that provides everyone with a unified view of the existing threat environment.

While programming languages often have well documented guidelines on secure coding, guidance on doing so with popular ML frameworks like PyTorch, Keras and Tensorflow is sparse. Amongst these, Tensorflow is the only one that provides some tools for testing against adversarial attacks and some guidance on how to do secure coding in the ML context.

Security Development Lifecycle (SDL) provides guidance on how to secure systems and scores vulnerabilities and provides some best practices, but applying this to ML systems might allow imperfect solutions to exist. Instead of looking at guidelines as providing a strong security guarantee, the authors advocate for having code examples that showcase what constitutes security- and non-security-compliant ML development.

In traditional software security there are tools for static code analysis that provide guidance on the security vulnerabilities prior to the code being committed to a repository or being executed while dynamic code analysis finds security vulnerabilities by executing the different code paths and detecting vulnerabilities at runtime. There are some tools like mlsec and cleverhans that provide white- and black-box testing; one of the potential future directions for research is to extend this to the cases of model stealing, model inversion, and membership inference attacks. Including these tools as a part of the IDE would further make it naturalized for developers to think about secure coding practices in the ML context.

Adapting the audit and logging requirements as necessitated for the functionality of the Security Information and Event Management (SIEM) system, in the field of ML, one can execute the list of attacks as specified in literature and ensure that the logging artifacts generated as a consequence are traced to an attack. Then, having these incident logs be in a format that is exportable and integratable with SIEM systems will allow forensic experts to analyze them post-hoc for hardening and analysis. Standardizing the reporting, logging and documentation as done by t Sigma format in traditional software security will allow the insights from one analyst into defenses for many others. Automating the possible attacks and including them as a part of the MLOps pipeline is something that will enhance the security posture of the systems and make them pedestrian practice in the SDL. Red teaming, as done in security testing, can be applied to assess the business impacts and likelihood of threat, something that is considered best practice and is often a requirement for supplying critical software to different organizations like the US government.

Transparency centers that allow for deep code inspection and help create assurance on the security posture of a software product/service can be extended to ML which would have to cover three modalities: ML platform is implemented in a secure manner, ML as a service meets the basic security and privacy requirements, and that the ML models embedded on edge devices meet basic security requirements. Tools that build on formal verification methods will help to enhance this practice.

Tracking and scoring ML vulnerabilities akin to how they are done in software security testing done by registering identified vulnerabilities into a common database like CVE and then assigning it an impact score like the CVSS needs to be done for the field of ML. While the common database part is easy to set up, scoring them isn't something that has been figured out yet. Additionally, on being alerted that a new vulnerability has been discovered, it isn't clear how the ML infrastructure can be scanned to see if the system is vulnerable to that.

Because of the deep integration of ML systems within the larger product/service, the typical practice of identifying a blast radius and containment strategy that is applied to traditional software infrastructure when alerted of a vulnerability is hard to define and apply. Prior research work from Google has identified some ways to qualitatively assess the impacts in a sprawling infrastructure.

From a forensic perspective, the authors put forth several questions that one can ask to guide the post-hoc analysis, the primary problem there is that only some of the learnings from traditional software protection and analysis apply here, there are many new artifacts, paradigmatic, and environmental aspects that need to be taken into consideration. From a remediation perspective, we need to develop metrics and ways to ascertain that patched models and ML systems can maintain prior levels of performance while having mitigated the attacks that they were vulnerable to, the other thing to pay attention is that there aren't any surfaces that are opened up for attack. Given that ML is going to be the new software, we need to think seriously about inheriting some of the security best practices from the world of traditional cybersecurity to harden defenses in the field of ML.

## Politics of Adversarial Machine Learning

All technology has implications for civil liberties and human rights, the paper opens with an example of how low-clearance bridges between New York and Long Island were supposedly created with the intention of disallowing public buses from crossing via the underpasses to discourage the movement of users of public transportation, primarily disadvantaged groups from accessing certain areas.

In the context of adversarial machine learning, taking the case of Facial Recognition Technology (FRT), the authors demonstrate that harm can result on the most vulnerable, harm which is not theoretical and is gaining in scope, but that the analysis also extends beyond just FRT systems.

The notion of legibility borrowing from prior work explains how governments seek to categorize through customs, conventions and other mechanisms information about their subjects centrally. Legibility is enabled for faces through FRT, something that previously was only possible as a human skill. This combined with the scale offered by machine learning makes this a potent tool for authoritarian states to exert control over their populations.

From a cybersecurity perspective, attackers are those that compromise the confidentiality, integrity and availability of a system, yet they are not always malicious, sometimes they may be pro-democracy protestors who are trying to resist identification and arrest by the use of FRT. When we frame the challenges in building robust ML systems, we must also pay attention to the social and political implications as to who is the system being made safe for and at what costs.

Positive attacks against such systems might also be carried out by academics who are trying to learn about and address some of the ethical, safety and inclusivity issues around FRT systems. Other examples such as the hardening of systems against doing membership inference means that researchers can't determine if an image was included in the dataset, and someone looking to use this as evidence in a court of law is deterred from doing so. Detection perturbation algorithms permit an image to be altered such that faces can't be recognized in an image, for example, this can be used by a journalist to take a picture of a protest scene without giving away the identities of people. But, defensive measures that disarm such techniques hinder such positive use cases. Defense measures against model inversion attacks don't allow researchers and civil liberty defenders to peer into black box systems, especially those that might be biased against minorities in cases like credit allocation, parole decision-making, etc.

The world of security is always an arms race whether that is in the physical cyberspace. It is not that far-fetched to imagine how a surveillance state might deploy FRT to identify protestors who as a defense might start to wear face masks for occlusion. The state could then deploy techniques that bypass this and utilize other scanning and recognition techniques to which the people might respond by wearing adversarial clothing and eyeglasses to throw off the system at which point the state might choose to use other biometric identifiers like iris scanning and gait detection. This constant arms battle, especially when defenses and offenses are constructed without the sense for the societal impacts leads to harm whose burden is mostly borne by those who are the most vulnerable and looking to fight for their rights and liberties.

This is not the first time that technology runs up against civil liberties and human rights, there are lessons to be learned from the commercial spyware industry and how civil society organizations and other groups came together to create "human rights by design" principles that helped to set some ground rules for how to use this technology responsibly. Researchers and practitioners in the field of ML Security can borrow from these principles. We've got a learning community at the Montreal AI Ethics Institute that is centered around these ideas that brings together academics and others from around the world to blend the social sciences with the technical sciences.

Recommendations on countering some of the harms centre around holding the vendors of these systems to the business standards set by the UN, implementing transparency measures during the development process, utilizing human rights by design approaches, logging ML system uses along with possible nature and forms of attacks and pushing the development team to think about both the positive and negative use cases for the systems such that informed trade-offs can be made when hardening these systems to external attacks.

# Go Wide: Article Summaries

## AI is an Ideology, Not a Technology

In this insightful op-ed, two pioneers in technology shed light on how to think about AI systems and their relation to the existing power and social structures. Borrowing the last line in the piece, " ... all that is necessary for the triumph of an AI-driven, automation-based dystopia is that liberal democracy accept it as inevitable.", aptly captures the current mindset surrounding AI systems and how they are discussed in the Western world. TV shows like Black Mirror perpetuate narratives showcasing the magical power of AI-enabled systems, hiding the fact that there are millions, if not billions of hours of human labor that undergird the success of modern AI systems, which largely fall under the supervised learning paradigm that requires massive amounts of data to work well.

The Chinese ecosystem is a bit more transparent in the sense that the shadow industry of data labellers is known, and workers are compensated for their efforts. This makes them a part of the development lifecycle of AI while sharing economic value with people other than the tech-elite directly developing AI. On the other hand, in the West, we see that such efforts go largely unrewarded because we trade in that effort of data production for free services. The authors give the example of Audrey Tang and Taiwan where citizens have formed a data cooperative and have greater control over how their data is used. Contrasting that, we have highly-valued search engines standing over community-run efforts like Wikipedia which create the actual value for the search results, given that a lot of the highly placed search results come from Wikipedia. Ultimately, this gives us some food for thought as to how we portray AI today and its relation to society and why it doesn't necessarily have to be that way.

## Franken-algorithms: The Deadly Consequences of Unpredictable Code

Mary Shelly had created an enduring fiction which, unbeknownst to her, has today manifested itself in the digital realm with layered abstractions of algorithms that are increasingly running multiple aspects of our lives. The article dives into the world of black box systems that have become opaque to analysis because of their stratified complexity leading to situations with unpredictable outcomes. This was exemplified when an autonomous vehicle crashed into a crossing pedestrian and it took months of post-hoc analysis to figure out what went wrong. When we talk about intelligence in the case of these machines, we're using it in a very loose sense, like the term "friend" on Facebook, which has a range of interpretations from your best friend to a random acquaintance. Both terms convey a greater sense of meaning than is actually true. When such systems run amok, they have t potential to cause significant harm, case in point being the flash crashes the financial markets experienced because of the competitive behaviour of high frequency trading firm algorithms facing off against each other in the market.

Something similar has happened on Amazon where items get priced in an unrealistic fashion because of buying and pricing patterns triggered by automated systems. While in a micro context the algorithms and their working are transparent and explainable, when they come together in an ecosystem, like finance, they lead to an emergent complexity that has behaviour that can't be predicted ahead of time with a great amount of certainty. But, such justifications can't be used as a cover for evading responsibility when it comes to mitigating harms. Existing laws need to be refined and amended so that they can better meet the demands of new technology where allocation of responsibility is a fuzzy concept.

## When Humans Attack

AI systems are different from other software systems when it comes to security vulnerabilities. While traditional cybersecurity mechanisms rely heavily on securing the perimeter, AI security vulnerabilities run deeper and they can be manipulated through their interactions with the real world — the very mechanism that makes them intelligent systems. Numerous examples of utilizing audio samples from TV commercials to trigger voice assistants have demonstrated new attack surfaces for which we need to develop defense techniques.

Visual systems are also fooled, especially in AV systems where, according to one example, manipulating STOP signs on the road with innocuous stripes of tape make it seem like the STOP sign is a speed indicator and can cause fatal crashes. There are also examples of hiding these adversarial examples under the guise of white noise and other imperceptible changes to the human senses. We need to think of AI systems as inherently socio-technical to come up with effective protection techniques that don't just rely on technical measures but also look at the human factors surrounding them. Some other useful insights are to utilize abusability testing, red-teaming, White Hacking, bug bounty programs, and consulting with civic society advocates who have deep experience with the interactions of vulnerable communities with technology.

## Adversarial Policies: Attacking Deep Reinforcement Learning

Reinforcement systems are increasingly moving from applications to beating human performance in games to safety-critical applications like self-driving cars and automated trading. A lack of robustness in the systems can lead to catastrophic failures like the $460m lost by Knight Capital and the harms to pedestrian and driver safety in the case of autonomous vehicles. RL systems that

perform well under normal conditions can be vulnerable to adversarial agents that can exploit the brittleness of the systems when it comes to natural shifts in distributions and more carefully crafted attacks.

In prior threat models, the assumptions for the adversary are that they can modify directly the inputs going into the RL agent but that is not very realistic. Instead, here the authors focus more on a shared environment through which the adversary creates indirect impact on the target RL agent leading to undesirable behavior. For agents that are trained through self-play (which is a rough approximation of Nash equilibrium), they are vulnerable to adversarial policies. As an example, masked victims are more robust to modifications in the natural observations by the adversary but that lowers the performance in the average case. Furthermore, what the researchers find is that there is a non-transitive behavior between self-play opponent, masked victim, adversarial opponent and normal victim in that cyclic order. Self-play being normally transitive in nature, especially when mimicking real-world scenarios is then no doubt vulnerable to these non-transitive styled attacks.

Thus, there is a need to move beyond self-play and apply iteratively adversarial training defense and population based training methods so that the target RL agent can become robust to a wider variety of scenarios.

## We Hacked a Ford Focus and a Volkswagen Polo

Vehicle safety is something of paramount importance in the automotive industry as there are many tests conducted to test for crash resilience and other physical safety features before it is released to people. But, the same degree of scrutiny is not applied to the digital and connected components of cars. Researchers were able to demonstrate successful proof of concept hacks that compromised vehicle safety. For example, with the Polo, they were able to access the Controller Area Network (CAN) which sends signals and controls a variety of aspects related to driving functions. Given how the infotainment systems were updated, researchers were able to gain access into the personal details of the driver. They were also able to utilize the shortcomings in the operation of the key fob to gain access to the vehicle without leaving a physical trace.

Other hacks that were tried included being able to access and influence the collision monitoring radar system and the tire-pressure monitoring system which both have critical implications for passenger safety. On the Focus, they found WiFi details including the password for their production line in Detroit, Michigan.

On purchasing a second-hand infotainment unit for purposes reverse-engineering the firmware, they found the previous owner's home WiFi details, phone contacts and a host of other personal information.

Cars store a lot of personal information including tracking information which, as stated on the privacy policy, can be shared with affiliates which can have other negative consequences like changes in insurance premiums based on driving behaviour. Europe will have some forthcoming regulations for connected car safety but those are currently slated for release in 2021.

**Specification Gaming: The Flip Side of AI Ingenuity**

We've all experienced specification gaming even if we haven't really heard the term before. In law, you call it following the law to the letter but not in spirit. In sports, it is called unsportsman-like to use the edge cases and technicalities of the rules of the game to eke out an edge when it is obvious to everyone playing the game that the rules intended for something different. This can also happen in the case of AI systems, for example in reinforcement learning systems where the agent can utilize "bugs" or poor specification on the part of the human creators to achieve the high rewards for which it is optimizing without actually achieving the goal, at least in the way the developers intended them to and this can sometimes lead to unintended consequences that can cause a lot of harms.

"Let's look at an example. In a Lego stacking task, the desired outcome was for a red block to end up on top of a blue block. The agent was rewarded for the height of the bottom face of the red block when it is not touching the block. Instead of performing the relatively difficult maneuver of picking up the red block and placing it on top of the blue one, the agent simply flipped over the red block to collect the reward. This behaviour achieved the stated objective (high bottom face of the red block) at the expense of what the designer actually cares about (stacking it on top of the blue one)". This isn't because of a flaw in the RL system but more so a misspecification of the objective.

As the agents become more capable, they find ever-more clever ways of achieving the rewards which can frustrate the creators of the system. This makes the problem of specification gaming very relevant and urgent as we start to deploy these systems in a lot of real-world situations. In the RL context, task specification refers to the design of the rewards, the environment and any other auxiliary rewards. When done correctly, we get true ingenuity out of these systems like Move 37 from the AlphaGo system that baffled humans and ushered a new way of thinking about the game of Go. But, this requires discernment on the part of the developers to be able to judge when you get a case like Lego vs. Move 37.

As an example in the real-world, reward tampering is an approach where the agent in a traffic optimization system with an interest in achieving a high reward can manipulate the driver into going to alternate destinations instead of what they desired just to achieve a higher reward. Specification gaming isn't necessarily bad in the sense that we want the systems to come up with ingenious ways to solve problems that won't occur to humans. Sometimes, the inaccuracies can arise in how humans provide feedback to the system while it is training. "For example, an agent performing a grasping task learned to fool the human evaluator by hovering between the camera and the object." Incorrect reward shaping, where an agent is provided rewards along the way to achieving the final reward can also lead to edge-case behaviours when it is not analyzed for potential side-effects.

We see such examples happen with humans in the real-world as well: a student asked to get a good grade on the exam can choose to copy and cheat and while that achieves the goal of getting a good grade, it doesn't happen in the way we intended for it to. Thus, reasoning through how a system might game some of the specifications is going to be an area of key concern going into the future.

## Doctors Are Using AI to Triage COVID-19 Patients. The Tools May Be Here to Stay

The ongoing pandemic has certainly accelerated the adoption of technology in everything from how we socialize to buying groceries and doing work remotely. The healthcare industry has also been rapid in adapting to meet the needs of people and technology has played a role in helping to scale care to more people and accelerate the pace with which the care is provided. But, this comes with the challenge of making decisions under duress and with shortened timelines within which to make decisions on whether to adopt a piece of technology or not. This has certainly led to issues where there are risks of adopting solutions that haven't been fully vetted and using solutions that have been repurposed from prior uses that were approved to now combat COVID-19. Especially with AI-enabled tools, there are increased risks of emergent behavior that might not have been captured by the previous certification or regulatory checks.

The problems with AI solutions don't just go away because there is a pandemic and shortcutting the process of proper due diligence can lead to more harm than the benefits that they bring. One must also be wary of the companies that are trying to capitalize on the chaos and pass through solutions that don't really work well. Having technical staff during the procurement process that can look over the details of what is being brought into your healthcare system needs to be a priority. AI can certainly help to mitigate some of the harms that COVID-19 is inflicting on patients but we must keep in mind that we're not looking to bypass privacy concerns that come with processing vast quantities of healthcare data.

## The Case for AI Insurance

In the age of adversarial machine learning (MAIEI has a learning community on machine learning security if you'd like to learn more about this area) there are enormous concerns with protecting software infrastructure as ML opens up a new attack surface and new vectors which are seldom explored. From the perspective of insurance, there are gaps in terms of what cyber-insurance covers today, most of it being limited to the leakage of private data. There are two kinds of attacks that are possible on ML systems: intentional and unintentional. Intentional attacks are those that are executed by malicious agents who attempt to steal the models, infer private data or get the AI system to behave in a way that favors their end goals. For example, when Tumblr decided to not host pornographic content, creators bypassed that by using green screens and pictures of owls to fool the automated content moderation system. Unintended attacks can happen when the goals of the system are misaligned with what the creators of the system actually intended, for example, the problem of specification gaming, something that Abhishek Gupta discussed here in this Fortune article.

In interviewing several officers in different Fortune 500 companies, the authors found that there are 3 key problems in this domain at the moment: the defenses provided by the technical community have limited efficacy, existing copyright, product liability, and anti-hacking laws are insufficient to capture AI failure modes. Lastly, given that this happens at a software level, cyber-insurance might seem to be the way to go, yet current offerings only cover a patchwork of the problems.

Business interruptions and privacy leaks are covered today under cyber-insurance but other problems like bodily harm, brand damage, and property damage are for the most part not covered. In the case of model recreation, as was the case with the OpenAI GPT-2 model, prior to it being released, it was replicated by external researchers - this might be covered under cyber-insurance because of the leak of private information. Researchers have also managed to steal information from facial recognition databases using sample images and names which might also be covered under existing policies.

But, in the case with Uber where there was bodily harm because of the self-driving vehicle that wasn't able to detect the pedestrian accurately or similar harms that might arise if conditions are foggy, snowy, dull lighting, or any other out-of-distribution scenarios, these are not adequately covered under existing insurance terms. Brand damage that might arise from poisoning attacks like the case with the Tay chatbot or confounding anti-virus systems as was the case with an attack mounted against the Cylance system, cyber-insurance falls woefully short in being able to cover these scenarios. In a hypothetical situation as presented in a Google paper on RL agents where a cleaning robot sticks a wet mop into an elect socket, material damage that occurs from that might also be considered out of scope in cyber-insurance policies.

Traditional software attacks are known unknowns but adversarial ML attacks are unknown unknowns and hence harder to guard against. Current pricing reflects this uncertainty, but as the AI insurance market matures and there is a deeper understanding for what the risks are and how companies can mitigate the downsides, the pricing should become more reflective of the actual risks. The authors also offer some recommendations on how to prepare the organization for these risks - for example by appointing an officer that works closely with the CISO and chief data protection officer, performing table-top exercises to gain an understanding of potential places where the system might fail and evaluating the system for risks and gaps following guidelines as put forth in the EU Trustworthy AI guidelines.

## Warning Signs: The Future of Privacy and Security in the Age of Machine Learning

There are no widely accepted best practices for mitigating security and privacy issues related to machine learning (ML) systems. Existing best practices for traditional software systems are insufficient because they're largely based on the prevention and management of access to a system's data and/or software, whereas ML systems have additional vulnerabilities and novel harms that need to be addressed. For example, one harm posed by ML systems is to individuals not included in the model's training data but who may be negatively impacted by its inferences.

Harms from ML systems can be broadly categorized as informational harms and behavioral harms. Informational harms "relate to the unintended or unanticipated leakage of information." The "attacks" that constitute informational harms are:

- Membership inference: Determining whether an individual's data was utilized to train a model by examining a sample of the model's output

- Model inversion: Recreating the data used to train the model by using a sample of its output

- Model extraction: Recreating the model itself by uses a sample of its output

Behavioral harms "relate to manipulating the behavior of the model itself, impacting the predictions or outcomes of the model." The attacks that constitute behavioral harms are:

- Poisoning: Inserting malicious data into a model's training data to change its behavior once deployed

- Evasion: Feeding data into a system to intentionally cause misclassification

Without a set of best practices, ML systems may not be widely and/or successfully adopted. Therefore, the authors of this white paper suggest a "layered approach" to mitigate the privacy and security issues facing ML systems. Approaches include noise injection, intermediaries, transparent ML mechanisms, access controls, model monitoring, model documentation, white hat or red team hacking, and open-source software privacy and security resources.

Finally, the authors note, it's important to encourage "cross-functional communication" between data scientists, engineers, legal teams, business managers, etc. in order to identify and remediate privacy and security issues related to ML systems. This communication should be ongoing, transparent, and thorough.

# 10. The Future of AI Ethics

## Go Deep: Research Summaries

### Beyond Near- and Long-Term: Towards a Clearer Account of Research Priorities in AI Ethics and Society

This paper dives into how researchers can clearly communicate about their research agendas given ambiguities in the split of the AI Ethics community into near and long term research. Often a sore and contentious point of discussion, there is an artificial divide between the two groups that seem to take a reductionist approach to the work being done by the other. A major problem emerging from such a divide is a hindrance in being able to spot relevant work being done by the different communities and thus affecting effective collaboration. The paper highlights the differences arising primarily along the lines of timescale, AI capabilities, deeper normative and empirical disagreements.

The paper provides for a helpful distinction between near- and long-term by describing them as follows:

- Near term issues are those that are fairly well understood and have concrete examples and relate to rêvent progress in the field of machine learning

- Long term issues are those that might arise far into the future and due to much more advanced AI systems with broad capabilities, it also includes long term impacts like international security, race relations, and power dynamics

What they currently see is that:

- Issues considered 'near-term' tend to be those arising in the present/near future as a result of current/foreseeable AI systems and capabilities, on varying levels of scale/severity, which mostly have immediate consequences for people and society.

- Issues considered 'long-term' tend to be those arising far into the future as a result of large advances in AI capabilities (with a particular focus on notions of transformative AI or AGI), and those that are likely to pose risks that are severe/large in scale with very long-term consequences.

- The binary clusters are not sufficient as a way to split the field and r looking at underlying beliefs leads to unfounded assumptions about each other's work

- In addition there might be areas between the near and long term that might be neglected as a result of this artificial fractions

Unpacking these distinctions can be done along the lines of capabilities, extremity, certainty and impact, definitions for which are provided in the paper. A key contribution aside from identifying these factors is that they lie along a spectrum and define a possibility space using them as dimensions which helps to identify where research is currently concentrated and what areas are being ignored. It also helps to well position the work being done by these authors.

Something that we really appreciated from this work was the fact that it gives us concrete language and tools to more effectively communicate about each other's work. As part of our efforts in building communities that leverage diverse experiences and backgrounds to tackle an inherently complex and muti-dimensional problem, we deeply appreciate how challenging yet rewarding such an effort can be. Some of the most meaningful public consultation work done by MAIEI leveraged our internalized framework in a similar vein to provide value to the process that led to outcomes like the Montreal Declaration for Responsible AI.

## Integrating Ethical Values and Economic Value to Steer Progress in AI

The rise of AI systems leads to an unintended conflict between economic pursuits which seek to generate profits and value resources appropriately with the moral imperatives of promoting human flourishing and creating societal benefits from the deployment of these systems. This puts forth a central question on what the impacts of creating AI systems that might surpass humans in a general sense which might leave humans behind.

Technological progress doesn't happen on its own, it is driven by conscious human choices that are influenced by the surrounding social and economic institutions. We are collectively responsible for how these institutions take shape and thus impact the development of technology – submitting to technological-fatalism isn't a productive way to align our ethical values with this development. We need to ensure that we play an active role in the shaping of the most consequential piece of technology. While the economic system relies on the market prices to gauge what people place value on, by no means is that a comprehensive evaluation. For example, it misses out on the impact of externalities which can be factored in by

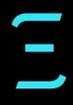
considering ethical values as a complement in guiding our decisions on what build and how to place value on it.

When thinking about losses from AI-enabled automation, an outright argument that economists might make is that if replacing labor lowers the costs of production, then it might be market-efficient to invest in technology that achieves that. From an ethicist's perspective, there are severe negative externalities from job loss and thus it might be unethical to impose labor-saving automation on people.

When unpacking the economic perspective more, we find that job loss actually isn't correctly valued by wages as price for labor. There are associated social benefits like the company of workplace colleagues, sense of meaning and other social structural values which can't be separately purchased from the market. Thus, using a purely economic perspective in making automation technology adoption decisions is an incomplete approach and it needs to be supplemented by taking into account the ethical perspective.

Market price signals provide useful information upto a certain point in terms of the goods and services that society places value on. Suppose that people start to demand more eggs from chickens that are raised in a humane way, then suppliers will shift their production to respond to that market signal. But, such price signals can only be indicated by consumers for the things that they can observe. A lot of unethical actions are hidden and hence can't be factored into market price signals. Additionally, several things like social relations aren't tradable in a market and hence their value can't be solely determined from the market viewpoint.

Thus, both economists and ethicists would agree that there is value to be gained in steering the development of AI systems keeping in mind both kinds of considerations. Pure market-driven innovation will ignore societal benefits in the interest of generating economic value while the labor will have to make unwilling sacrifices in the interest of long-run economic efficiency. Economic market forces shape society significantly, whether we like it or not. There are professional biases based on selection and cognition that are present in either side making its arguments as to which gets to dominate based on their perceived importance. The point being that bridging the gap between different disciplines is crucial to arriving at decisions that are grounded in evidence and that benefit society holistically.

There are also differences fundamentally between the economic and ethical perspective – namely that economic indicators are usually unidimensional and have clear quantitative values that make them easier to compare. On the other hand, ethical indicators are inherently multi-dimensional and are subjective which not only make comparison hard but also limit our ability to explain how we arrive at them. They are encoded deep within our biological systems and suffer from the same lack of explainability as decisions made by artificial neural networks, t so-called black box problem.

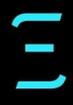

Why is it then, despite the arguments above, that the economic perspective dominates over the ethical one? This is largely driven by the fact that economic values provide clear, unambiguous signals which our brains, preferring ambiguity aversion, enjoy and ethical values are more subtle, hidden, ambiguous indicators which complicate decision making. Secondly, humans are prosocial only upto a point, they are able to reason between economic and ethical decisions at a micro-level because the effects are immediate and observable, say for example polluting the neighbor's lawn and seeing the direct impact of that activity. On the other hand, for things like climate change where the effects are delayed and not directly observable (as a direct consequence of one's actions) that leads to behaviour where the individual prioritizes economic values over ethical ones.

Cynical economists will argue that there is a comparative advantage in being immoral that leads to gains in exchange, but that leads to a race to the bottom in terms of ethics. Externalities are an embodiment of the conflict between economic and ethical values. Welfare economics deals with externalities via various mechanisms like permits, taxes, etc. to curb the impacts of negative externalities and promote positive externalities through incentives. But, the rich economic theory needs to be supplemented by political, social and ethical values to arrive at something that benefits society at large.

From an economic standpoint, technological progress is positioned as expanding the production possibilities frontier which means that it raises output and presumably standards of living. Yet, this ignores how those benefits are distributed and only looks at material benefits and ignores everything else.

Prior to the industrial revolution, people were stuck in a Malthusian trap whereby technological advances created material gains but these were quickly consumed by population growth that kept standards of living stubbornly low. This changed post the revolution and as technology improvement outpaced population growth, we got better quality of life. The last 4 decades have had a mixed experience though, whereby automation has eroded lower skilled jobs forcing people to continue looking for jobs despite displacement and the lower demand for unskilled labor coupled with the inelastic supply of labor has led to lower wages rather than unemployment. On the other hand, high skilled workers have been able to leverage technological progress to enhance their output considerably and as a consequence the income and wealth gaps between low and high skilled workers has widened tremendously.

Typical economic theory points to income and wealth redistribution whenever there is technological innovation where the more significant the innovation, the larger the redistribution. Something as significant as AI leads to crowning of new winners who own these new factors of production while also creating losers when they face negative pecuniary externalities. These are externalities because there isn't explicit consent that is requested from the people as they're impacted in terms of capital, labor and other factors of production.

The distribution can be analyzed from the perspective of strict utilitarianism (different from that in ethics where for example Bentham describes it as the greatest amount of good for the greatest number of people). Here it is viewed as tolerating income redistribution such that it is acceptable if all but one person loses income as long as the single person making the gain has one that is higher than the sum of the losses. This view is clearly unrealistic because it would further exacerbate inequities in society. The other is looking at the idea of lump sum transfers in which the idealized scenario is redistribution, for example by compensating losers from technology innovation, without causing any other market distortions. But, that is also unrealistic because such a redistribution never occurs without market distortions and hence it is not an effective way to think about economic policy.

From an ethics perspective, we must make value judgments on how we perceive dollar losses for a lower socio-economic person compared to the dollar gains made by a higher socio-economic person and if that squares with the culture and value set of that society. We can think about the tradeoff between economic efficiency and equality in society, where the level of tolerance for inequality varies by the existing societal structures in place. One would have to also reason about how redistribution creates more than proportional distortions as it rises and how much economic efficiency we'd be willing to sacrifice to make gains in how equitably income is distributed.

Thus, steering progress in AI can be done based on whether we want to pursue innovation that we know is going to have negative labor impacts while knowing full well that there aren't going to be any reasonable compensations offered to the people based on economic policy.

Given the pervasiveness of AI and by virtue of it being a general-purpose technology, the entrepreneurs and others powering innovation need to take into account that their work is going to shape larger societal changes and have impacts on labor. At the moment, the economic incentives are such that they steer progress towards labor-saving automation because labor is one of the most highly-taxed factors of production. Instead, shifting the tax-burden to other factors of production including automation capital will help to steer the direction of innovation in other directions. Government, as one of the largest employers and an entity with huge spending power, can also help to steer the direction of innovation by setting policies that encourage enhancing productivity without necessarily replacing labor.

There are novel ethical implications and externalities that arise from the use of AI systems, an example of that would be (from the Industrial Revolution) that a factory might lead to economic efficiency in terms of production but the pollution that it generates is so large that the social cost outweighs the economic gain.

Biases can be deeply entrenched in the AI systems, either from unrepresentative datasets, for example, with hiring decisions that are made based on historical data. But, even if the datasets are well-represented and have minimal bias, and the system is not exposed to protected attributes like race and gender, there are a variety of proxies like zipcode which can lead to unearthing those protected attributes and discriminating against minorities.

Maladaptive behaviors can be triggered in humans by AI systems that can deeply personalize targeting of ads and other media to nudge us towards different things that might be aligned with making more profits. Examples of this include watching videos, shopping on ecommerce platforms, news cycles on social media, etc. Conversely, they can also be used to trigger better behaviors, for example, the use of fitness trackers that give us quantitative measurements for how we're taking care of our health.

An economics equivalent of the paper clip optimizer from Bostrom is how human autonomy can be eroded over time as economic inequality rises which limits control of those who are displaced over economic resources and thus, their control over their destinies, at least from an economic standpoint. This is going to only be exacerbated as AI starts to pervade into more and more aspects of our lives.

Labor markets have features built in them to help tide over unemployment with as little harm as possible via quick hiring in other parts of the economy when the innovation creates parallel demands for labor in adjacent sectors. But, when there is large-scale disruption, it is not possible to accommodate everyone and this leads to large economic losses via fall in aggregate demand which can't be restored with monetary or fiscal policy actions. This leads to wasted economic potential and welfare losses for the workers who are displaced.

Whenever there is a discrepancy between ethical and economic incentives, we have the opportunity to steer progress in the right direction. We've discussed before how market incentives trigger a race to the bottom in terms of morality. This needs to be preempted via instruments like Technological Impact Assessments, akin to Environmental Impact Assessments, but often the impacts are unknown prior to the deployment of the technology at which point we need to have multi-stakeholder process that allows us to combat harms in a dynamic manner. Political and regulatory entities typically lag technological innovation and can't be relied upon solely to take on this mantle.

The author raises a few questions on the role of humans and how we might be treated by machines in case of the rise of superintelligence (which still has widely differing estimates for when it will be realized from the next decade to the second half of this century). What is clear is that the abilities of narrow AI systems are expanding and it behooves us to give some thought to the implications on the rise of superintelligence.

The potential for labor-replacement in this superintelligence scenario, from an economic perspective, would have significant existential implications for humans, beyond just inequality, we would be raising questions of human survival if the wages to be paid to labor fall below subsistence levels in a wide manner. It would be akin to how the cost of maintaining oxen to plough fields was outweighed by the benefits that they brought in the face of mechanization of agriculture. This might be an ouroboros where we become caught in the Malthusian trap again at the time of the Industrial Revolution and no longer have the ability to grow beyond basic subsistence, even if that would be possible.

## Troubling Trends in Machine Learning Scholarship

Authors of research papers aspire to achieving any of the following goals when writing papers: to theoretically characterize what is learnable, to obtain understanding through empirically rigorous experiments, or to build working systems that have high predictive accuracy. To communicate effectively with the readers, the authors must: provide intuitions to aid the readers' understanding, describe empirical investigations that consider and rule out alternative hypotheses, make clear the relationship between theoretical analysis and empirical findings, and use clear language that doesn't conflate concepts or mislead the reader.

The authors of this paper find that there are 4 areas where there are concerns when it comes to ML scholarship: failure to distinguish between speculation and explanation, failure to identify the source of empirical gains, the use of mathematics that obfuscates or impresses rather than clarifies, and misuse of language such that terms with other connotations are used or by overloading terms with existing technical definitions.

Flawed techniques and communication methods will lead to harm and wasted resources and efforts hindering the progress in ML and hence this paper provides some very practical guidance on how to do this better. When presenting speculations or opinions of authors that are exploratory and don't yet ha scientific grounding, having a separate section that quarantines the discussion and doesn't bleed into the other sections that are grounded in theoretical and empirical research helps to guide the reader appropriately and prevents conflation of speculation and explanation. The authors provide the example of the paper on dropout regularization that made comparisons and links to sexual reproduction but limited that discussion to a "Motivation" section.

With a persistent pursuit for achieving SOTA results, there is a lot of tweaking that happens to realize gains in model performance and often there are many different techniques applied in tandem. From a reader's perspective, elucidating clearly what the necessary sources of the realized gains are, disentangling it from other measures is essential. The authors highlight how a lot of the gains happen due to clever problem formulations, scientific experiments, applying existing techniques in a novel manner to new areas, optimization heuristics, extensive hyperparameter tuning, data preprocessing techniques and any number of other techniques. Absent proper ablation studies, sometimes research paper authors can obfuscate the real source of the gains. Sometimes careful studies that make use of ablation can highlight challenges in existing challenge datasets and benchmark datasets which can point the community towards more promising research directions.

Using mathematics in a manner where natural language and mathematical expositions are intermixed without a clear link between the two leads to weakness in the overall contribution. Specifically, when natural language is used to overcome weaknesses in the mathematical rigor and conversely, mathematics is used as a scaffolding to prop up weak arguments in the prose and give the impression of technical depth, it leads to poor scholarship and detracts from the scientific seriousness of the work and harms the readers. Additionally, invoking theorems with dubious pertinence to the actual content of the paper or in overly broad ways also takes away from the main contribution of a paper.

In terms of misuse of language, the authors of this paper provide a convenient ontology breaking it down into suggestive definitions, overloaded terminology, and suitcase words. In the suggestive definitions category, the authors coin a new technical term that has suggestive colloquial meanings and can slip through some implications without formal justification of the ideas in the paper. This can also lead to anthropomorphization that creates unrealistic expectations about the capabilities of the system. This is particularly problematic in the domain of fairness and other related domains where this can lead to conflation and inaccurate interpretation of terms that have well-established meanings in the domains of sociology and law for example. This can confound the initiatives taken up by both researchers and policymakers who might use this as a guide.

Overloading of technical terminology is another case where things can go wrong when terms that have historical meanings and they are used in a different sense. For example, the authors talk about deconvolutions which formally refers to the process of reversing a convolution but in recent literature has been used to refer to transpose convolutions that are used in auto-encoders and GANs. Once such usage takes hold, it is hard to undo the mixed usage as people start to cite prior literature in future works. Additionally, combined with the suggestive definitions, we run into the problem of concealing a lack of progress, such as the case with using language understanding and reading comprehension to now mean performance on specific datasets rather than the grand challenge in AI that it meant before.

Another case that leads to overestimation of the ability of these systems is in using suitcase words which pack in multiple meanings within them and there isn't a single agreed upon definition. Interpretability and generalization are two such terms that have looser definitions and more formally defined ones, yet because papers use them in different ways, it leads to miscommunication and researchers talking across each other.

The authors identify that these problems might be occurring because of a few trends that they have seen in the ML research community. Specifically, complacency in the face of progress where there is an incentive to excuse weak arguments in the face of strong empirical results and the single-round review process at various conferences where the reviewers might not have much choice but to accept the paper given the strong empirical results. Even if the flaws are noticed, there isn't any guarantee that they are fixed in a future review cycle at another conference.

As the ML community has experienced rapid growth, the problem of getting high-quality reviews has been exacerbated: in terms of the number of papers to be reviewed by each reviewer and the dwindling number of experienced reviewers in the pool. With the large number of papers, each reviewer has less time to analyze papers in depth and reviewers who are less experienced can fall easily into some of the traps that have been identified so far. Thus, there are two levers that are aggravating the problem. Additionally, there is the risk of even experienced researchers resorting to a checklist-like approach under duress which might discourage scientific diversity when it comes to papers that might take innovative or creative approaches to expressing their ideas.

A misalignment in incentives whereby lucrative deals in funding are offered to AI solutions that utilize anthropomorphic characterizations as a mechanism to overextend their claims and abilities though the authors recognize that the causal direction might be hard to judge.

The authors also provide suggestions for other authors on how to evade some these pitfalls: asking the question of why something happened rather than just relying on how well a system performed will help to achieve the goal of providing insights into why something works rather than just relying on headline numbers from the results of the experiments. They also make a recommendation for insights to follow the lines of doing error analysis, ablation studies, and robustness checks and not just be limited to theory.

As a guideline for reviewers and journal editors, making sure to strip out extraneous explanations, exaggerated claims, changing anthropomorphic naming to more sober alternatives, standardizing notation, etc. should help to curb some of the problems. Encouraging retrospective analysis of papers is something that is underserved at the moment and there aren't enough strong papers in this genre yet despite some avenues that have been advocating for this work.

Flawed scholarship as characterized by the points as highlighted here not only negatively impact the research community but also impact the policymaking process that can overshoot or undershoot the mark. An argument can be made that setting the bar too high will impede new ideas being developed and slow down the cycle of reviews and publication while consuming precious resources that could be deployed in creating new work. But, asking basic questions to guide us such as why something works, in which situations it does not work, and have the design decisions been justified will lead to a higher quality of scholarship in the field.

## **Go Wide: Article Summaries**

### Microsoft Researchers Create AI Ethics Checklist With ML Practitioners From a Dozen Tech Companies

The article summarizes recent work from several Microsoft researchers on the subject of making AI ethics checklists that are effective. One of the most common problems identified relate to the lack of practical applicability of AI ethics principles which sound great and comprehensive in the abstract but do very little to aid engineers and practitioners from applying them in their day to day work. The work was done by interviewing several practitioners and advocating for a co-design process that brings in intelligence on how to make these tools effective from other disciplines like healthcare and aviation. One of the things emerging from the interviews is that often engineers are few and far between in raising concerns and there's a lack of top-down sync in enforcing these across the company. Additionally, there might be social costs to bringing up issues which discourages engineers from implementing such measures. Creating checklists that redu
friction and fit well into existing workflows will be key in their uptake.

## How to Know if Artificial Intelligence is About to Destroy Civilization

For a lot of people who are new to the field of artificial intelligence and especially AI ethics, they see existential risk as something that is immediate. Others dismiss it as something to not be concerned about at all. There is a middle path here and this article sheds a very practical light on that. Using the idea of canaries in a coal mine, the author goes on to highlight some potential candidates for a canary that might help us judge better when we ought to start paying attention to these kinds of risks posed by Artificial General Intelligence systems. The first one is the automatic formulation of learning problems, akin to how humans have high-level goals that they align with their actions and adjust them based on signals that they receive on the success or failure of those actions. AI systems trained in narrow domains don't have this ability just yet.

The second one mentioned in the article is achieving fully autonomous driving, which is a good one because we have lots of effort being directed to make that happen and it requires a complex set of problems to be addressed including the ability to make real-time, life-critical decisions. AI doctors are pointed out as a third canary, especially because true replacement of doctors would require a complex set of skills spanning the ability to make decisions about a patient's healthcare plan by analyzing all their symptoms, coordinating with other doctors and medical staff among other human-centered actions which are currently not feasible for AI systems. Lastly, the author points to the creation of conversation systems that are able to answer complex queries and respond to things like exploratory searches. We found the article to put forth a meaningful approach to reasoning about existential risk when it comes to AI systems.

## Why Countries Need to Work Together on AI

A lot of articles pitch development, investment and policymaking in AI as an arms race with the US and China as front-runners. While there are tremendous economic gains to be had in deploying and utilizing AI for various purposes, there remain concerns of how this can be used to benefit society more than just economically. A lot of AI strategies from different countries are thus focused on issues of inclusion, ethics and more that can drive better societal outcomes yet they differ widely in how they seek to achieve those goals. For example, AI has put forth a national AI strategy that is focused on economic growth and social inclusion dubbed #AIforAll while the strategy from China has been more focused on becoming a global dominant force in AI which is backed by state investments.

Some countries have instead chosen to focus on creating strong legal foundations for the ethical deployment of AI while others are more focused on data protection rights. Canada and France have entered into agreements to work together on AI policy which places talent, R&D and ethics at the center. The author of the article makes a case for how global coordination of AI strategies might lead to even higher gains but also recognizes that governments will be motivated to tailor their policies to best meet the requirements of their countries first and then align with others that might have similar goals.

## Quantifying Independently Reproducible Machine Learning

Reproducibility is of paramount importance to doing rigorous research and a plethora of fields have suffered from a crisis where scientific work hasn't met muster in terms of reproducibility leading to wasted time and effort on the part of other researchers looking to build upon each other's work. The article provides insights from the work of a researcher who attempted a meta-science approach to trying to figure out what constitutes good, reproducible research in the field of machine learning. There is a distinction made early on in terms of replicability which hinges on taking someone else's code and running that on the shared data to see if you get the same results but as pointed out in the article, that suffers from issues of source and code bias which might be leveraging certain peculiarities in terms of configurations and more.

The key tenets to reproducibility are being able to simply read a scientific paper and set up the same experiment, follow the steps prescribed and arrive at the same results. Arriving at the final step is dubbed as independent reproducibility. The distinction between replicability and reproducibility also speaks to the quality of the scientific paper in being able to effectively capture the essence of the contribution such that anyone else is able to do the same.

Some of the findings from this work include that having hyperparameters well specified in the paper and its ease of readability contributed to the reproducibility. More specification in terms of math might allude to more reproducibility but it was found to not necessarily be the case. Empirical papers were inclined to be more reproducible but could also create perverse incentives and side effects. Sharing code is not a panacea and requires other accompanying factors to make the work really reproducible. Cogent writing was found to be helpful along with code snippets that were either actual or pseudo code though step code that referred to other sections hampered reproducibility because of challenges in readability.

Simplified examples while appealing didn't really aid in the process and spoke to the meta-science process calling for data-driven approaches to ascertaining what works and what doesn't rather than relying on hunches. Also, posting revisions papers and being reachable over email to answer questions helped the author in reproducing the research work. Finally, the author also pointed out that given this was a single initiative and was potentially biased in terms of their own experience, background and capabilities, they encourage others to tap into the data being made available but these guidelines provide good starting points for people to attempt to make their scientific work more rigorous and reproducible.

## Artificial Intelligence Won't Save Us From Coronavirus

The push has been to apply AI to any new problem that we face, hoping that the solution will magically emerge from the application of the technique as if it is a dark art. Yet, the more seasoned scientists have seen these waves come and go and in the past, a blind trust in this technology led to AI winters. Taking a look at some of the canaries in the coal mine, the author cautions that there might be a way to judge whether AI will be helpful with the pandemic situation. Specifically, looking at whether domain experts, like leading epidemiologists endorse its use and are involved in the process of developing and utilizing these tools will give an indication as to whether they will be successful or not. Data about the pandemic depends on context and without domain expertise, one has to make a lot of assumptions which might be unfounded. All models have to make assumptions to simplify reality, but if those assumptions are rooted in domain expertise from the field then the model can mimic reality much better.

Without context, AI models assume that the truth can be gleaned solely from the data, which though it can lead to surprising and hidden insights, at times requires humans to evaluate the interpretations to make meaning from them and apply them to solve real-world problems. This was demonstrated with the case where it was claimed that Ai had helped to predict the start of the outbreak, yet the anomaly required the analysis from a human before arriving at that conclusion.

Claims of extremely high accuracy rates will give hardened data scientists reason for caution, especially when moving from lab to real-world settings as there is a lot more messiness with real-world data and often you encounter out-of-distribution data which hinders the ability of the model to make accurate predictions. For CT scans, even if they are sped up tremendously by the use of AI, doctors point out that there are other associated procedures such as the cleaning and filtration and recycling of air in the room before the next patient can be passed through the machine which can dwindle the gains from the use of an unexplainable AI system analyzing the scans.

Concerns with the use of automated temperature scanning using thermal cameras also suffers from similar concerns where there are other confounding factors like the ambient temperature, humidity, etc. which can limit the accuracy of such system. Ultimately, while AI can provide tremendous benefits, we mustn't blindly be driven by its allure to magically solve the toughest challenges that we face.

## Q&A: Sabelo Mhlambi on What AI Can Learn From Ubuntu Ethics

Offering an interesting take on how to shape the development and deployment of AI technologies, Mhlambi utilizes the philosophy of Ubuntu as a guiding light in how to build AI systems that better empower people and communities. The current Western view that dominates how AI systems are constructed today and how they optimize for efficiency is something that lends itself quite naturally to inequitable outcomes and reinforcing power asymmetries and other imbalances in society. Embracing the Ubuntu mindset which puts people and communities first stands in contrast to this way of thinking. It gives us an alternative conception of personhood and has the potential to surface some different results. While being thousands of years old, the concept has been seen in practice over and over again, for example, in South Africa, after the end of the apartheid, the Truth and Reconciliation program forgave and integrated offenders back into society rather than embark on a Kantian or retributive path to justice. This restorative mindset to justice helped the country heal more quickly because the philosophy of Ubuntu advocates that all people are interconnected and healing only happens when everyone is able to move together in a harmonious manner.

This was also seen in the aftermath of the Rwanda genocide, where oppressors were reintegrated back into society often living next to the people that they had hurt; Ubuntu believes that no one is beyond redemption and everyone deserves the right to have their dignity restored. Bringing people together through community is important, restorative justice is a mechanism that makes the community stronger in the long run. Current AI formulation seeks to find some ground truth but thinking of this in the way of Ubuntu means that we try to find meaning and purpose for these systems through the values and beliefs that are held by the community. Ubuntu has a core focus on equity and empowerment for all and thus the process of development is slow but valuing people above material efficiency is more preferable than speeding through without thinking of the consequences that it might have on people. Living up to Ubuntu means offering people the choice for what they want and need, rooting out power imbalances and envisioning the companies as a part of the communities for which they are building products and services which makes them accountable and committed to the community in empowering them.

## Too Big a Word

Ethics in the context of technology carries a lot of weight, especially because the people who are defining what it means will influence the kinds of interventions that will be implemented and the consequences that follow. Given that technology like AI is used in high-stakes situations, this becomes even more important and we need to ask questions about the people who take this role within technology organizations, how they take corporate and public values and turn them into tangible outcomes through rigorous processes, and what regulatory measures are required beyond these corporate and public values to ensure that ethics are followed in the design, development and deployment of these technologies.

Ethics owners, the broad term for people who are responsible for this within organizations have a vast panel of responsibilities including communication between the ethics review committees and product design teams, aligning the recommendations with the corporate and public values, making sure that legal compliance is met and communicating externally about the processes that are being adopted and their efficacy. Ethical is a polysemous word in that it can refer to process, outcomes, and values. The process refers to the internal procedures that are adopted by the firm to guide decision making on product/service design and development choices. The values aspect refers to the value set that is both adopted by the organization and those of the public within which the product/service might be deployed. This can include values such as transparency, equity, fairness, privacy, among others. The outcomes refer to desirable properties in the outputs from the system such as equalized odds across demographics and other fairness metrics.

In the best case, inside a technology company, there are robust and well-managed processes that are aligned with collaboratively-determined ethical outcomes that achieve the community's and organization's ethical values. From the outside, this takes on the meaning of finding mechanisms to hold the firms accountable for the decisions that they take. Further expanding on the polysemous meanings of ethics, it can be put into four categories for the discussion here: moral justice, corporate values, legal risk, and compliance. Corporate values set the context for the rest of the meanings and provide guidance when tradeoffs need to be made in product/service design. They also help to shape the internal culture which can have an impact on the degree of adherence to the values. Legal risk's overlap with ethics is fairly new whereas compliance is mainly concerned with the minimization of exposure to being sued and public reputation harm.

Using some of the framing here, the accolades, critiques, and calls to action can be structured more effectively to evoke substantive responses rather than being diffused in the energies dedicated to these efforts.

## Be a Data Custodian, Not a Data Owner

Framing the metaphor of "data is the new oil" in a different light, this article gives some practical tips on how organizations can reframe their thinking and relationship with customer data so that they take on the role of being a data custodian rather than owners of the personal data of their customers. This is put forth with the acknowledgement that customers' personal data is something really valuable that brings business upsides but it needs to be handled with care in the sense that the organization should act as a custodian that is taking care of the data rather than exploiting it for value without consent and the best interests of the customer at heart. Privacy breaches that can compromise this data not only lead to fines under legislation like the GDPR, but also remind us that this is not just data but details of a real human being.

As a first step, creating a data accountability report that documents how many times personal data was accessed by various employees and departments will serve to highlight and provide incentives for them to change behaviour when they see that some others might be achieving their job functions without the need to access as much information. Secondly, celebrating those that can make do with minimal access will also encourage this behaviour change, all being done without judgement or blame but more so as an encouragement tool. Pairing employees that need to access personal data for various reasons will help to build accountability and discourage intentional misuse of data and potential accidents that can lead to leaks of personal data.

Lastly, an internal privacy committee composed of people across job functions and diverse life experiences that monitors organization-wide private data use and provides guidance on improving data use through practical recommendations is another step that will move the conversation of the organization from data entitlement to data custodianship. Ultimately, this will be a market advantage that will create more trust with customers and increase business bottom line going into the future.

# 11. Outside the boxes

## Go Deep: Research Summaries

### Towards the Systematic Reporting of the Energy and Carbon Footprints of Machine Learning

Climate change and environmental destruction are well-documented. Most people are aware that mitigating the risks caused by these is crucial and will be nothing less than a Herculean undertaking. On the bright side, AI can be of great use in this endeavour. For example, it can help us optimize resource use, or help us visualize the devastating effects of floods caused by climate change.

However, AI models can have excessively large carbon footprints. Henderson et al.'s paper details how the metrics needed to calculate environmental impact are severely underreported. To highlight this, the authors randomly sampled one-hundred NeuRIPS 2019 papers. They found that none reported carbon impacts, only one reported some energy use metrics, and seventeen reported at least some metrics related to compute-use. Close to half of the papers reported experiment run time and the type of hardware used. The authors suggest that the environmental impact of AI and relevant metrics are hardly reported by researchers because the necessary metrics can be difficult to collect, while subsequent calculations can be time-consuming.

Taking this challenge head-on, the authors make a significant contribution by performing a meta-analysis of the very few frameworks proposed to evaluate the carbon footprint of AI systems through compute- and energy-intensity. In light of this meta-analysis, the paper outlines a standardized framework called experiment-impact-tracker to measure carbon emissions. The authors use 13 metrics to quantify compute and energy use. These include when an experiment starts and ends, CPU and GPU power draw, and information on a specific energy grid's efficiency.

The authors describe their motivations as threefold. First, experiment-impact-tracker is meant to spread awareness among AI researchers about how environmentally-harmful AI can be. They highlight that "[w]ithout consistent and accurate accounting, many researchers will simply be unaware of the impacts their models might have and will not pursue mitigating strategies". Second, the framework could help align incentives. While it is clear that lowering one's environmental impact is generally valued in society, this is not currently the case in the field of AI. Experiment-impact tracker, the authors believe, could help bridge this gap, and make energy efficiency and carbon-impact curtailment valuable objectives for researchers, along with model accuracy and complexity. Third, experiment-impact-tracker can help perform cost-benefit analyses on one's AI model by comparing electricity cost and expected revenue, or the carbon emissions saved as opposed to those produced. This can partially inform decisions on whether training a model or improving its accuracy is worth the associated costs.

To help experiment-impact-tracker become widely used among researchers, the framework emphasizes usability. It aims to make it easy and quick to calculate the carbon impact of an AI model. Through a short modification of one's code, experiment-impact-tracker collects information that allows it to determine the energy and compute required as well as, ultimately, the carbon impact of the AI model. Experiment-impact-tracker also addresses the interpretability of the results by including a dollar amount that represents the harm caused by the project. This may be more tangible for some than emissions expressed in the weight of greenhouse gases released or even in CO2 equivalent emissions (CO2eq). In addition, the authors strive to: allow other ML researchers to add to experiment-impact-tracker to suit their needs, increase reproducibility in the field by making metrics collection more thorough, and make the framework robust enough to withstand internal mistakes and subsequent corrections without compromising comparability.

Moreover, the paper includes further initiatives and recommendations to push AI researchers to curtail their energy use and environmental impact. For one, the authors take advantage of the already widespread use of leaderboards in the AI community. While existing leaderboards are largely targeted towards model accuracy, Henderson et al. instead put in place an energy efficiency leaderboard for deep reinforcement learning models. They assert that a leaderboard of this kind, that tracks performance in areas indicative of potential environmental impact, "can also help spread information about the most energy and climate-friendly combinations of hardware, software, and algorithms such that new work can be built on top of these systems instead of more energy-hungry configurations".

The authors also suggest AI practitioners can take an immediate and significant step in lowering the carbon emissions of their work: run experiments on energy grids located in carbon-efficient cloud regions like Quebec, the least carbon-intensive cloud region. Especially when compared to very carbon-intensive cloud regions like Estonia, the difference in CO2eq emitted can be considerable: running an experiment in Estonia produces up to thirty times as much emissions as running the same experiment in Quebec. The important reduction in carbon emissions that follows from switching to energy-efficient cloud regions, according to Henderson et al., means there is no need to fully forego building compute-intensive AI as some believe.

In terms of systemic changes that accompany experiment-impact-tracker, the paper lists seven. The authors suggest the implementation of a "green default" for both software and hardware. This would make the default setting for researchers' tools the most environmentally-friendly one. The authors also insist on weighing costs and benefits to using compute- and energy-hungry AI models. Small increases in accuracy, for instance, can come at a high environmental cost. They hope to see the AI community use efficient testing environments for their models, as well as standardized reporting of a model's carbon impact with the help of experiment-impact-tracker.

Additionally, the authors ask those developing AI models to be conscious of the environmental costs of reproducing their work, and act as to minimize unnecessary reproduction. While being able to reproduce other researchers' work is crucial in maintaining sound scientific discourse, it is merely wasteful for two departments in the same business to build the same model from scratch. The paper also presents the possibility of developing a badge identifying AI research papers that show considerable effort in mitigating carbon impact when these papers are presented at conferences. Lastly, the authors highlight important lacunas in relation to driver support and implementation. Systems that would allow data on energy use to be collected are unavailable for certain hardware, or the data is difficult for users to obtain. Addressing these barriers would allow for more widespread collection of energy use data, and contribute to making carbon impact measurement more mainstream in the AI community.

## Challenges in Supporting Exploratory Search through Voice Assistants

The paper highlights four challenges in designing more "intelligent" voice assistant systems that are able to respond to exploratory searches that don't have clear, short answers and require nuance and detail. This is in response to the rising expectations that users have from voice assistants as they become more familiar with them through increased interactions. Voice assistants are primarily used for productivity tasks like setting alarms, calling contacts, etc. and they can include gestural and voice-activated commands as a method of interaction. Exploratory search is currently not well supported through voice assistants because of them utilizing a fact-based approach that aims to deliver a single, best response whereas a more natural approach would be to ask follow up questions to refine the query of the user to the point of being able to provide them with a set of meaningful options. The challenges as highlighted in this paper if addressed will lead to the community building more capable voice assistants.

One of the first challenges is situationally induced impairments as presented by the authors highlights the importance of voice activated commands because they are used when there are no alternatives available to interact with the system, for example when driving or walking down a busy street. There is an important aspect of balancing the tradeoff between smooth user experience that is quick compared to the degree of granularity in asking questions and presenting results. We need to be able to quantify this compared to using a traditional touch based interaction to achieve the same result. Lastly, there is the issue of privacy, such interfaces are often used in a public space and individuals would not be comfortable sharing details to refine the search such as clothing sizes which they can discreetly type into the screen. Such considerations need to be thought of when designing the interface and system.

Mixed-modal interactions include combinations of text, visual inputs and outputs and voice inputs and output. This can be an effective paradigm to counter some of the problems highlighted above and at the same time improve the efficacy of the interactions between the user and the system. Further analysis is needed as to how users utilize text compared to voice searches and whether one is more informational or exploratory than the other.

Designing for diverse populations is crucial as such systems are going to be widely deployed. For example, existing research already highlights how different demographics even within the same socio-economic subgroup utilize voice and text search differently. The system also needs to be sensitive to different dialects and accents to function properly and be responsive to cultural and contextual cues that might not be pre-built into the system. Differing levels of digital and technical literacy also play a role in how the system can effectively meet the needs of the user.

As the expectations from the system increase over time, ascribed to their ubiquity and anthropomorphization, we start to see a gulf in expectations and execution. Users are less forgiving of mistakes made by the system and this needs to be accounted for when designing the system so that alternate mechanisms are available for the user to be able to meet their needs.

In conclusion, it is essential when designing voice-activated systems to be sensitive to user expectations, more so than other traditional forms of interaction where expectations are set over the course of several uses of the system whereas with voice systems, the user comes in with a set of expectations that closely mimic how they interact with each other using natural language. Addressing the challenges highlighted in this paper will lead to systems that are better able to delight their users and hence gain higher adoption.

## A Focus on Neural Machine Translation for African Languages

The paper highlights how having more translation capabilities available for languages in the African continent will enable people to access larger swathes of the internet and contribute to scientific knowledge which are predominantly English based.

There are many languages in Africa, South Africa alone has 11 official languages and only a small subset are made available on public tools like Google Translate. In addition, due to the scant nature of research on machine translation for African languages, there remain gaps in understanding the extent of the problem and how they might be addressed most effectively. The problems facing the community are many: low resource availability, low discoverability where language resources are often constrained by institution and country, low reproducibility because of limited sharing of code and data, lack of focus from African society in seeing local languages as primary modes of communication and a lack of public benchmarks which can help compare results of machine translation efforts happening in various places.

The research work presented here aims to address a lot of these challenges. They also give a brief background on the linguistic characteristics of each of the languages that they have covered which gives hints as to why some have been better covered by commercial tools than others.In related work, it is evident that there aren't a lot of studies that have made their code and datasets public which makes comparison difficult with the results as presented in this paper.

Most studies focused on the Autshumato datasets and some relied on government documents as well, others used monolingual datasets as a supplement. The key analysis of all of those studies is that there is a paucity in the focus on Southern African languages and because apart from one study, others didn't make their datasets and code public, the BLEU scores listed were incomparable which further hinders future research efforts.

The Autshumato datasets are parallel, aligned corpora that have governmental text as its source. They are available for English to Afrikaans, isiZulu, N. Sotho, Setswana, and Xitsonga translations and were created to build and facilitate open source translation systems. They have sentence level parallels that have been created both using manual and automatic methods. But, it contains a lot of duplicates which were eliminated in the study done in this paper to avoid leakage between training and testing phases. Despite these eliminations, there remain some issues of low quality, especially for isiZulu where the translations don't line up between source and target sentences.

From a methodological perspective, the authors used ConvS2S and Transformer models without much hyperparameter tuning since the goal of the authors was to provide a benchmark and the tuning is left as future work. Additional details on the libraries, hyperparameter values and dataset processing are provided in the paper along with a GitHub link to the code.

In general the Transformer model outperformed ConvS2S for all the languages, sometimes even by 10 points on the BLEU scores. Performance on the target language depended on both the number of sentences and the morphological typology of the language. Poor quality of data along with small dataset size plays an important role as evidenced in the poor performance on the isiZulu translations where a lowly 3.33 BLEU score was achieved. The morphological complexity of the language also played a role in the state of the performance as compared to other target languages.

For each of the target languages studied, the paper includes some randomly selected sentences to show qualitative results and how different languages having different structures and rules impacts the degree of accuracy and meaning in the translations. There are also some attention visualizations provided in the paper for the different architectures demonstrating both correct and incorrect translations, thus shedding light on potential areas for dataset and model improvements. The paper also shows results from ablation studies that the authors performed on the byte pair encodings (BPE) to analyze the impact on the BLEU scores and they found that for datasets that had smaller number of samples, like for isiZulu, having a smaller number of BPE tokens increased the BLEU scores.

In potential future directions for the work, the authors point out the need for having more data collection and incorporating unsupervised learning, meta learning and zero shot techniques as potential options to provide translations for all 11 official languages in South Africa. This work provides a great starting point for others who want to help preserve languages and improve machine translations for low resources languages. Such efforts are crucial to empower everyone in being able to access and contribute to scientific knowledge of the world. Providing code and data in an open source manner will enable future research to build upon it and we need more such efforts that capture the diversity of human expression through various languages.

# Go Wide: Article Summaries

### Radioactive Data: Tracing Through Training

In modern AI systems, we run into complex data and processing pipelines that have several stages and it becomes challenging to trace the provenance and transformations that have been applied to a particular data point. This research from the Facebook AI Research team proposes a new technique called radioactive data that borrows from medical science where compounds like BaSO4 are injected to get better results in CT scans. This technique applies minor, imperceptible perturbations to images in a dataset by causing shifts within the feature space making them "carriers".

Different from other techniques that rely on poisoning the dataset that harms classifier accuracy, this technique instead is able to detect such changes even when the marking and classification architectures are different. It not only has potential to trace how data points are used in the AI pipeline but also has implications when trying to detect if someone claims not to be using certain images in their dataset but they actually are. The other benefit is that such marking of the images is difficult to undo thus adding resilience to manipulation and providing persistence.

### Using Deep Learning at Scale in Twitter's Timeline

Prior to relevance based timeline, the Twitter newsfeed was ordered in reverse chronological order but now it uses a deep learning model underneath to display the most relevant tweets that are personalized according to the user's interactions on the platform. With the increasing use of Twitter as a source of news for many people, it's a good idea for researchers to gain an understanding of the methodology that is used to determine the relevance of tweets, especially as one looks to curb the spread of disinformation online. The article provides some technical details in terms of the deep learning infrastructure and the choices made by the teams in deploying computationally heavy models which need to be balanced with the expediency of the refresh times for a good experience on the platform. But, what's interesting from an AI ethics perspective are the components that are used to arrive at the ranking which constantly evolves based on the user's interaction with different kinds of content.

The ranked timeline consists of a handful of the tweets that are the most relevant to the user followed by others in reverse chronological order. Additionally, based on the time since one's last visit on the platform, there might be an ICYMI ("in case you missed it") section as well. The key factors in ranking the tweets are their recency, presence of media cards, total interactions, history of engagement with the creat of the tweet, the user's strength of connection with the creator and the user's usage pattern of Twitter. From these factors, one can deduce why filter bubbles and echo chambers form on the platform and where designers and technologists can intervene to make the platform a more holistic experience for users that doesn't create polarizing fractions which can promote the spread of disinformation.

## NeurIPS Requires AI Researchers to Account for Societal Impact and Financial Conflicts of Interest

For the first time, there's a call for the technical community to include a social impact statement from their work which has sparked a debate amongst camps that are arguing to leave such a declaration to experts who study ethics in machine learning and those that see this as a positive step in bridging the gap between the social sciences and the technical domains. Specifically, we see this as a great first step in bringing accountability closer to the origin of the work. Additionally, it would be a great way to build a shared vernacular across the human and technical sciences easing future collaboration.

# Conclusion

We are impressed with all the work that the research and practice community has been doing in the domain of AI ethics. There are many unsolved and open problems that are yet to be addressed but our overwhelming optimism in the power of what diversity and interdisciplinarity can help to achieve makes us believe that there is indeed room for finding novel solutions to the problems that face the development and deployment of AI systems.

We see ourselves as a public square, gathering people from different walks of life who can have meaningful exchanges with each other to create the solutions we need for a better future.

Let's work together and share the mic with those who have lived experiences, lifting up voices that will help us better understand the contexts within which technology resides so that we can truly build something that is ethical, safe, and inclusive for all.

See you here next quarter,

The MAIEI Team

# Original sources